\documentclass[10 pt]{article}
\usepackage{graphicx} 
\graphicspath{{./images/}}
\usepackage[margin=1in]{geometry}
\usepackage{color,soul}
\renewcommand{\arraystretch}{1.5}
\usepackage[backend=biber,maxnames=6,sorting=none,giveninits=true]{biblatex}
\usepackage{amssymb,amsthm, paralist,algpseudocode,bm}
\usepackage{cancel}
\usepackage{algorithm}
\usepackage{url}
\usepackage{upgreek}
\usepackage{fancyvrb}
\usepackage{mathrsfs}
\usepackage{multirow}
\usepackage{hhline}
\usepackage{booktabs}
\usepackage[table]{xcolor}
\usepackage{tikz}
\usepackage{matlab-prettifier}
\usepackage{setspace}
\usepackage[export]{adjustbox}
\usepackage{wrapfig}
\usepackage{listings}
\usepackage{enumitem}
\usepackage{caption}
\usepackage{subcaption}
\usepackage{amsmath}
\usepackage{indentfirst}
\usepackage{mathdots}
\usepackage{authblk}
\usepackage{longtable}
\usepackage{tabularx}
\usepackage{array}
\usepackage[title]{appendix}
\newcolumntype{P}[1]{>{\centering\arraybackslash}p{#1}}

\title{Multi-scale simulation of red blood cell trauma in large-scale high-shear flows after Norwood operation}

\author[1,\thanks{Corresponding author: sm2436@cornell.edu}]{Saba Mansour}
\author[1]{Emily Logan}
\author[2]{James F. Antaki}
\author[1]{Mahdi Esmaily} 
\affil[1]{Sibley School of Mechanical and Aerospace Engineering, Cornell University, Ithaca, NY, USA}
\affil[2]{Meinig School of Biomedical Engineering, Cornell University, Ithaca, NY, USA}

\begin{document}
\maketitle

\begin{abstract} 

Cardiovascular surgeries and mechanical circulatory support devices create non-physiological blood flow conditions that can be detrimental, especially for pediatric patients. A source of complications is mechanical red blood cell (RBC) damage induced by the localized supraphysiological shear fields. To understand such complications, we introduce a multi-scale numerical model to predict the risk of hemolysis in a set of idealized anatomies. We employed our in-house CFD solver coupled with Lagrangian tracking and cell-resolved fluid-structure interaction to measure flow-induced stresses and strains on the RBC membrane. The Norwood procedure, well-known to be associated with high mortality rate, is selected for its importance in the survival of the single-ventricle population. We simulated three anatomies including 2.5mm and 4.0mm diameter modified Blalock-Taussig (BT) shunts and a 2.5mm central shunt (CS), with hundreds of RBCs in each case for statistical analysis. The results show that the conditions created by these surgeries can elongate RBCs by more than two-fold (3.1\% of RBCs for 2.5mm BT shunt, 1.4\% for 4mm BT shunt, and 8.8\% for CS). Shear and areal strain metrics also reveal that the central shunt creates the greatest deformations on the RBCs membrane, indicating it is a more hemolytic procedure in comparison to the BT shunt. Between the two BT shunts, the smaller diameter is slightly more prone to hemolysis. These conclusions are confirmed when strain history and different damage thresholds are considered. The spatial damage maps produced based on these metrics highlighted hot zones that match the clinical images of shunt thrombosis.

\end{abstract}

\section{Introduction}

Hypoplastic left heart syndrome (HLHS) is a critical congenital heart defect where the left side of the heart is underdeveloped. Patients with this malformation suffer from a range of conditions including atresia or stenosis of mitral and aortic valves, and hypoplasia of the ascending aorta, aortic arch, and left ventricle \cite{Noonan1958}. In the United States, more than 1 in every 4000 neonates are born with HLHS \cite{Mai2019}. Infants born with this fatal defect require multiple surgeries for survival. The Norwood (or stage I) procedure is a palliative surgery performed on newborns with HLHS as well as other complex single ventricle defects that share the same physiologic features \cite{Jacobs1995}. This operation is often followed by Glenn (or stage II) and Fontan (or stage III) operations, respectively.

During the Norwood operation, a new aorta is created and connected to the right ventricle, and a shunt connecting the new aorta to pulmonary arteries is inserted to maintain the blood flow to the lungs. Modified Blalock–Taussig (BT) shunt, central shunt, and Sano shunt also called right ventricle to pulmonary artery (RV-PA) shunt, are different shunt configurations used in the stage I procedure. All these configurations are inherently flawed as they create a parallel circulation between systemic and pulmonary circuits that overloads the single ventricle. Furthermore, they produce hypoxia due to the mixture of oxygenated and de-oxygenated blood in the right atrium. Complications such as renal failure \cite{Chamberlain2022}, blockage of the shunt by means of clot formation or thrombosis, cardiac arrest \cite{Rajab2021}, and blood trauma are also common among Norwood patients. Thus, despite improvements since the introduction of this procedure in 1983, the in-hospital mortality rate remains as high as approximately 16\% in the Norwood population \cite{Mascio2019, Murni2019, Meza2023}; which is among the greatest in common congenital heart procedures.
Thus, there exists a pressing need to enhance the survival rate of the Norwood procedure, as the first stage in the treatment of HLHS. 

Some of the mortality in Norwood patients can be attributed to the blockage of the shunt, which is caused by the abnormal flow environment within the shunt. Hemolysis has been also observed secondary to Norwood procedure with modified BT shunt reducing patient's hematocrit from 55\% to 37\% within two weeks after the operation \cite{Ryerson2007}. Accumulating evidence suggests that mechanical RBC damage, i.e., blood trauma, might be the central underlying reason for secondary complications \cite{McNamee2023} such as renal and liver failure, thrombosis, and exacerbated hypoxia. RBC damage and destruction, increase the risk of thrombosis by promoting clotting pathways, e.g.,  von Willebrand factor (vWF) and platelet activation \cite{Faghih2019}. Damage and loss of healthy RBCs also further reduce oxygen delivery. 
Even without complete rupture, and through the pores formed on the membrane of highly deformed RBC, hemoglobin and other RBC contents can be released to plasma. Free hemoglobin is toxic and can lead to acute kidney injury \cite{DVANAJSCAK20191400}. However, these processes are highly complicated to model computationally. As a result, to gain some insight into the effect of shunt flow environment on the blood constituents, we take a preliminary step and look at RBC response to altered flow in this study as it pertains to other mentioned complications. 

Blood is a colloidal suspension of cells immersed in plasma. The cellular nature of blood is the underlying reason for many of its fascinating characteristics. The average hematocrit (Ht), i.e., the volume fraction of RBCs (erythrocytes), is roughly 45\%; White blood cells and platelets combined form around 1\% of blood's volume and the remaining is plasma \cite{DeHaan2018}. Hence, the properties of RBCs, the most numerous of blood cells, dictate the behavior of blood. Red blood cells have a biconcave disk shape with an approximate diameter of $8 \mu m$, and thickness of $2 \mu m$, respectively. When RBCs are exposed to high non-physiologically hemodynamic stresses, such as those created with reconstructive surgeries and in medical devices, they might rupture in a process called hemolysis. This lethal form of blood damage is usually purposefully averted in the recent generation of circulatory assist devices thanks to our understanding of the underlying reasons, i.e., higher than normal magnitudes of and prolonged exposure to shear stress. Nonetheless, circumventing such extreme conditions is not adequate. As even mechanical stresses well below the identified limits for hemolysis, are strong enough to cause structural and functional alterations to RBC, namely sublethal RBC damage \cite{McNamee2023}. 

Unlike hemolysis that can be easily identified by \textit{in vitro} and \textit{in vivo} studies through measuring the concentration of plasma free hemoglobin, detecting sublethal blood trauma is difficult due to the lack of a direct test \cite{Olia2017}. Conversely, computational fluid dynamics (CFD) enable us to not only simulate the global transport of blood flow, but also, microscopic phenomena at the cellular level, such as RBC deformation, aggregation, and coagulation, to gain valuable insight into the challenges associated with blood-wetted devices and cardiovascular surgeries. In light of this, cell-resolved computational techniques have been developed and used to replicate blood cell dynamics, albeit for small-scale (sub-millimeter) simulations. Dissipative particle dynamics (DPD) \cite{Fedosov2010,Liu2024}, lattice Boltzmann method (LBM), and smoothed particle hydrodynamics (SPH) \cite{Hosseini2012} are particle-based methods \cite{Tsubota2006} generally used for simulation of RBCs. For instance, Javadi et al. \cite{Javadi2021} utilized DPD to study the effect of different biophysical factors on blood's viscosity in a cubic simulation box of edge size equal to $50 \mu m$, where they were able to simulate up to $Ht = 55\%$. Takeishi et al. \cite{Takeishi2019} numerically analyzed the rheology of a suspension of RBCs in blood plasma for wall-bounded shear flows using LBM coupled with IBM. For a domain size of $\approx 10 \mu m$ in each direction, they solved for volume fractions up to $41\%$. 

As opposed to the meshless particle-based methods that have to fill the whole space with particles, immersed boundary methods (IBM) benefit from a combination of CFD grids and a collection of immersed particles representing an immersed boundary; which have been used for the simulation of blood cells and platelets and understanding their behavior \cite{Yazdani2011, Kassen2022}. On the other hand, boundary integral methods are another type of continuum models that only solve for the velocity of a surface such as a blood cell membrane. Once the membrane velocity is obtained, one can calculate the solution at any point inside or outside of the closed membrane, if required. Assumptions of thin membrane and Stokes flow in the vicinity of blood cells permit the use of boundary element methods which is considerably cheaper compared to other methods. This method, which is adopted here, has been used in several earlier studies \cite{Ramanujan1998, Pozrikidis2003, Zhao2010, Sinha2015}. For example, Pozrikidis simulated the flow-induced deformation of RBC under simple shear flow \cite{Pozrikidis2003}. In another study, Zhao et al. \cite{Zhao2010} utilized the same method to capture the deformation of RBCs in complex geometries with spectral accuracy facilitated by the use of spherical harmonics. Also, different modes of RBC motion have been well studied with the use of boundary element methods \cite{Sinha2015}.

Fully coupled cell-resolved simulation of RBC deformation and damage in organ-scale flow is prohibitively expensive due to the orders of magnitude difference between the size of RBCs and arteries. It is for this reason that blood cell damage in large-scale flows has been historically predicted using stress-based models in conjunction with empirical equations (e.g., \cite{Li2022, Giersiepen1990}). 
As opposed to stress-based models that rely only on instantaneous stresses of flow, strain-based models coupled with Lagrangian particle tracking methods account for RBC motion and deformation to give a rather accurate prediction of blood damage in large-scale flows. Arora et al. \cite{Arora2004} first proposed the idea of accounting for RBC deformation along their path lines to estimate RBC damage through a tensor-based, i.e., strain-based, model. They simulated RBCs as ellipsoidal liquid droplets and employed their method to study blood trauma in a 2D pump. More recently, the same idea was utilized and improved to resolve the shape of a handful of blood cells in 3D geometries such as a narrow neck \cite{Zhao2010} and mechanical prostheses heart valve \cite{Ezzeldin2015}, and in 3D turbulent flows \cite{Rydquist2024}. Moreover, reduced-order models of RBCs \cite{Porcaro2023,Porcaro2024} have been recently developed to approximately obtain RBC deformation at a lower cost compared to cell-resolved models. However, these methods cannot fully replicate the intricate cellular-level physics.

To avoid the high cost of direct numerical simulation of RBCs in large-scale flows, we use a multi-scale cell-resolved Lagrangian model to produce cell-scale accurate measures of RBC deformation in large-scale flow. The results of such simulation provide insight into mechanical load on cells that can be used for relative comparison of different geometries \cite{Gusenbauer2018,Porcaro2023}. The goal of this study, thus, is to obtain a reliable statistical description of the cell-scale mechanical behavior of red blood cells in different designs for the Norwood procedure. For this purpose, we consider modified BT shunt (mBTS) and central shunt (CS) and compare them in terms of various mechanical factors that are linked to sublethal blood trauma. 
It should be noted that the overall computational framework that is utilized here is generic and can be applied to other cardiovascular surgeries and devices to compare them in terms of in-bulk RBC damage due to medium flow-induced stresses. Moreover, the methodology is designed with special attention paid to limiting the overall computational time which would be beneficial for individual risk assessment by building digital twins.

The paper is organized as follows. The second section covers the description and validation of the numerical methods used for the simulation of the flow field, tracking of red blood cells, and blood cell membrane dynamics. The third section is dedicated to the independence studies of numerical parameters along with the choice of material properties. The main results and their interpretation are presented in the fourth section. Finally, the main conclusions of this study are reported in the fifth section, respectively.


\section{Methods}

\begin{figure} [ht!]
\centering
\includegraphics[width=1\linewidth]{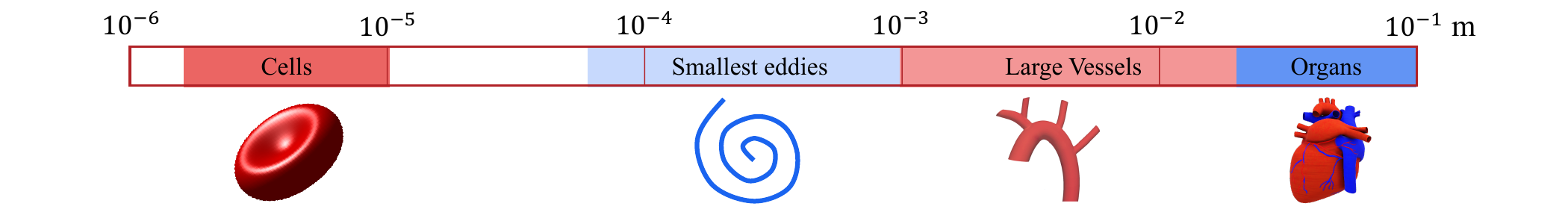}
\vspace{-8mm}
\caption{Scale separation}
\label{fig:Scale}
\end{figure}

The present section introduces the modeling approach, procedures, and governing equations. Moreover, the validation of the new components of the solver is covered in this section.
Direct simulation of RBCs in organ-scale flows is prohibitively expensive due to multiple order separation of scales between blood cells and organs as observed in Fig. \ref{fig:Scale}.

Hence, a multi-scale cell-resolved approach, introduced in a previous work \cite{Rydquist2022}, is adapted here, where a one-way coupled approach is used between RBCs and surrounding flow. This means that the effect of flow on RBCs is simulated while the effect of individual RBCs on macro-scale flow is taken into account using effective medium properties, representing whole blood. The accuracy of this modeling approach has been demonstrated in an earlier study, which showed simulating a single cell using an effective viscosity accurately reproduces results obtained from simulation of whole blood, thereby effectively capturing the RBCs interactions through an adjustment of plasma viscosity \cite{Rydquist2023}. The RBCs are much smaller than the characteristic length of large-scale flow and almost immediately relax to the velocity of surrounding flow due to their short relaxation time. Thus, the far-field boundary condition imposed at the cell-scale is obtained from the macro-scale flow field according to
\begin{equation}
    \bm{u}^\infty(\bm{x},t) = \bm{U}(\bm{x}_c,t) + (\bm{x}-\bm{x}_c)\cdot\nabla \bm{u}^\infty(\bm{x}_c,t),
    \label{eqn:decompose}
\end{equation}
where $\bm{u}^\infty$ is the far field fluid velocity, $\bm{U}$ is the fluid velocity at the location of RBC (i.e., its center point), $\nabla \bm{u}$ is the velocity gradient, and $\bm{x}_c$ is the cell location. 
This means that we assume the flow curvature, before disturbance introduced by the RBC, is negligible at the length scale of a RBC. Hence, RBC follows the flow with uniform velocity, while the velocity gradient part of the flow field deforms the cell. 

\begin{figure} [ht!]
\centering
\includegraphics[width=1\linewidth]{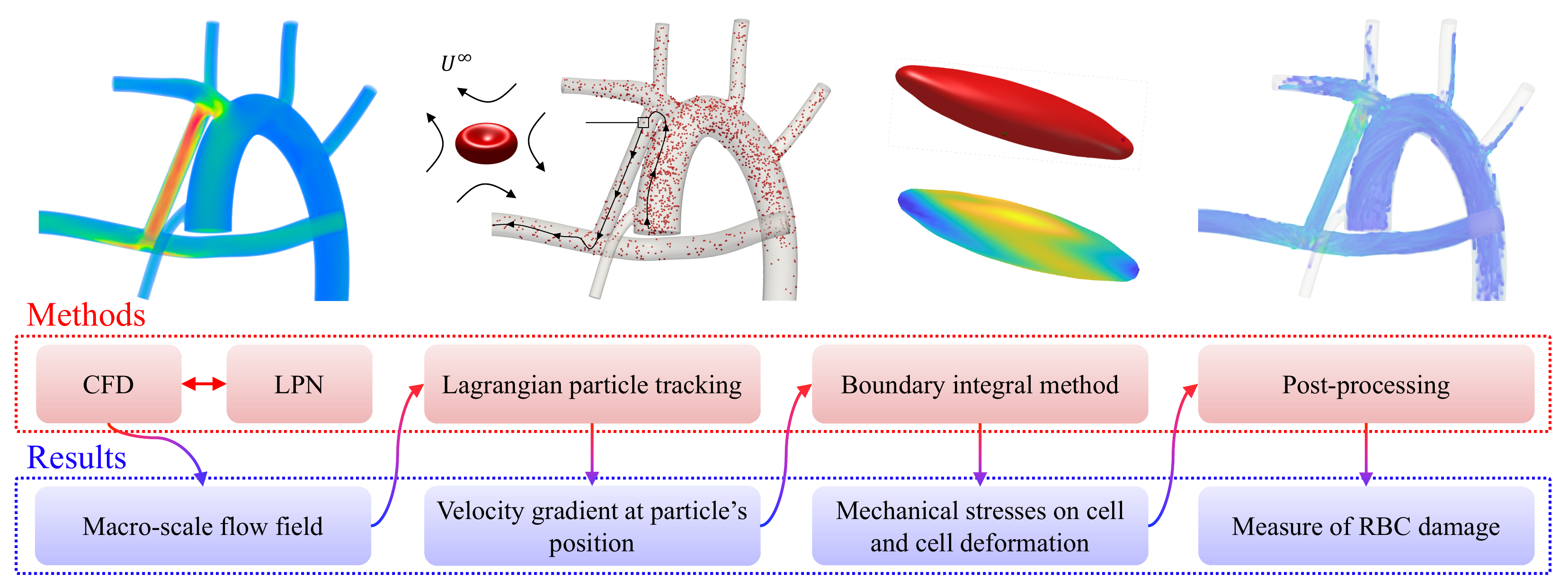}
\caption{Methodological summary and outcomes of each stage}
\label{fig:FlowChart}
\end{figure}

From a computational standpoint, three separate solvers are utilized in this study for the multi-scale cell-resolved simulation of RBCs. First, CFD simulation of large-scale flow is done with the help of a lumped parameter network (LPN). This is to resolve blood flow behavior at the anatomical level. Second, a Lagrangian particle tracking algorithm specifically designed for fast simulation of particles in time-periodic flows is introduced and used to track red blood cells and acquire velocity gradient data along their trajectories. Last but not least, the deformation of the membrane of red blood cells due to the fluid-structure interaction (FSI) with the fluid inside and outside of the membrane is captured using a boundary integral method (BIM). The flow chart of all utilized methods, data input/output, and a simplified visualization of the outcomes of this multi-scale approach which is the deformation of red blood cells tracked through an organ-scale flow are illustrated in Fig. \ref{fig:FlowChart}. 
In the following sections, we discuss each of these three components in detail.

\subsection{CFD Simulations}
To model blood flow, we solve the incompressible Navier-Stokes equations that are given by
\begin{equation}
    \nabla \cdot \bm{u} = 0,
\label{eqn:cont}
\end{equation}
\begin{equation}
    \rho (\dot{\bm{u}} + \bm{u} \cdot \nabla \bm{u}) = - \nabla p + \nabla \cdot \bm{\tau}, \qquad \forall \bm{x} \in \Omega,
\label{eqn:NS}
\end{equation}
\begin{equation}
    \bm{\tau} = \mu (\nabla \bm{u} + \nabla \bm{u} ^ T),
\label{eqn:Tau}
\end{equation}
\begin{equation}
    \bm{u} = \bm{g}, \qquad \bm{x} \in \Gamma_g,
\label{eqn:BCDir}
\end{equation}
\begin{equation}
    \bm{\tau}\cdot\bm{n} - p \bm{n} = \bm{h}, \qquad \bm{x} \in \Gamma_h,
\label{eqn:BCNeu}
\end{equation}
where $\bm{u}(\bm{x},t)$, $\dot{\bm{u}}(\bm{x},t)$, $\rho$, $p(\bm{x},t)$, $\bm{\tau}(\bm{x},t)$, and $\bm{n}$ are velocity vector, time derivative of velocity vector, fluid density, pressure, stress tensor, and wall-normal vector, respectively. They are a function of position $\bm x$ and time $t$. $\Omega$ is the computational domain. Equations \ref{eqn:BCDir} and \ref{eqn:BCNeu}, represent Dirichlet and Neumann boundary conditions (BCs) applied to corresponding boundaries, i.e., $\Gamma_g$ and $\Gamma_h$, respectively.

Equations 2-6 are solved using a verified in-house multi-physics finite element solver (MUPFES) \cite{EsmailyMoghadam2013,Esmaily-Moghadam2013}.
This solver utilizes upwinding and pressure-stabilization techniques to ensure the stability of result in strong convection and permit the use of equal order shape functions for velocity and pressure \cite{BROOKS1982,Jia2023}. 
A specialized iterative algorithm, preconditioner, and parallelization strategy are implemented for an efficient and scalable solution of the linear system of equations \cite{Esmaily-Moghadam2013,Esmaily_Partition2015}.
The solver uses an implicit generalized-$\alpha$ time integration scheme \cite{JANSEN2000305}, where at each time step, several Newton–Raphson iterations are performed to drop the residual by over four orders of magnitude. At each Newton–Raphson iteration, a linear system is solved using the generalized minimal residual (GMRES) method \cite{SaadGMRES}. This solver is parallelized using a message-passing interface (MPI) using domain partitioning. All computations are performed on a cluster of AMD Opteron$^{TM}$ 6378 processors that are interconnected via a QDR Infiniband.

\begin{figure} [ht!]
\centering
\tabskip=0pt
\valign{#\cr
  \hbox{%
    \begin{subfigure}[b]{.65\textwidth}
    \centering
\includegraphics[width=\textwidth]{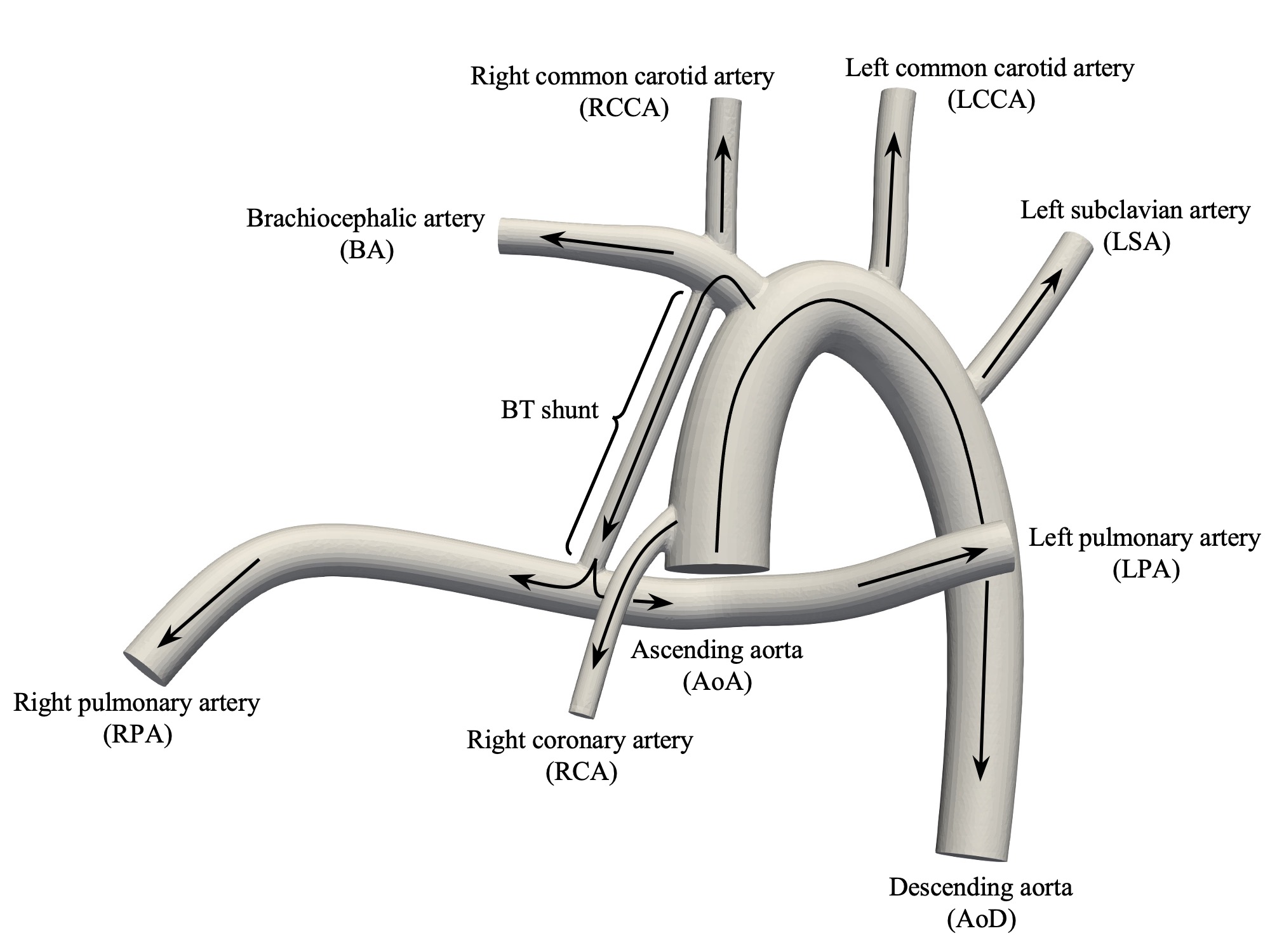}
    \caption{2.5mm modified BT shunt}
    \end{subfigure}%
  }\cr
  \noalign{\hfill}
  \hbox{%
    \begin{subfigure}{.28\textwidth}
    \centering
\includegraphics[width=\textwidth]{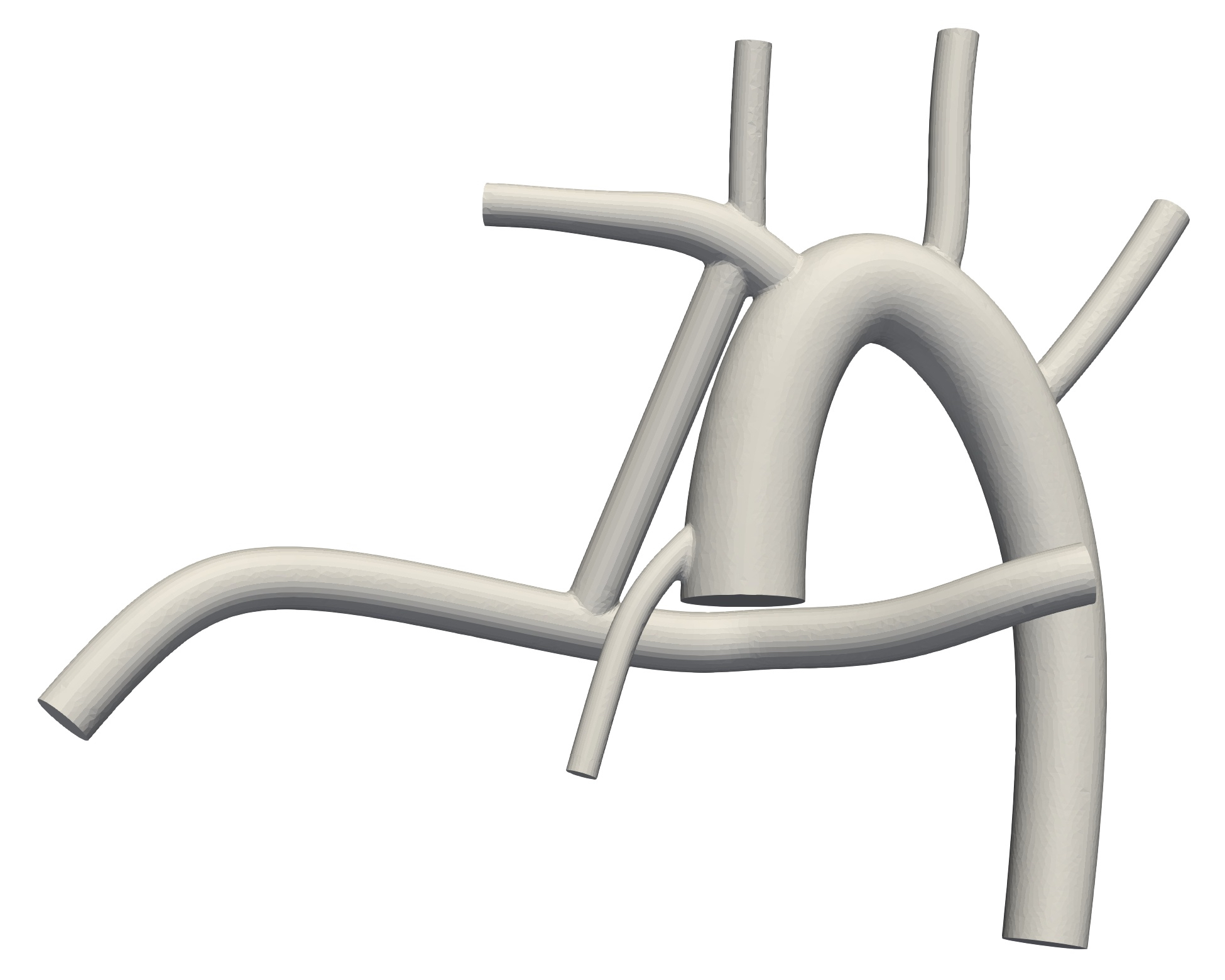}
    \caption{4.0mm modified BT shunt}
    \end{subfigure}%
  }\vfill
  \hbox{%
    \begin{subfigure}{.28\textwidth}
    \centering
\includegraphics[width=\textwidth]{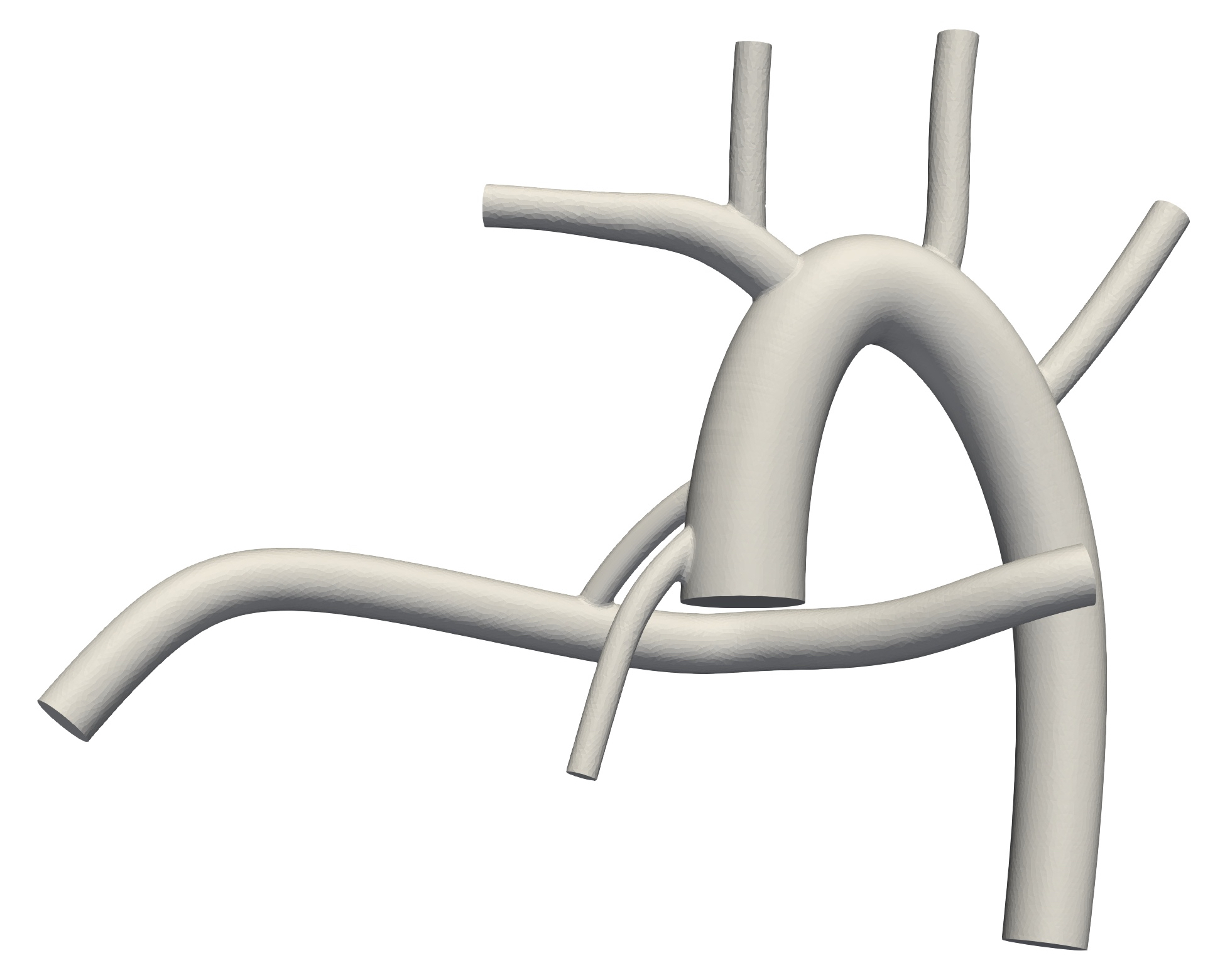}
    \caption{2.5 mm central shunt}
    \end{subfigure}%
  }\cr
}
\caption{The three post-Norwood anatomies adopted for blood flow simulation}
\label{fig:LabeledBT}
\end{figure}

The idealized geometries adopted for macro-scale blood flow simulations are based on earlier studies involving Norwood operation \cite{Jia2021,Esmaily-Moghadam2015,Moghadam2012}. The adopted anatomy and direction of blood flow in various branches are shown in Fig. \ref{fig:LabeledBT}(a). Briefly, blood flow coming from the heart enters the circulatory system from ascending aorta (AoA). This flow goes toward the lower body through the descending aorta (AoD), and toward the upper body through the brachiocephalic artery (BA), right common carotid artery (RCCA), left common carotid artery (LCCA), and left subclavian artery (LSA). Flows from the upper and lower parts of the body will return to the heart before reentering AoA. The insertion of BT or central shunt, allows blood flow to go toward the lungs through the right pulmonary artery (RPA) and left pulmonary artery (LPA), before returning to the heart. The right coronary artery (RCA) is responsible for supplying blood to the heart muscles. The three simulated geometries, including a 2.5mm modified BT shunt (2.5BT), a 4.0mm modified BT shunt (4.0BT), and a 2.5mm central shunt (2.5CS), respectively, are shown in Figure \ref{fig:LabeledBT}(a)-(c).

\subsubsection{Boundary Conditions}

The effect of the portion of the circulatory system not included in 3D models shown in Figure \ref{fig:LabeledBT} is captured through a 0D lumped parameter network, the details of which are included in Appendix \ref{appendix:LPN}. This LPN is based on previous studies \cite{Migliavacca2001,LAGANA20051129,Moghadam2012,Jia2021}. Hence, readers are referred to those articles for detailed explanations of the LPN formulation and 0D-3D coupling strategies. In brief, to capture the complex response of the circulatory system, a system of ordinary differential equations (ODEs), is solved in the 0D domain to compute the pressure and flow rate for the Neumann and Dirichlet BCs in the 3D domain. The components appearing in the LPN, using an electrical circuit analogy (taking flow and pressure as current and voltage, respectively), are resistors, capacitors, inductors, and diodes to recreate physiological behaviors such as resistance to flow, vessel wall distensibility, blood flow inertia, and heart valves, respectively. We use both Dirichlet and Neumann BCs at the 0D/3D interface. The Dirichlet BC, which is imposed on AoA, requires passing pressure data to the 0D solver in exchange for receiving flow rate data. In contrast, the Neumann BC, which is imposed on remaining branches, requires passing the flow rate to the 0D solver in exchange for the pressure. The values of different parameters of this LPN are reported in Appendix \ref{appendix:LPN}. These values recreate the working condition of a newborn's heart after the Norwood operation with 120 beats per minute.

\subsubsection{Adaptive mesh refinement}
\begin{figure} [ht!]
\centering
\begin{subfigure}{.3\textwidth}
  \centering
  \includegraphics[width=1\linewidth]{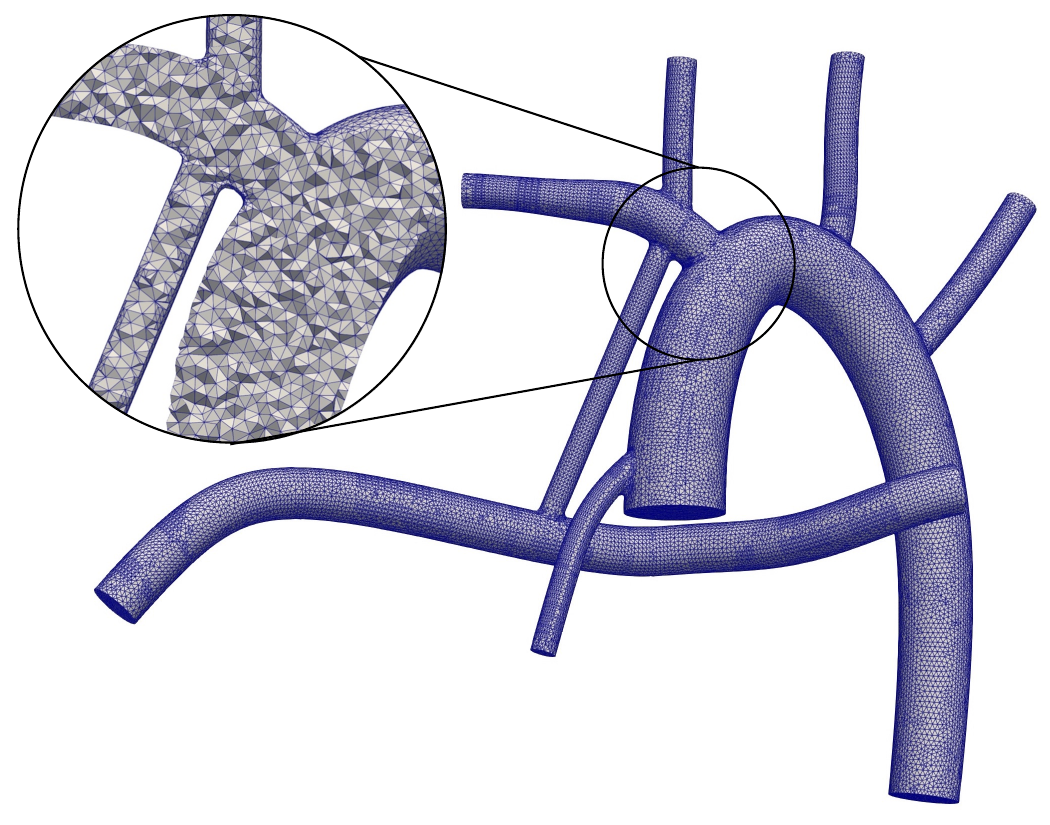}
  \caption{1\textsuperscript{st} iteration}
  \label{fig_mr:sub1b}
\end{subfigure}%
\begin{subfigure}{.3\textwidth}
  \centering
  \includegraphics[width=1\linewidth]{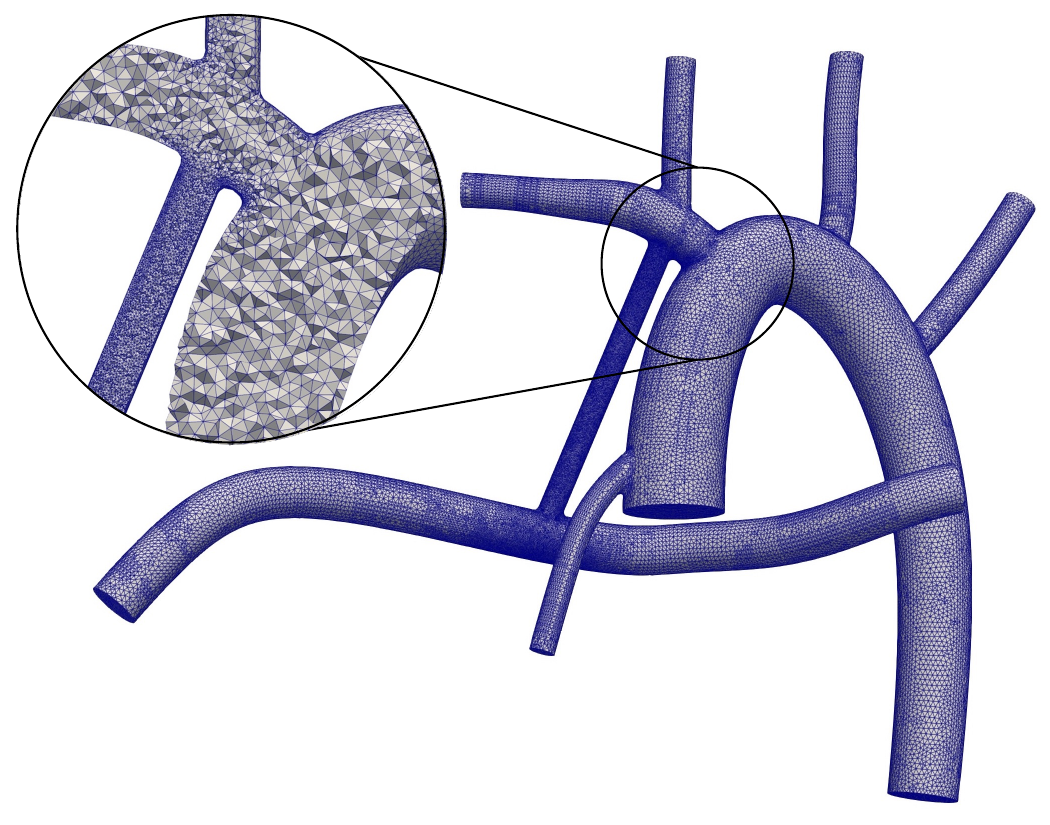}
  \caption{3\textsuperscript{rd} iteration}
  \label{fig_mr:sub2b}
\end{subfigure}%
\begin{subfigure}{.3\textwidth}
  \centering
  \includegraphics[width=1\linewidth]{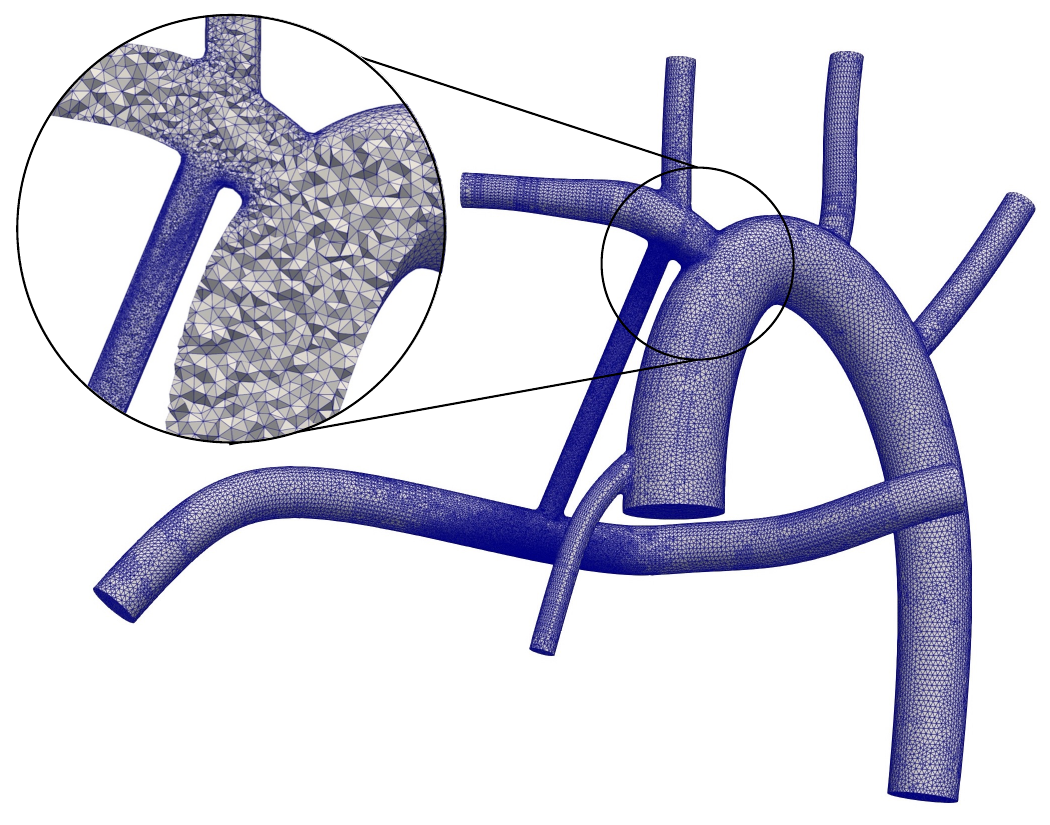}
  \caption{5\textsuperscript{th} iteration}
  \label{fig_mr:sub3b}
\end{subfigure}%
\caption{Adaptive mesh refinement}
\label{fig:3meshrefine}
\end{figure}

We use tetrahedral elements for discretizing the solution (both velocity and pressure) in space. 
To ensure grid-independency of our results while keeping these computations affordable, we perform successive mesh adaptation. In total, five grids with descending average element sizes are generated and compared for each configuration. The mesh adaptation process is guided by the $L_2$-norm of error, which for linear shape functions used in this study, is proportional to \cite{Jia2021}
\begin{equation}
    E  \propto h^2 \| \nabla^2 \bm{u} \|_{L_2},
    \label{eqn:meshadapt}
\end{equation}
 where $h$ is the element size. This criterion produces a finer grid in areas of flow with a high velocity curvature, such as the shunt. Figure \ref{fig:3meshrefine} illustrates iterations 1, i.e., base mesh, 3, and 5 of the five-step adaptive mesh refinement and their magnified cross sections, for the $2.5$mm mBTS, where the first, i.e., uniform, mesh is generated using TetGen \cite{Si2010}. The number of elements varies from around 250,000 to 3,300,000 going from the base mesh to the final mesh. Detailed information on the generated grids for the mesh independence study is available in Section \ref{meshstudy}.
 Note that the mesh refinement is done for steady-state cases with BCs obtained from LPN at a specific time in the cardiac cycle.

\subsection{Particle tracking}

Lagrangian particle tracking is used to advance particles due to its superiority over Eulerian methods in predicting the transient transport of particles \cite{Zhang2007,Sutera1975}. Differently put, predicting RBC damage based on average shear rate or shear stresses alone is not sufficient, and the flow field imposed on cells should be incorporated to acquire reliable data.
Since the relaxation time of RBCs is much shorter than that of the flow, they are modeled as mass-less particles as they flow through the simulated anatomies shown in Fig. \ref{fig:LabeledBT}. 
That entails solving 
\begin{equation}
    \frac{d\bm{x}_p}{dt} = \bm{u}(\bm{x}_p,t),
    \label{eqn:particleadvance}
\end{equation}
where $\bm{x}_p$ is the particle's position and $\bm{u}(\bm{x}_p,t)$ is the velocity of the fluid at the position of the particle at a given time. 
Provided that flow is stored on an unstructured grid, finding $\bm{u}(\bm{x}_p,t)$ can become computationally demanding, especially for a large number of elements and particles.
In a brute-force approach, the cost of finding $\bm{u}(\bm{x}_p,t)$ scales as $\mathcal{O}(N_{p,tot}N_{el})$ with $N_{p,tot}$ and $N_{el}$ being the total number of particles and number of CFD grid elements.
To reduce this particle tracking cost,
 we adopt a search-box algorithm, which is explained in previous work in detail \cite{Rydquist2020}.
In brief, at the beginning of the particle tracking simulation, a coarse structured Cartesian grid, referred to as search boxes, is built on top of the CFD mesh, and elements overlapping with each search box are identified. After this initial step, at each time step, the element hosting a specific particle is found by searching only through the elements associated with the easy-to-identify associated search box. 
Note that as explained in \cite{Rydquist2020}, there exists an optimal number of search boxes based on factors such as number of particles ($N_p$), number of elements ($N_e$), number of time steps ($N_t$), element type, etc. that is adopted in this study.

A second factor influencing the cost of particle tracking is the unsteady nature of the flow. 
Through the cardiac cycle, heart muscles relax and contract periodically during the diastole and systole phases, respectively, which creates a time-periodic flow inside the veins.
Therefore, particle tracking is done as a post-process step by incorporating the periodicity of the flow into the solver which further reduces the computational cost of the particle tracking algorithm.
Instead of simultaneously solving for the flow field and particles during multiple cardiac cycles, the flow field can be obtained for a single cycle before starting the particle tracking algorithm. Once the flow field is obtained, we can use the generated data to track particles going through our desired geometry for multiple cardiac cycles and simulation instances without having to solve for the flow field over and over again.

In practice, CFD results are first saved, albeit every few time steps. The gap between time steps is selected to preserve temporal fluctuations in the flow field. This way we avoid reading too many large data files, which is highly time and memory-consuming. To map this coarse sampling of data to finer time gaps, we use the fast Fourier transform (FFT). 
We verify that the original CFD results are preserved with a downsampling rate of two, with a maximum error of less than 2\%, which is adopted in this study.

The validation of the implemented particle tracking algorithm is done using a Hagen–Poiseuille flow, for which the analytical solution is known. Particles' velocity obtained by the solver is compared against the analytical solution; The maximum error among 10000 particles for this case is $0.1\%$, proving the correctness of the implemented algorithm.

\subsubsection{Random seeding of particles}
\begin{figure} [ht!]
\centering
\begin{subfigure}{.005\textwidth}
\begin{tikzpicture}
\node[anchor=south west] at (0,0) {};
\node[overlay] at (0pt,110pt) {(a)};
\end{tikzpicture}
\end{subfigure}
\begin{subfigure}{0.48\textwidth}
  \centering
  \includegraphics[width=1\linewidth]{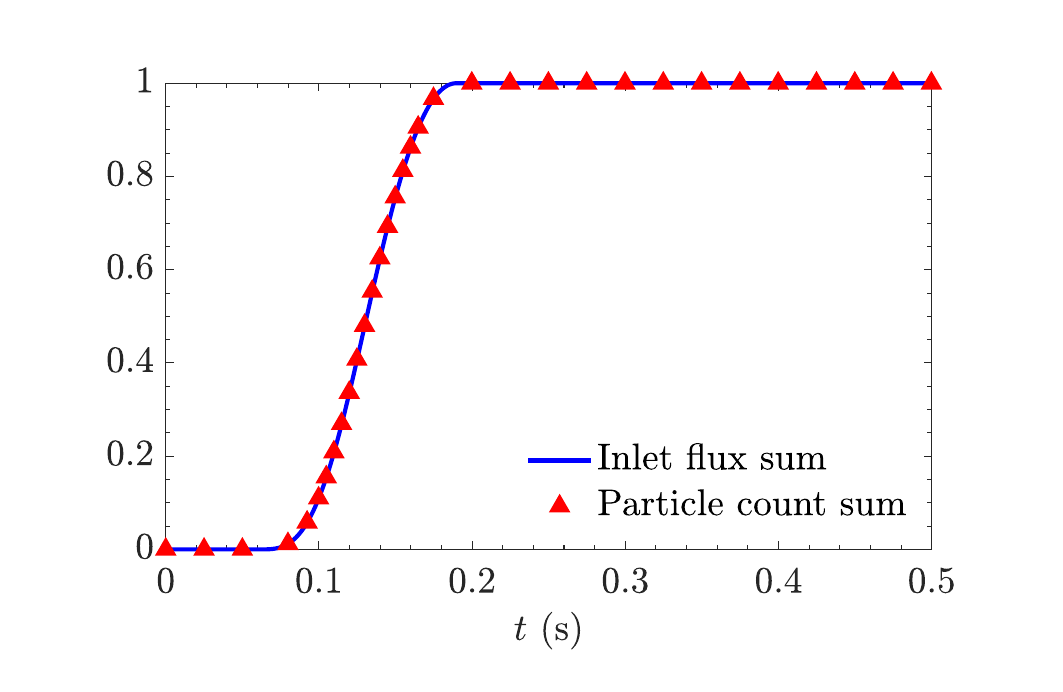}
\vspace*{-8mm}
\label{fig:PrtsPerStep}
\end{subfigure}
\begin{subfigure}{.005\textwidth}
\begin{tikzpicture}
\node[anchor=south west] at (0,0) {};
\node[overlay] at (0pt,110pt) {(b)};
\end{tikzpicture}
\end{subfigure}
\begin{subfigure}{.48\textwidth}
  \centering
  \includegraphics[width=0.95\linewidth]{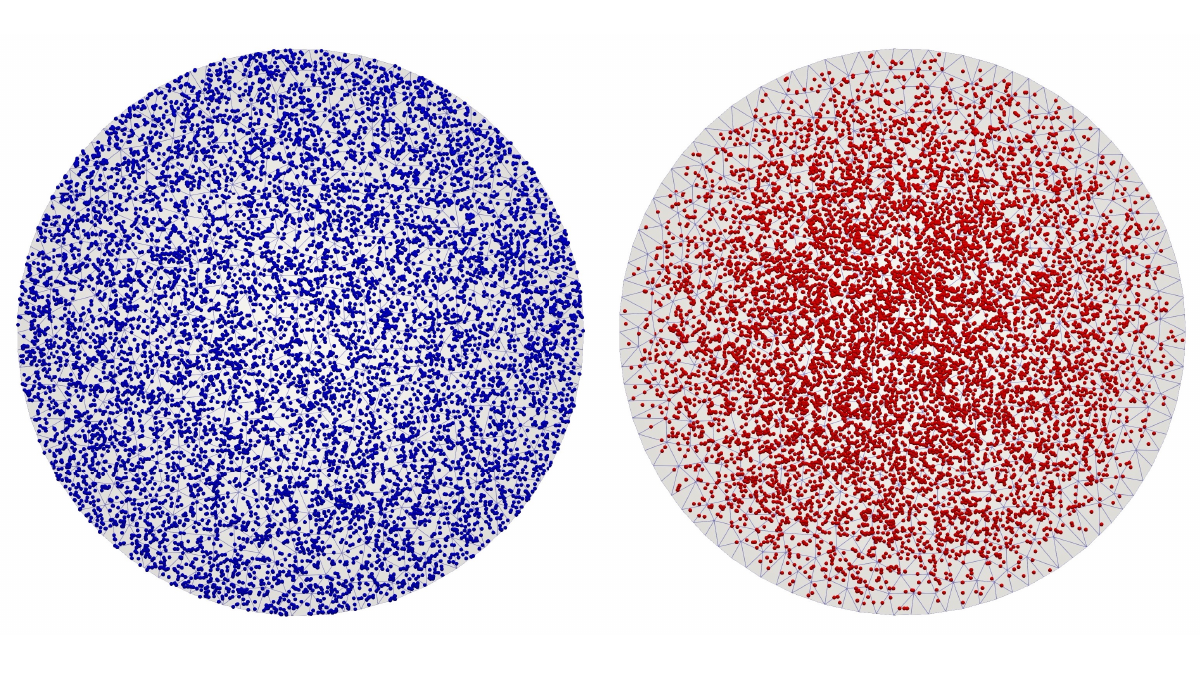}
  \label{fig_rel:sub2}
\end{subfigure}
\caption{Random seeding of particles at the inlet; (a) Accumulative normalized particle count and inlet flux for one cardiac cycle, and (b) Uniform distribution (shown with blue) versus distribution with local fluid velocity effect (shown with red)}
\label{fig:Release}
\end{figure}
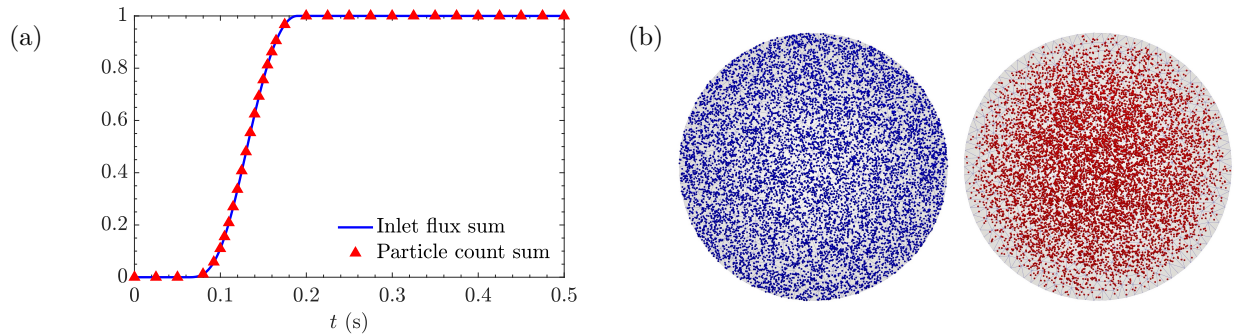

Particles are introduced into the computational domain through the AoA inlet via a random seeding algorithm that respects the local fluid velocity. Namely, this algorithm seeds more particles per unit time in regions of the inlet where the flow is entering the domain faster. This adjustment is to ensure the overall particle volume density remains constant, which must hold given that RBC number density (hematocrit) is uniform in the major blood vessels excluding the micron-size layer near the lumen.  Computationally, this is accomplished by distributing the total number of particles released at a cardiac cycle among different time steps according to their volumetric flux, i.e., 
\begin{equation}
    N_p(t_i) =  \frac{N_{p,tot}}{V(t_{cycle})} \left[ V(t_{i+1}) - V(t_i) \right],
    \label{eqn:n_particle_dt}
\end{equation}
\begin{equation}
    V(t_i) = \int_{t_0}^{t_i} \int_{\Gamma_{inlet}} \bm{u}\cdot\bm{n} \textit{ }  d\Gamma  \textit{ }  dt,
\label{eqn:V_dt}
\end{equation}
where $N_{p,tot}$ is the prescribed total number of particles to be released in a cardiac cycle, $V(t_i)$ is the volume of flow entering the inlet (AoA) from the beginning of the cardiac cycle until time step $t_i$. Note that $N_p(t_i)$ will be set to zero during a bulk backflow when $V(t_{i+1}) - V(t_i) < 0$. For instance, Fig \ref{fig:Release}(a) compares a normalized summation of released particles, i.e., the fraction of particles released so far over the requested number of particles, using Eq. \ref{eqn:n_particle_dt} with the sum of inlet flux at AoA normalized by the cardiac input associated with one cycle. The sum of released particles at all time steps in this case is 5000 particles.

Furthermore, to randomly locate particles on the inlet at each time step, a probability proportional to the fluid velocity is incorporated. Figure \ref{fig:Release}(b) shows the random positioning of 10000 particles with and without consideration of the volumetric flow rate for the inlet of a pipe flow. Compared to a random release of particles without consideration of fluid velocity, this criteria further maintains a fixed Ht level. Differently put, if at each time step, particles are released with a uniform distribution, after some time, the particles will become over-populated near the vessel walls where the velocity is smaller.

\subsubsection{Particle wall collision}
To prevent particles from passing through the wall during Lagrangian particle tracking, it is important to first identify such instances. During particle tracking, the element containing a particle called the host element, will be identified for obtaining local fluid velocity.
If a particle's host element cannot be located, it means that the particle is outside the domain. In such instances, it is crucial to determine whether the particle exited the domain through the wall or an outlet (in practice we only check the outlets since on aggregate they contain far fewer elements than the wall).
In cases where a particle exits through the wall due to numerical error, it is returned back to the computational domain through the following process.

For a particle passing the wall, the boundary element on the wall that the particle passed through is located. 
This is efficiently done by finding the closest boundary node to the particle and identifying the immediate neighbors of that vertex, i.e., elements having that node as a vertex. This is followed by checking for intersections between those elements and particle trajectory from the previous time step to the current one. 
If the mentioned line does not intersect with any of the immediate boundary elements, a second layer of elements, i.e., elements that share a vertex with first-layer elements, can be checked. Having said that, in the simulations performed in this study, the intersecting element is always found within the first layer, due to the choice of tracking time step size.
Once the element is identified, the particle, instead of exiting through the wall, is reflected back into the domain. The angle of reflection is set equal to the angle of incidence and the total displacement of the particle is kept the same.

\subsection{Red blood cell dynamics}

The red cells are formed from a membrane that encapsulates the cytoplasm and hemoglobin. RBCs can survive in the blood circulation system and pass through narrow capillaries due to the durable and flexible membrane structure consisting of the lipid bilayer and the protein cytoskeleton \cite{Lux2016}. This section focuses on the equations used for the simulation of RBC membrane motion as a result of stresses applied by the surrounding flow. 
For the sake of brevity, the utilized methodology for simulation of RBC dynamics that is thoroughly described in \cite{Rydquist2022}, is only briefly explained here. 
This solver is validated and used in a series of earlier studies \cite{Rydquist2022,Rydquist2023,Rydquist2024}. 

The RBC membrane is thin compared to its size and thus is assumed to be a continuum 2D surface in 3D. Since the cell has a smooth topology, spherical harmonic basis functions are employed to parameterize the membrane's surface, yielding spectral accuracy. The hydrodynamic traction jump over the membrane is composed of the traction jumps caused by in-plane tension and out-of-plane bending. This traction jump ($\Delta \bm{f}$) should be in equilibrium with the internal membrane load ($\bm{p}$) since the membrane inertia is assumed to be negligible \cite{Pozrikidis2001},
\begin{equation}
    \Delta \bm{f} = \Delta \bm{f}_{tension} + \Delta \bm{f}_{bending} = - \bm{p}.
    \label{eqn:traction}
\end{equation}
Choosing proper constitutive models plays an integral role in the proper simulation of RBC mechanics as studied by Barthes-Biesel et al. \cite{Barthes-Biesel2002}. They compared Hooke, 
Mooney–Rivlin, and Skalak's laws for the simulation of RBC membrane mechanics, and showed that unlike small deformations, for which all laws produce the same result, large deformations are highly dependent on the utilized constitutive law. The Skalak model \cite{Skalak1973}, particularly developed for RBCs and accounting for their strain-hardening behavior, is utilized in this work to reflect the tension. Results obtained by this model are shown to match with the experimental results from ektacytometry and optical tweezers experiments \cite{Sinha2015,Rydquist2022}. The introduced strain energy function of the RBC membrane is
\begin{equation}
    W_T = \frac{G}{4}\left[(I_1^2+2I_1-2I_2)+CI_2^2\right],
    \label{eqn:skalak}
\end{equation}
where $G$ and $C$ are shear elastic modulus and dilatation ratio. $I_1$ and $I_2$ are strain invariants defined as
\begin{equation}
    \begin{array}{c}
    I_1 = \lambda_1^2 + \lambda_2^2 - 2, \\
    I_2 = \lambda_1^2 \lambda_2^2 -1,
    \end{array}
    \label{eqn:invarient}
\end{equation}
with $\lambda_1$ and $\lambda_2$ being the principal strains. $\lambda_1$ and $\lambda_2$ are representatives of the ratios associated with the dimensions of the local area element in the current, i.e., deformed state, to its reference, i.e., undeformed state. Anti-aliasing of tension calculations is done by using a finer grid for nonlinear operations and inverse spherical harmonic transform to interpolate calculated values to the coarse grid. Furthermore, the resistance to bending is represented using the Helfrich bending model \cite{Helfrich1973}. This model describes the bending energy of the membrane based on the bending modulus, $E_B$, the mean curvature, $\kappa$, and the spontaneous curvature $c_0$ as 
\begin{equation}
    W_B = \frac{E_B}{2} \int(2\kappa- c_0)^2 dS.
    \label{eqn:helfrich}
\end{equation}

The fluid-structure interaction of membrane mechanics with the surrounding flow field is captured using the boundary integral method. The fluid velocity on the membrane's surface of the RBC is solved under the assumption of Stokes flow at the cell-scale owing to the fact that Reynolds numbers corresponding to flow around RBCs are significantly smaller than one. The flow outside of the particle is then derived using the divergence and reciprocal theorems to integrate Green's functions arising from the Stokes equations, with a point force having the strength $\bm{g}$,
\begin{equation}
    \begin{array}{c}
    -\nabla P + \mu \nabla^2 \bm{u} + \bm{g} \delta (\bm{x} - \bm{x_0}) = 0,\\
    \nabla\cdot\bm{u} = 0.
    \end{array}
    \label{eqn:stokes}
\end{equation}

Combining with the equations corresponding to interior fluid, the velocity at point $\bm{x_0}$ on the membrane will satisfy \cite{Pozrikidis1992}
\begin{align}
\begin{split}
    u_j(\bm{x_0}) & = \frac{2}{1+\lambda}
    u^{\infty}_j(\bm{x_0}) - \frac{1}{4\pi\mu (1+\lambda)} \int_S \Delta f_i(\bm{x}) G_{ij}(\bm{x},\bm{x_0}) dS(\bm{x}) \\   
    & + \frac{1-\lambda}{4\pi(1+\lambda)} \int_S u_i(\bm{x}) T_{ijk}(\bm{x},\bm{x_0}) n_k(\bm{x}) dS(\bm{x}),
\end{split}
    \label{eqn:BIM}
\end{align}
where $\lambda$ in the viscosity ratio of inside to surrounding fluids, $G_{ij}$ is the Stokes flow Green's function for the velocity, and $T_{ijk}$ is for the stress.

\section{Independence studies and utilized parameters}
To ensure the validity of the results, in this section, we explore their dependence on various numerical parameters that do not have a physical significance. That includes but is not limited to mesh independence and cycle-to-cycle convergence for the flow solver, time step size independence and particle count independence for the particle tracking solver, and time step size independence for the BIM solver. Additionally, all the parameters used to simulate RBC deformation are discussed in this section.

\subsection{Flow solver}
\subsubsection{Mesh independence study} \label{meshstudy}
As stated before, five grids are generated for the means of grid independence study for each geometry. The number of elements generated for each of the 3 studied geometries, using adaptive mesh refinement, are reported in Table \ref{table:MeshDetails}.

\begin{figure} [ht!]
\centering
\begin{subfigure}{.48\textwidth}
  \centering
  \includegraphics[width=1\linewidth]{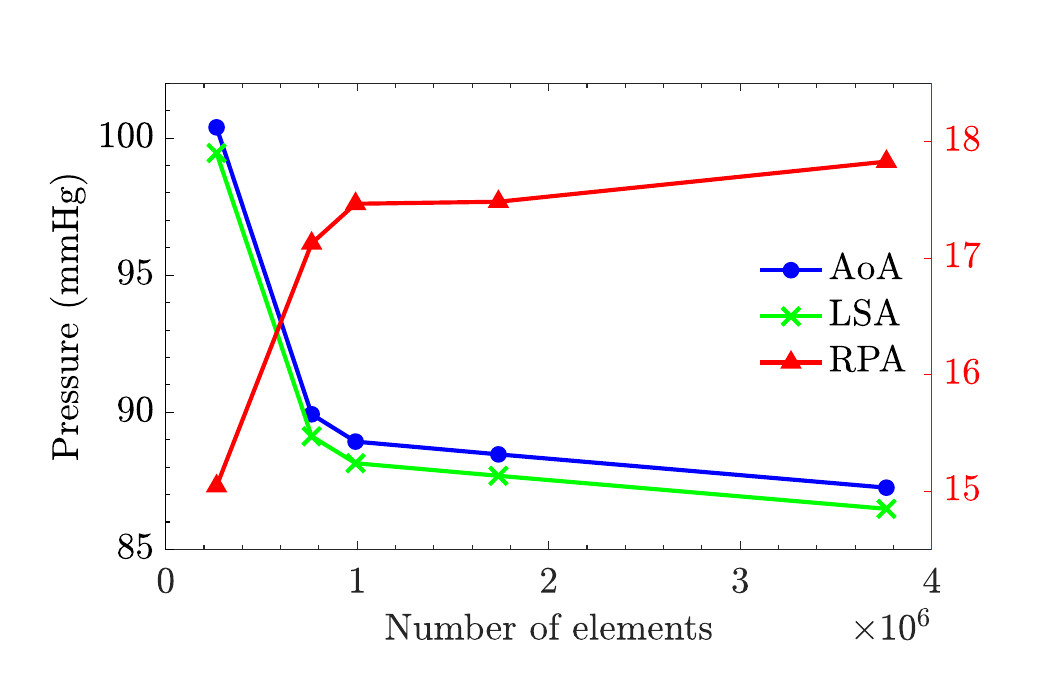}
  \vspace*{-8mm}
  \caption{Steady}
  \label{fig_ms:sub1}
\end{subfigure}
\begin{subfigure}{.48\textwidth}
  \centering
  \includegraphics[width=1\linewidth]{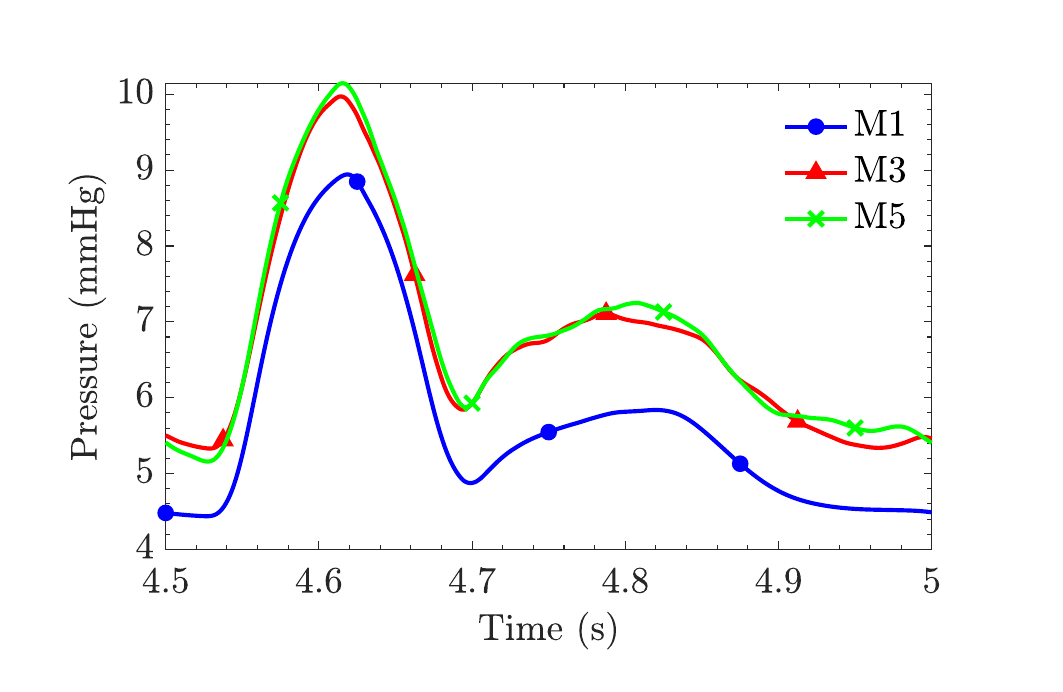}
  \vspace*{-8mm}
  \caption{Transient}
  \label{fig_ms:sub2}
\end{subfigure}
\caption{Mesh independence study}
\label{fig:MeshStudy}
\end{figure}

\begin{table}[ht!]
\renewcommand{\arraystretch}{1.0}
\centering
\caption{Number of elements}
\begin{tabular}{ p{1cm} p{1.9cm} p{1.9cm} p{1.6cm} }
\hline
\textbf{Mesh} & \textbf{2.5BT} & \textbf{4.0BT} & \textbf{2.5CS} \\
\hline
 M1 & $265,732$ & $246,567$ & $248,753$\\
 M2 & $762,202$ & $510,754$ & $645,129$ \\
 M3 & $991,612$ & $811,992$ & $838,894$ \\
 M4 & $1,737,669$ & $1,490,588$ & $1,453,038$ \\
 M5 & $3,763,681$ & $3,187,933$ & $3,041,257$ \\
\hline
\end{tabular}
\label{table:MeshDetails}
\end{table}
The generated grids are then compared, both in steady and transient simulations to identify the coarsest mesh that produces a mesh-independent solution, hence ensuring both accuracy and efficiency. Take for instance the 2.5mm mBTS; Fig. \ref{fig:MeshStudy}(a) shows the spatial average of steady pressure at three faces, i.e., AoA, LSA, and RPA. Furthermore, Fig. \ref{fig:MeshStudy}(b) shows the spatial average of transient RPA pressure, at 10\textsuperscript{th} cycle, on three meshes, i.e., 1, 3, and 5, for the same geometry. 
For this geometry, the maximum grid convergence index (GCI) \cite{Celik2008} observed on all boundaries at steady state, using meshes 1, 3, and 5, is equal to $0.4311\%$. Moreover, the average root mean square (RMS) difference, observed on all boundaries, between the transient pressure of meshes 3, and 5, at 10\textsuperscript{th} cycle, is $0.3138$ mmHg. Using the same approach, maximum GCI values of $0.2084\%$ and $0.3976\%$, and average RMS values of $0.3348$ mmHg and $0.1818$ mmHg are calculated for 4.0mm mBTS and 2.5mm CS, respectively.
Thus, the 3\textsuperscript{rd} iteration mesh of all geometries, is chosen for the remainder of this study.

\subsubsection{Cycle-to-cycle convergence}
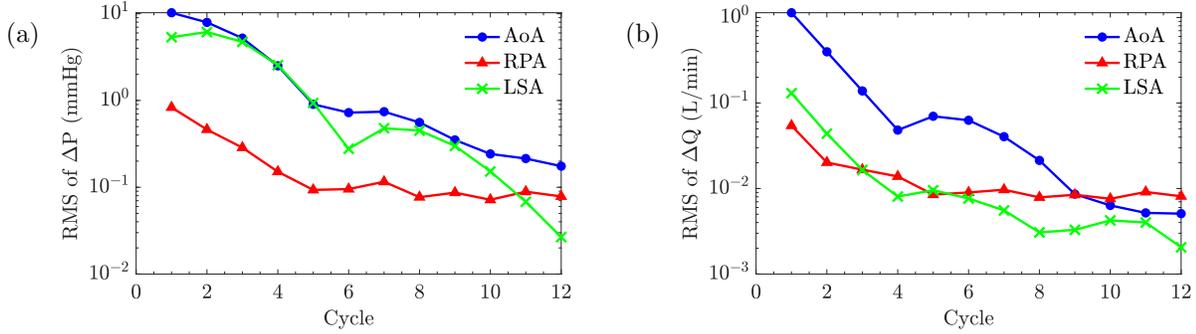
\begin{figure} [ht!]
\centering
\vspace*{-4mm}
\begin{subfigure}{.005\textwidth}
\begin{tikzpicture}
\node[anchor=south west] at (0,0) {};
\node[overlay] at (0pt,110pt) {(a)};
\end{tikzpicture}
\end{subfigure}
\begin{subfigure}{.48\textwidth}
  \centering
  \includegraphics[width=1\linewidth]{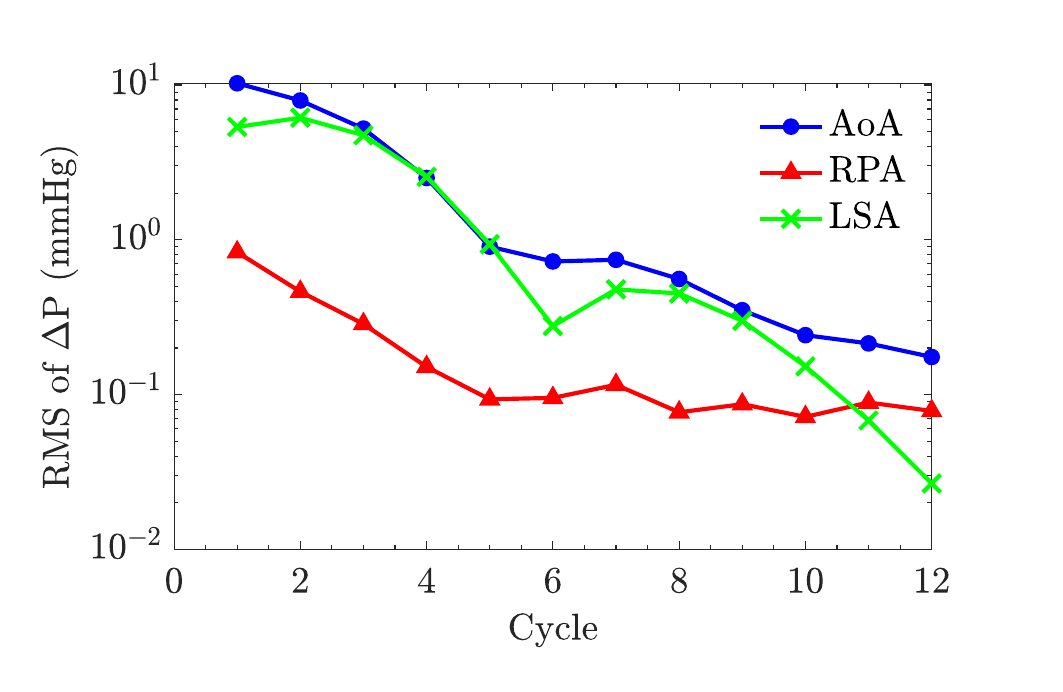}
  \vspace*{-8mm}
  \label{fig_cc:sub1}
\end{subfigure}
\begin{subfigure}{.005\textwidth}
\begin{tikzpicture}
\node[anchor=south west] at (0,0) {};
\node[overlay] at (0pt,110pt) {(b)};
\end{tikzpicture}
\end{subfigure}
\begin{subfigure}{.48\textwidth}
  \centering
  \includegraphics[width=1\linewidth]{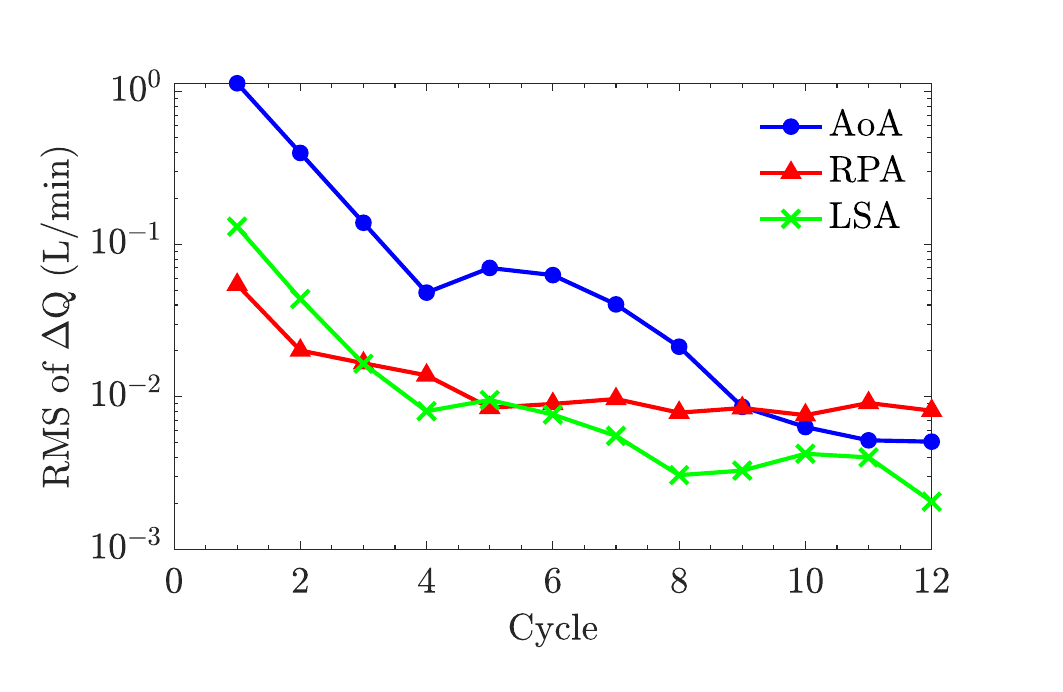}
  \vspace*{-8mm}
  \label{fig_cc:sub2}
\end{subfigure}
\caption{Cycle-to-cycle convergence of (a) Pressure and (b) Volume flow rate}
\label{fig:CycleConv}
\end{figure}

Transient simulations done in this study are all initialized with zero velocity and pressure fields. Therefore, 13 cardiac cycles were simulated for each model to achieve cycle-to-cycle convergence. The RMS difference of pressure and volume flow rate between each cycle and the last cycle in a logarithmic scale can be seen in Fig. \ref{fig:CycleConv}. The RMS difference between cycles 12 and 13, with a maximum of 0.199 mmHg and 0.0167 L/min, corresponding to a maximum of less than 3\% when normalized with mean value, over all faces, is negligible. Hence, there exists no significant difference between cycles after this point. Consequently, all the reported results below correspond to the 14\textsuperscript{th} cycle, with the starting time shown as 0 s, unless stated otherwise. 

\subsubsection{Flow solver time step size independence}
\begin{figure} [ht!]
\centering
\begin{subfigure}{.005\textwidth}
\begin{tikzpicture}
\node[anchor=south west] at (0,0) {};
\node[overlay] at (0pt,110pt) {(a)};
\end{tikzpicture}
\end{subfigure}
\begin{subfigure}{.48\textwidth}
  \centering
  \includegraphics[width=1\linewidth]{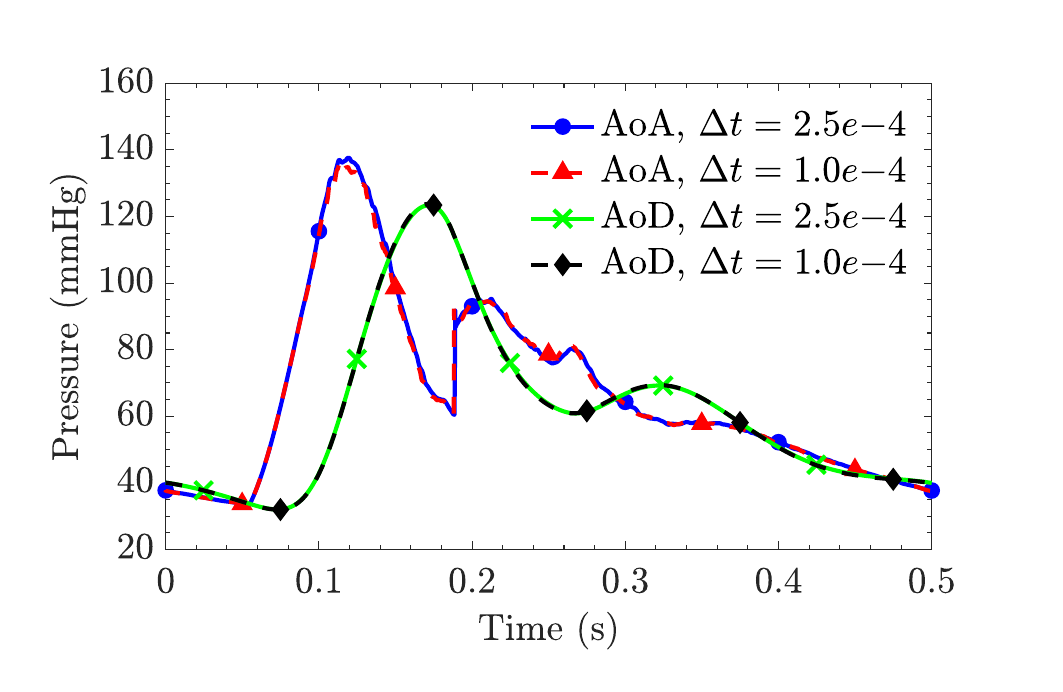}
  \vspace*{-8mm}
  \label{fig_ts:sub1}
\end{subfigure}
\begin{subfigure}{.005\textwidth}
\begin{tikzpicture}
\node[anchor=south west] at (0,0) {};
\node[overlay] at (0pt,110pt) {(b)};
\end{tikzpicture}
\end{subfigure}
\begin{subfigure}{.48\textwidth}
  \centering
  \includegraphics[width=1\linewidth]{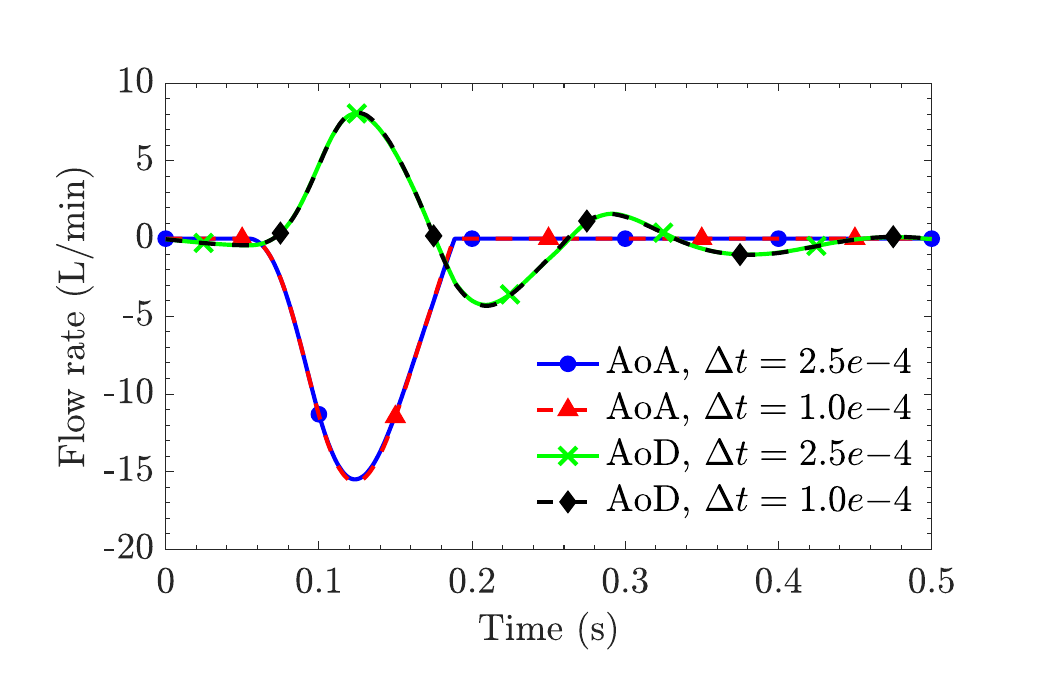}
  \vspace*{-8mm}
  \label{fig_ts:sub2}
\end{subfigure}
\caption{The effect of time step size on (a) Pressure and (b) Volume flow rate}
\label{fig:TimeStep}
\end{figure}
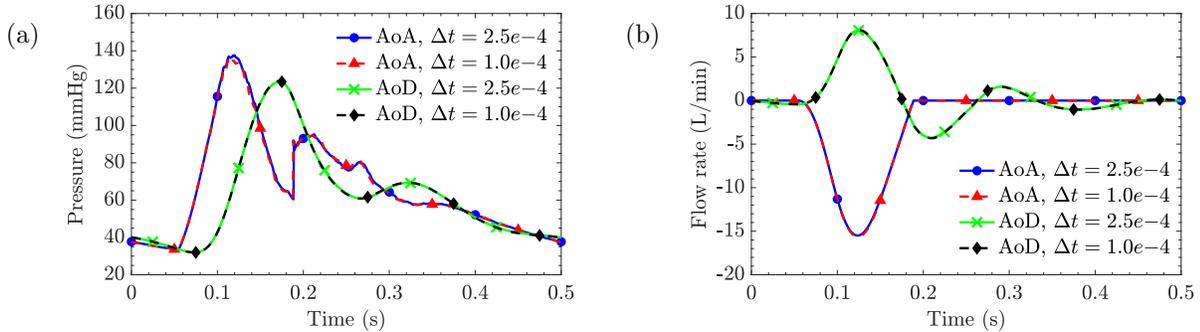

As was shown in  \cite{Jia2023}, conventional formulation of stabilization parameter can produce results that diverge as time-step size is dropped below a certain threshold. To ensure our results are in fact time-step size-independent, we perform three simulations with $\Delta t = 5 \times 10^{-4} s$, $\Delta t = 2.5 \times 10^{-4} s$ and $\Delta t = 1 \times 10^{-4} s$. Figure \ref{fig:TimeStep} shows average pressure and flow rate results on AoA and AoD faces, for the two smaller time-steps. Notice that the results are almost identical. Based on these findings, $\Delta t = 2.5 \times 10^{-4} s$, which corresponds to 2000 time steps per cardiac cycle, is chosen for all remaining simulations. 

\subsection{Red blood cell tracking}
\subsubsection{Particle tracking time step size independence}
\begin{figure} [ht!]
\centering
\includegraphics[width=0.48\linewidth]{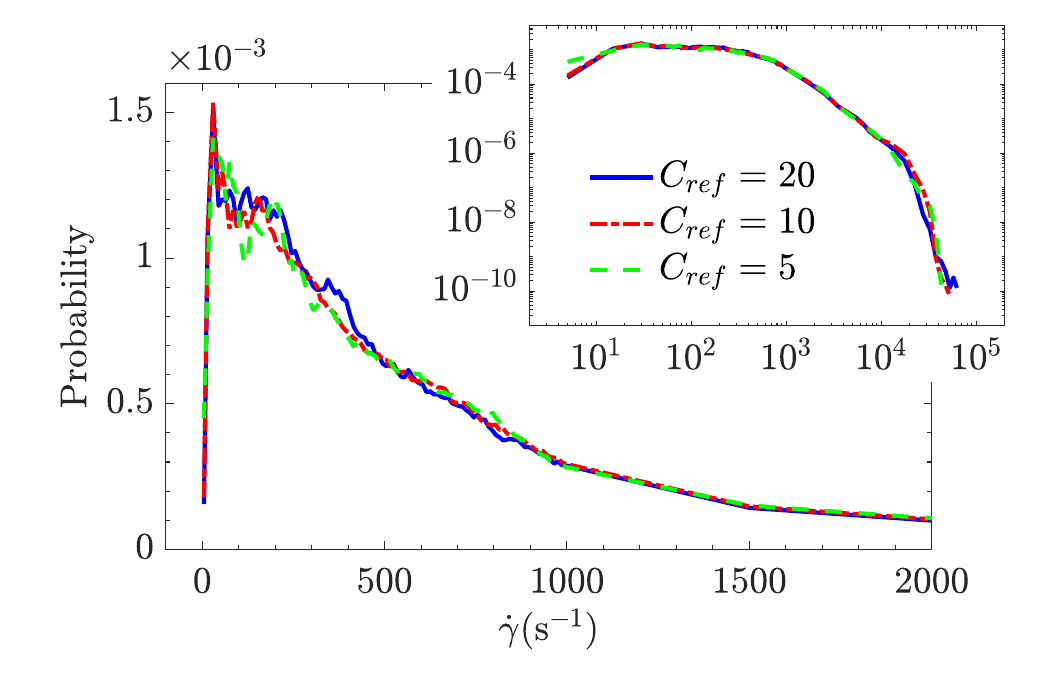}
  \vspace*{-6mm}
\caption{PDF of $\dot{\gamma}$ for different refinement factors. Inset shows the same plot in a logarithmic scale.}
\label{fig:RefFac}
\end{figure}

To make sure the particle tracking time-step size is sufficiently small and does not affect our results, the statistical distribution of 
\begin{equation}
    \dot{\gamma} = \sqrt{{\partial_j u_i}{\partial_j u_i}},
    \label{eqn:gamma_dot}
\end{equation}
which is the 2-norm of the velocity gradient tensor, also called shear rate, is calculated for all particles at all times. 
Particle tracking time step is refined by a factor of $C_{ref}=$5, 10, and 20 relative to the sampled data points, and the resulting probability density functions (PDFs) are compared against each other in Fig. \ref{fig:RefFac} for the 4.0mm mBTS geometry. These samples are recorded only for those particles that pass through the shunt out of 2000 released particles.  $C_{ref} = 10$ is chosen for the remaining calculations provided that it has less than 5\% difference with the highest refinement factor.

We propose running particle tracking computations after and not simultaneously with the flow simulation. 
On the hardware used in this study, large-scale flow simulation of one cardiac cycle on a mesh of $\approx10^6$ elements, with 2000 time steps, takes $\approx350$ core-hours (CH) to complete.  With 10,000 time steps per cardiac cycle and 5,000 particles, the proposed particle tracking solver will take $\approx70$ CH for 4 cardiac cycles in addition to the cost of large-scale flow simulation. On the other hand, the same simulation with coupled particle tracking would take $\approx1800$ CH when using a time step size equal to what was used for the particle tracking solver. Thus, performing particle tracking as a post-processing step allows for the reuse of same flow data for multiple calculations and thereby reduces the simulation costs by two orders of magnitude.

\subsubsection{Particle count independence}
The tails of PDFs in Fig. \ref{fig:RefFac} represent the instances of high shear where RBC damage can occur. Since a small percentage of ensembles falls within these tails, their accurate statistical representation will require simulation of a large number of particles.  
By systematically increasing the number of particles in a series of simulations, we observed less than 1.5\% difference between velocity gradient magnitude for the release of 5,000 and 15,000 particles per cycle. Owing to this statistical convergence and the accuracy of single-cell results, far fewer RBC simulations can be performed compared to considering average Ht, especially in large-scale flows. Therefore, for all the following simulations, the number of particles released during each cardiac cycle is set to 5,000.

\subsubsection{Cardiac cycle count independence}
A proper statistical sampling of RBCs requires particles to be uniformly distributed in the computational domain. Since particles are gradually introduced into the domain at the AoA inlet, one may opt to continue seeding particles at the inlet for many cardiac cycles to obtain a relatively uniform spatial distribution of RBCs throughout the domain. That approach, however, is costly as it requires tracking many particles for many cycles. Instead, we opt for a less expensive yet statistically equivalent approach. It involves releasing particles consistently at the inlet through one cardiac cycle and continuing the simulation long enough so that all RBCs exit the computational domain. Table \ref{table:NoCardiacCycles} shows the percentage of particles that pass through the shunt and exit the computational domain at each cardiac cycle normalized with that of all cycles, for a continuous release of particles in the first cycle. Based on the observed trend, for the remainder of this study, results of the first 3 cycles are considered, so that almost all the particles passing through the shunt have enough time to exit the domain.

\begin{table}[ht!]
\renewcommand{\arraystretch}{1.0}
\centering
\caption{Percentage of particles passing through the shunt and exiting domain per cardiac cycle}
\begin{tabular}{ p{1.4cm} p{1.8cm} p{1.8cm} p{1.4cm} }
\hline
\textbf{Cycle} & \textbf{2.5BT} & \textbf{4.0BT} & \textbf{2.5CS} \\
\hline
 1 & 61.13 & 83.96 & 75.37 \\
 2 & 38.05 & 15.68 & 24.12 \\
 3 & 0.82 & 0.35  & 0.52 \\
 4 & 0 & 0  & 0 \\
 \hline
 Total & 100 & 99.99  & 100.01 \\

\hline
\end{tabular}
\label{table:NoCardiacCycles}
\end{table}

\begin{figure} [ht!]
\centering
\begin{subfigure}{.30\textwidth}
  \centering
  \includegraphics[width=1\linewidth]{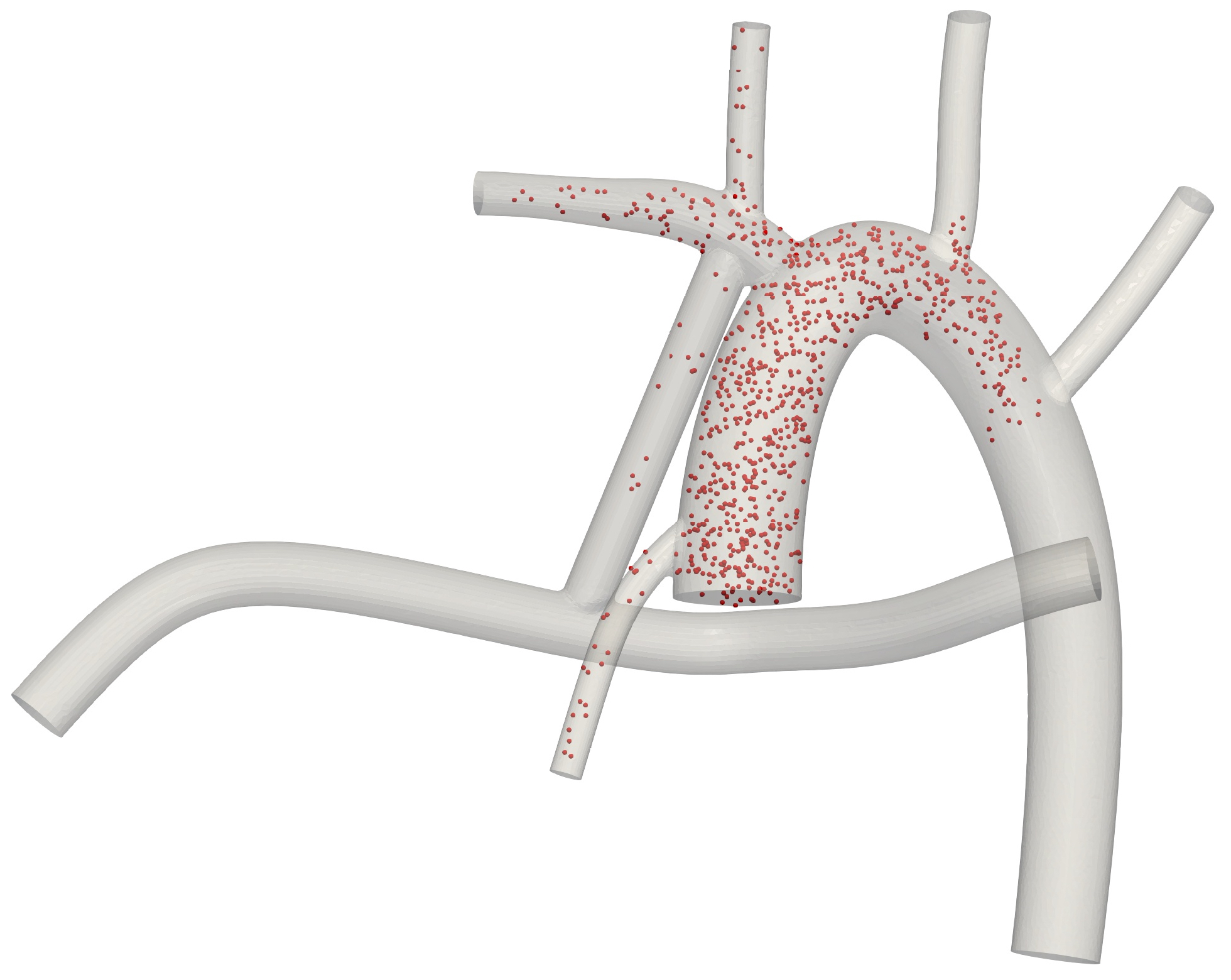}
  \caption{20\% into 1\textsuperscript{st} cycle}
  \label{fig_t:sub1}
\end{subfigure}
\begin{subfigure}{.30\textwidth}
  \centering
  \includegraphics[width=1\linewidth]{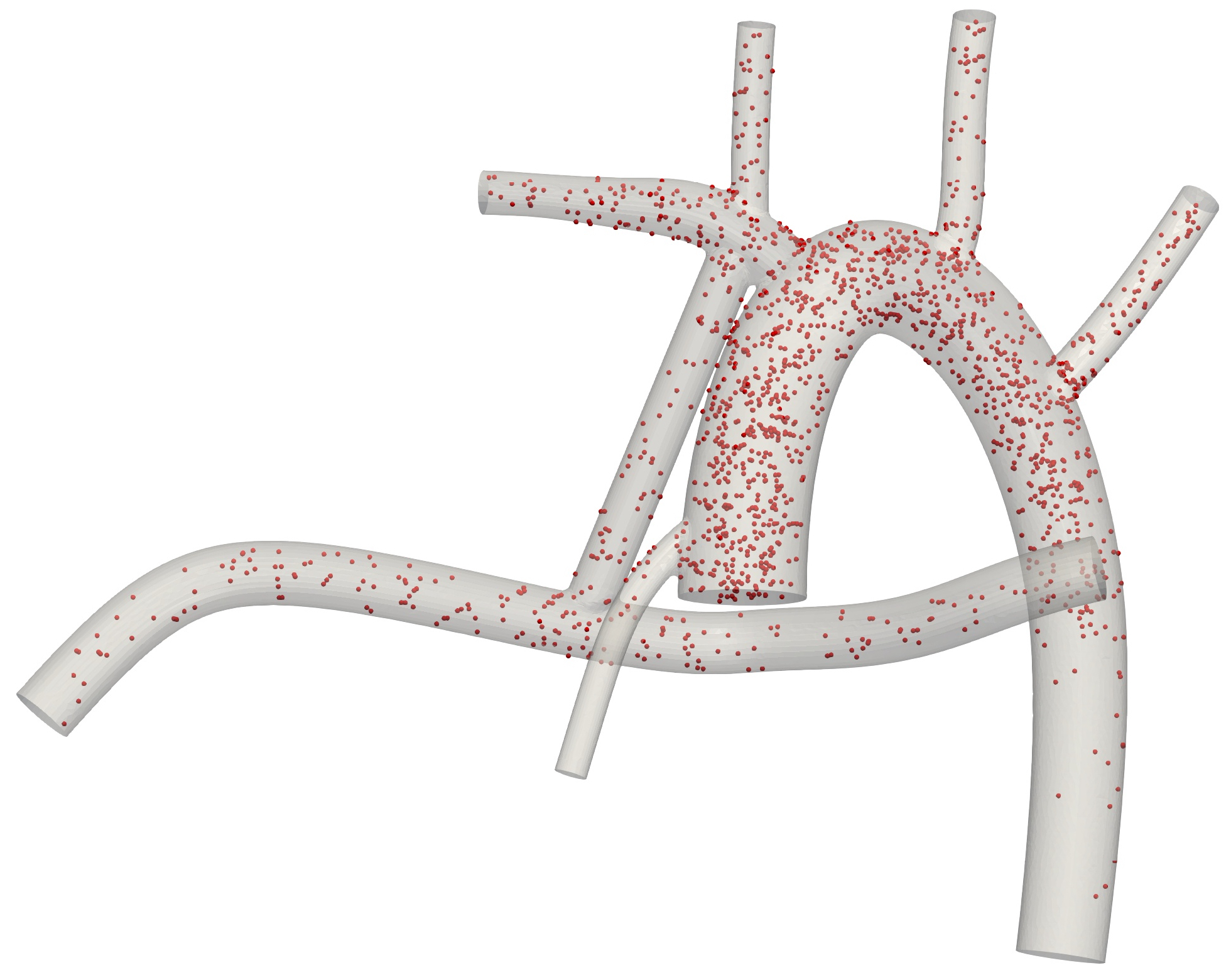}
  \caption{40\% into 1\textsuperscript{st} cycle}
  \label{fig_t:sub2}
\end{subfigure}
\begin{subfigure}{.30\textwidth}
  \centering
  \includegraphics[width=1\linewidth]{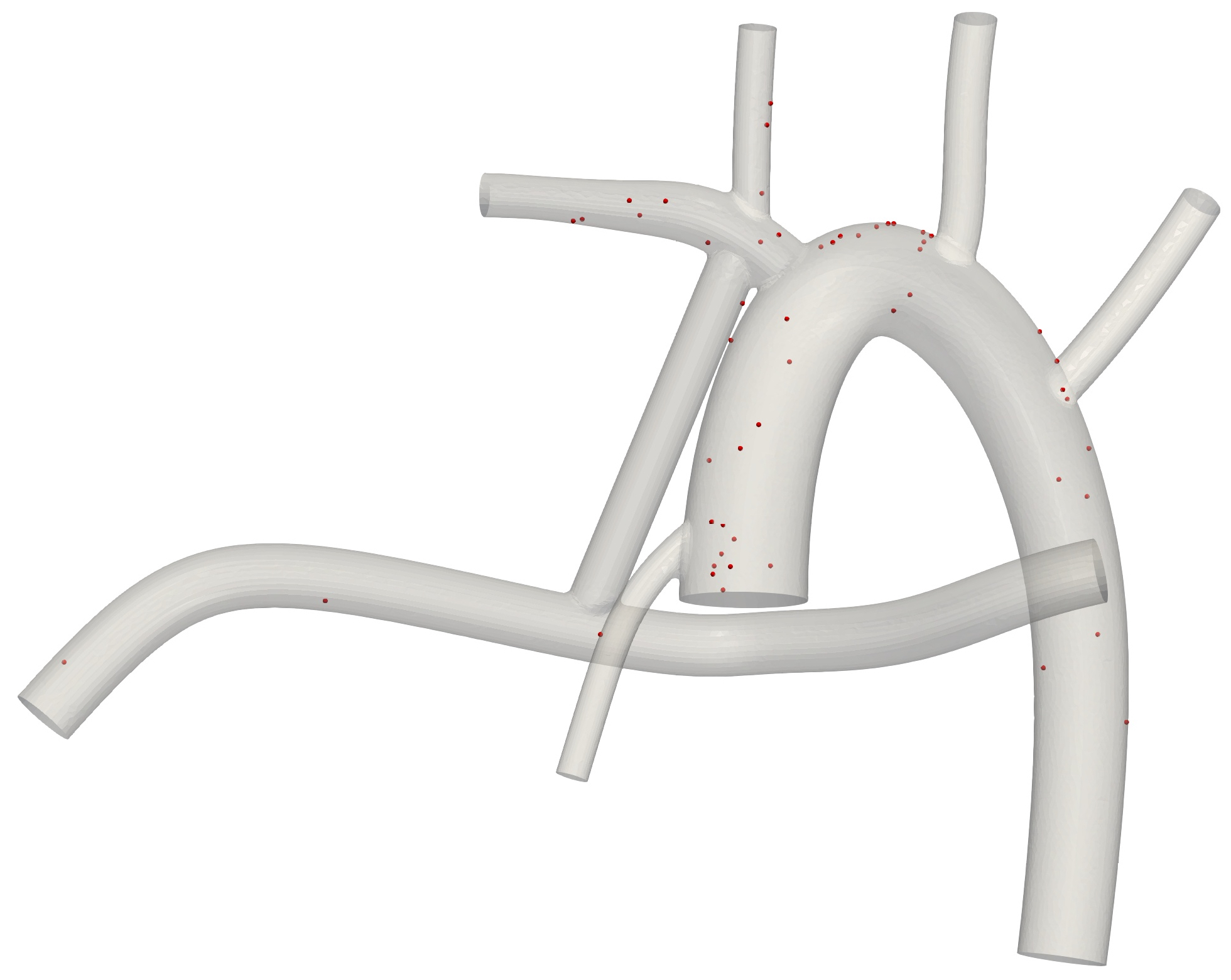}
  \caption{Beginning of 3\textsuperscript{rd} cycle}
  \label{fig_t:sub3}
\end{subfigure}%
\caption{Particle tracking snapshot at different time points}
\label{fig:track}
\end{figure}

 Locations of the particles at $t=0.1$ s, $t=0.2$ s, and $t=1.0$ s (the end of 2\textsuperscript{nd} cardiac cycle) after the release of 5000 particles during the first cycle are illustrated in Fig. \ref{fig:track} for 4.0mm mBTS geometry. The process of particles entering the domain and filling the geometry can be seen in Fig. \ref{fig:track}(a,b), while Fig. \ref{fig:track}(c) shows the beginning of 3\textsuperscript{rd} cardiac cycle where few RBCs are left in the domain in accordance with the results reported in Table \ref{table:NoCardiacCycles}.

\subsection{Red blood cell deformation }
\subsubsection{Parameters}
Utilized parameters for simulation of RBC dynamics are summarized in Table \ref{table:HarmsParameter}. Shear elastic modulus is an important parameter characterizing the motion of RBCs. Early studies reported values 2 to 10 $\mu$N/m using optical tweezers and micropipette aspiration experiments \cite{Henon1999, Chien1978}. However, these values are obtained at low shear rates. Only more recently, have the shear modulus of RBC membrane, $G$, of RBCs been evaluated under high shear rates. Mancuso and Ristenpart \cite{Mancuso2017} were able to accurately reproduce experimental results under extensional flow with shear rates as high as 2,000 s$^{-1}$ using the Skalak model and $G \approx 60$ $\mu$N/m. They showed that as the shear rate increases, the shear modulus also increases, non-linearly, from the well-known value of $\approx 6$ $\mu$N/m by an order of magnitude. The increase in shear rate causes a decrease in membrane surface viscosity, significantly reducing the ratio of viscous effects to that of shear. Similarly, an increase in elastic shear modulus with the extension ratio, age of RBCs \cite{Qiang2019}, and fluid stress \cite{Tsubota2021} has been reported. A fixed value of $G = 60$ $\mu$N/m is used in this study due to the high average shear rates of $\approx10^3$ s$^{-1}$ observed along the RBC trajectories. For a stronger strain-hardening behavior, Skalak constitutive law can be modified to account for the increase in shear modulus with increased fluid stress \cite{Mancuso2017,Tsubota2021}. Note that the values of shear modulus from different constitutive laws such as strain-softening/hardening Yeoh law, or strain-softening neo-Hookean law correspond to smaller values for the strain-hardening Skalak law, with the ratio increasing at larger deformations \cite{Dimitrakopoulos2012}. 

\begin{table}[ht!]
\renewcommand{\arraystretch}{1.0}
\centering
\caption{Parameters used for cell-resolved simulations}
\begin{tabular}{ p{2cm} p{5cm} p{1.8cm} p{1.5cm} p{2cm} }
\hline
\textbf{Parameter} & \textbf{Description} & \textbf{Value} & \textbf{Unit} & \textbf{References} \\
\hline
 $G$ & Shear elastic modulus & $6\times10^{-5}$ & N/m & \cite{Mancuso2017} \\
 $E_B$ & Bending modulus & $0.95\times10^{-18}$  & J & \cite{Poz2005,Evans2008}  \\
 $\mu$ & Blood viscosity & 4 & cP & \cite{Chien1970,Nader2019} \\
 $\mu_{int}$ & Cytoplasm viscosity & 6 & cP & \cite{Pozrikidis2003,Zhao2010} \\
 $a$ & Characteristic radius & $2.82 \times 10^{-6}$ & m & \cite{Zhao2010} \\
 $c_0$ & Spontaneous curvature & $-2/a$ & m$^{-1}$ & \cite{Cordasco2014} \\
 $\Delta t$ & Time step & $1 \times 10^{-5}$ & s & \\
 $C$ & Dilatation ratio & $15$ & - & \cite{Sinha2015,Barthes-Biesel2002}\\
 $\lambda$ & Viscosity ratio: $\mu_{int}/\mu$ & 1.5 & - & \\
\hline
\end{tabular}
\label{table:HarmsParameter}
\end{table}

As for the bending modulus, values ranging from $\approx10^{-19}$ \cite{Mohandas1994} to $\approx 0.7 \sim 2 \times 10^{-18}$ \cite{Poz2005,Evans2008} are reported. A dimensionless value of $E_B^* = \frac{E_B}{a^2 G} = 0.002$ is used in our study which corresponds to $E_B \approx 0.95 \times 10^{-18}$ J, where $a$ is the characteristic radius of a sphere having the same volume as a RBC. Blood is simulated as a Newtonian fluid with an effective medium viscosity of 4 cP = 0.004 Pa.s and density of 1060 kg/m$^3$. Using whole blood (rather than plasma) viscosity permits us to model cell-cell interactions in our single RBC simulations \cite{Rydquist2023}. The selected value for the viscosity corresponds to high shear rates considering that the blood is a non-Newtonian shear-thinning fluid whose viscosity is high at low shear rates and decreases to a plateau at high shear rates. Based on these values, capillary number $Ca = \frac{\mu \dot{\gamma} a}{G}$ varies from 0.188 to 7.52 as $\dot{\gamma} $ changes from $1,000$ s$^{-1}$ to $40,000$ s$^{-1}$. 

Spontaneous curvature, which is selected to match the resting shape of RBCs to its regular biconcave shape \cite{Peng2014}, is $-2/a$. 
The dilatation ratio, which corresponds to the ratio of the dilatation modulus $E_D$ to the shear elastic modulus, captures the cell's resistance to local area alterations.  
The physical value of the dilatation ratio is $C_{ph} \approx 10^4$ \cite{Yazdani2011} (e.g., $C \approx 5000$ for $E_D \approx 0.3$ N/m \cite{Evans1976}). This large value results in costly computations due to the numerical stiffness. It has been demonstrated that choosing a lower value of $C$, even by orders of magnitude so long as it is larger than 10, results in well-converged large-scale RBC dynamics that are consistent with various experimental data \cite{Barthes-Biesel2002,Sinha2015}. Following that practice in the literature \cite{Yazdani2011,Pozrikidis2003}, a lower value $C=15$ is employed in the current work. 
Furthermore, a spherical harmonics order of 12 and an upsampling rate of 4 were used in the current study to capture RBC dynamics, which were shown to produce accurate results in earlier studies \cite{Rydquist2022,Rydquist2023}. 
 
According to turbulent shear flow experimental data of Sutera and Mehrjardi \cite{Sutera1975}, shear stress in the range of $100-250$ Pa is associated with medium hemolysis, i.e., lower than 10\% of RBCs. That percentage, however, rapidly rises once the shear stress increases beyond that threshold. 
In our computations, we observe instability near that threshold. More specifically, our simulations remain stable up to $Ca = 10.85$, which corresponds to a shear rate of 57,700 s$^{-1}$ and a shear stress of 230 Pa. The simulation instability for values close to and above that range may be attributed to the cell's rupture given that the RBC experiences very large deformation. From a numerical point of view, this failure might be a consequence of highly deformed shapes of these RBCs that could no longer be captured with the specified harmonics order. Such unresolved scales can affect the resolved scales solution and hence cause instability. Resultingly, further increasing the area dilatation ratio and spherical harmonics order, while decreasing the time step size might result in a more stable condition. That said, we are unsure as to whether the sources of these instabilities are physical or numerical. Hence, The unstable RBC simulations ($\approx 5.6\%$ of all simulated RBCs) are excluded in the results presented below.  

\subsubsection{Particle deformation time step size independence}

The effect of time step size on the computed maximum and average shear and areal strains is evaluated for a single RBC.
Computations performed using $\Delta t = 10^{-5}$s showed less than 2\% error compared to $10^{-6}$s, hence that time step size is used for the rest of this study. 
The run time for cell-resolved simulation of a single RBC for 1 cardiac cycle with a time step size of $\Delta t = 10^{-5} s$, is roughly 50 CH. Thus, the overall runtime of all simulations performed in this study is approximately 60,000 CH (15,000 for CFD, 200 for particle tracking, and 43,000 for cell-resolved simulations). This translates to a total computational time of approximately 1.6 days using 32 nodes for personalized outcome prediction of each shunt configuration.

\section{Results and Discussions}
\subsection{Large-scale flow characteristics}
\begin{figure} [ht!]
\centering
\begin{subfigure}{.08\textwidth}
  \centering
  \begin{tikzpicture}
\node[anchor=south west] at (0,0) {\includegraphics[width=0.8\linewidth]{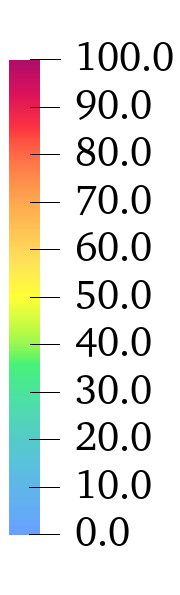}};
\node[overlay] at (20pt,105pt) {WSS (Pa)};
\end{tikzpicture}
\end{subfigure}%
\hspace{1mm}
\begin{subfigure}{.28\textwidth}
  \centering
  \includegraphics[width=1\linewidth]{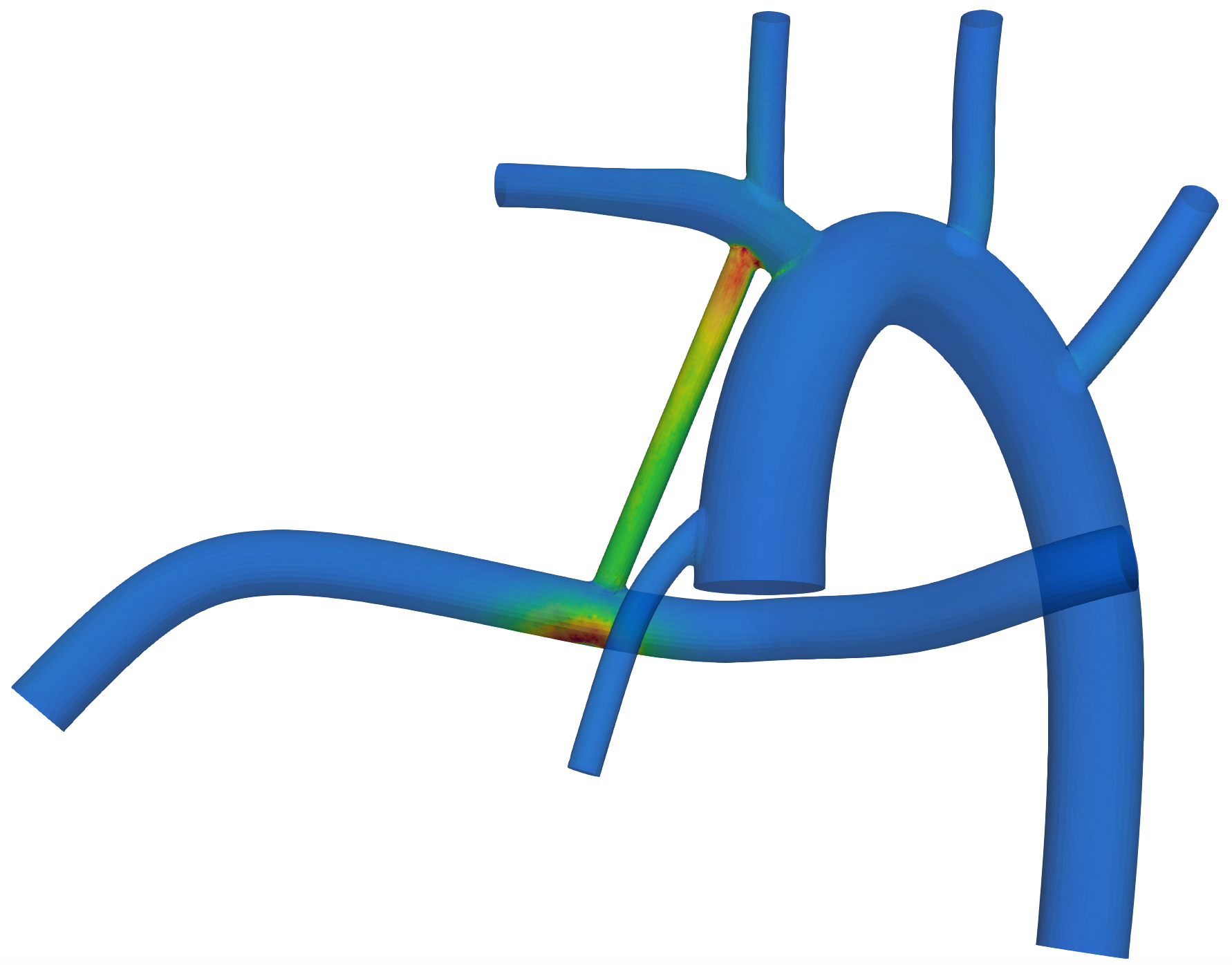}
  \caption{2.5mm modified BT shunt}
  \label{fig_wss:sub1}
\end{subfigure}
\begin{subfigure}{.28\textwidth}
  \centering
  \includegraphics[width=1\linewidth]{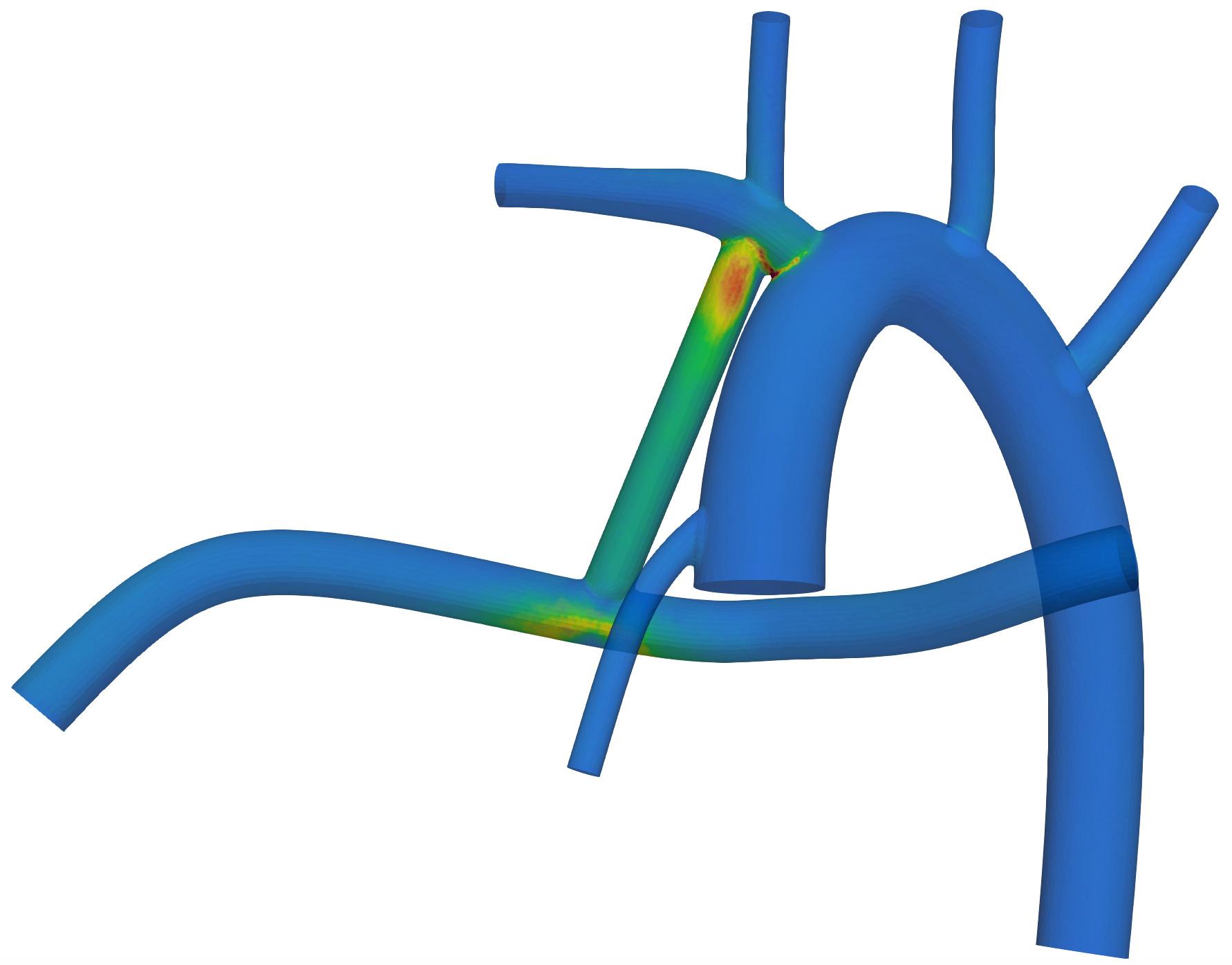}
  \caption{4.0mm modified BT shunt}
  \label{fig_wss:sub2}
\end{subfigure}
\begin{subfigure}{.28\textwidth}
  \centering
  \includegraphics[width=1\linewidth]{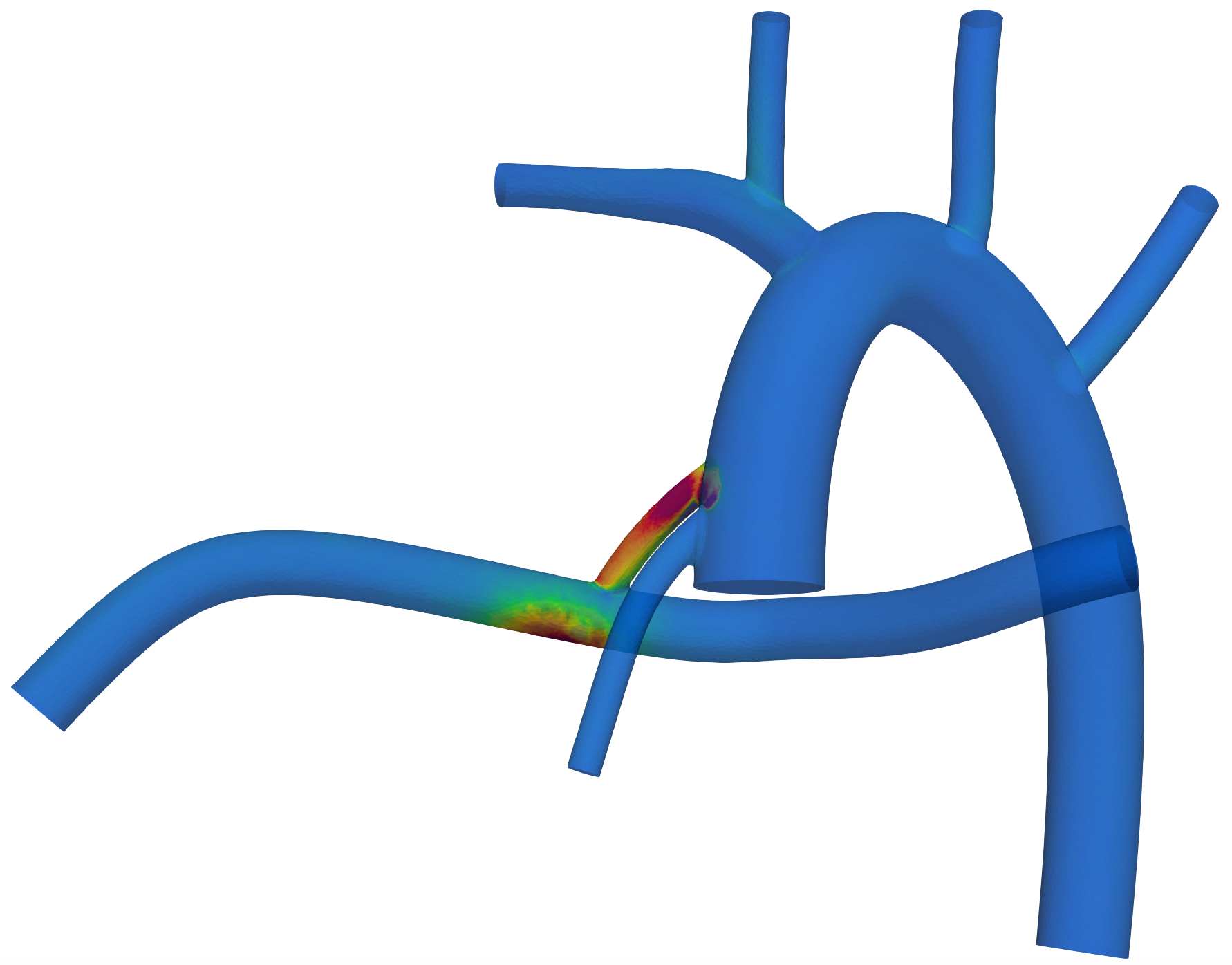}
  \caption{2.5mm central shunt}
  \label{fig_wss:sub3}
\end{subfigure}%

\caption{Time-averaged wall shear stress}
\label{fig:wss}
\vspace{-4mm}
\end{figure}

Traditionally, one relies on the characteristics of the large-scale flow field, such as the spatial distribution of velocity gradient tensor, to evaluate various surgical options. To reproduce those results, the time average wall shear stress (WSS) magnitudes for three anatomies, are illustrated here in Fig. \ref{fig:wss}. Comparing these results with our strain-based cell-resolved analysis could be beneficial in understanding the differences and similarities between the two evaluation methods.
In all three cases, the highest values of WSS are observed in the shunt and at its anastomosis with the pulmonary artery, where the flow impinges the lumen and quickly changes direction. We closely examine these high regions of WSS as they may contribute to the complications associated with Norwood surgery.

The maximum WSS magnitude recorded during the cardiac cycle is $\approx$220 Pa for the 2.5mm mBTS, $\approx$330 Pa for the 4.0mm mBTS, and $\approx$290 Pa for the 2.5mm CS. These values suggest that between these geometries, the 4.0mm mBTS is a riskier option compared to the other two configurations. 
However, temporal averaging suggests that the CS creates higher WSS in the shunt region, followed by the 2.5BT and 4.0BT, respectively (Fig. \ref{fig:wss}). 

Other relevant factors in selecting a surgical design, such as the average arterial and pulmonary pressures, cardiac output (CO), flow distribution between systemic circulation (sys) and pulmonary arteries (PAs), heart load (HL), and oxygen delivery (OD) for each configuration are summarized and compared against literature in Table \ref{table:AveP&Q}. 
The results are also in good agreement with another study that reported a flow rate of 1.254 L/min for PAs and an average pulmonary pressure of 12.9 mmHg for 3.5mm modified BT shunts \cite{Esmaily-Moghadam2015}. The present results are also in agreement with the reported clinical data for mBTS that are $P_{\rm AoA} = 70\pm9$ mmHg, $P_{\rm PAs} = 12\pm4$ mmHg, and $Q_{\rm sys}/_{\rm PAs} = 0.7\pm0.2$ \cite{Ryerson2007}.
In accordance with \cite{Moghadam2012,Jia2021}, the HL is calculated as the
integral of the aortic pressure times flow rate. Similarly, OD is computed as $Q_{\rm sys}C_{\rm PAs} - (Q_{\rm sys}/Q_{\rm PAs}) V_{\rm O_2}$, where $C_{\rm PAs}=0.22$ mL$_{O_2}$/mL and $V_{O_2}=0.874$ mL$_{O_2}$/s are oxygen concentration in PAs and total oxygen consumption, respectively.

%
\begin{table}[ht!]
\renewcommand{\arraystretch}{1.}
\centering
\caption{Bulk flow predictions}
\begin{tabular}{ p{2.5cm} p{1.8cm} p{1.8cm} p{1.6cm} p{2.cm} }
\hline
 & \textbf{2.5BT} & \textbf{4.0BT} & \textbf{2.5CS} & \textbf{3.5BT \cite{Jia2021}} \\
\hline
$P_{\rm AoA}$ (mmHg) & 67.0 & 67.0 & 67.0 & 67.1 \\
$P_{\rm PAs}$ (mmHg) & 6.44 & 12.4 & 7.83 & 11.0 \\
$CO$ (L/min) & 1.59 & 2.21 & 1.76 & 2.08 \\
$Q_{\rm sys}$ (L/min) & 0.98 & 0.91 & 0.99 & 0.94 \\
$Q_{\rm PAs}$ (L/min) & 0.61 & 1.3 & 0.77 & 1.14 \\
$Q_{\rm PAs}/Q_{\rm sys}$  & 0.62 & 1.43 & 0.71 & 1.21 \\
$HL$ (N.m/min) & 10.1 & 16.4  & 11.3 & 29.5 \\
$OD$ (mL$_{\rm O_2}$/s) & 2.19 & 2.73 & 2.51 & 2.72 \\
\hline
\end{tabular}
\label{table:AveP&Q}
\end{table}

\subsection{RBC dynamics, deformation and damage}

In what follows, we only consider the RBCs that pass through the shunt to evaluate various surgical configurations as it is the shunt insertion that contributes to the complications happening after the Norwood surgery. In doing so, the entire history of RBC deformation as it passes through the shunt is recorded provided that hemolysis is known to depend on both the shear rate and exposure time.
After performing each cell-resolved simulation using the BIM solver, multiple parameters are extracted for each RBC. Those include the area, volume, length, width, and height of the RBC, as well as the maximum and average values of (local) areal and shear strains. In extracting these parameters, we treat a given time point and a single RBC as one ensemble. Note that $\lambda_1 \lambda_2$ represents areal strain on the surface of RBC, and $\lambda_1/\lambda_2$ represents shear strain. $\lambda_1 \lambda_2 = 1$ and $\lambda_1 /\lambda_2 = 1$ represent an initial unstrained configuration. Using these metrics, we identify incidents of extreme RBC deformation and subsequently their damage to compare different surgical configurations.

\subsubsection{Red blood cell dynamics}

\begin{figure} [ht!]
\centering
\begin{subfigure}{.48\textwidth}
  \centering
  \includegraphics[width=1\linewidth]{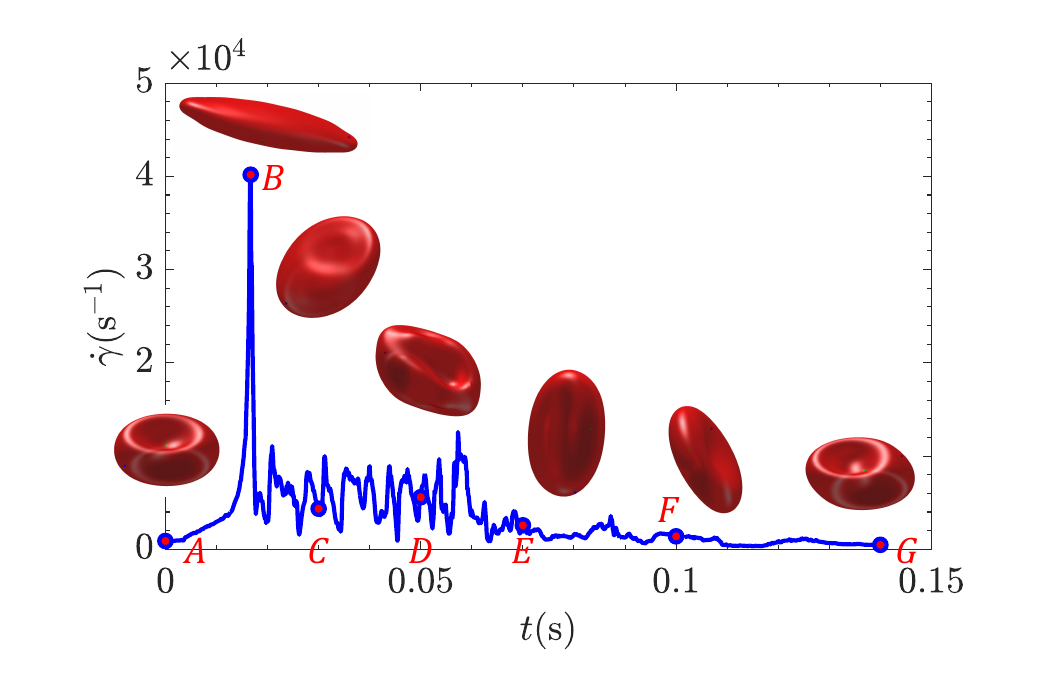}
  \vspace*{-8mm}
  \caption{Deformation superimposed on shear rate}
  \label{fig_g:sub1}
\end{subfigure}
\begin{subfigure}{.48\textwidth}
  \centering
  \includegraphics[width=0.9\linewidth]{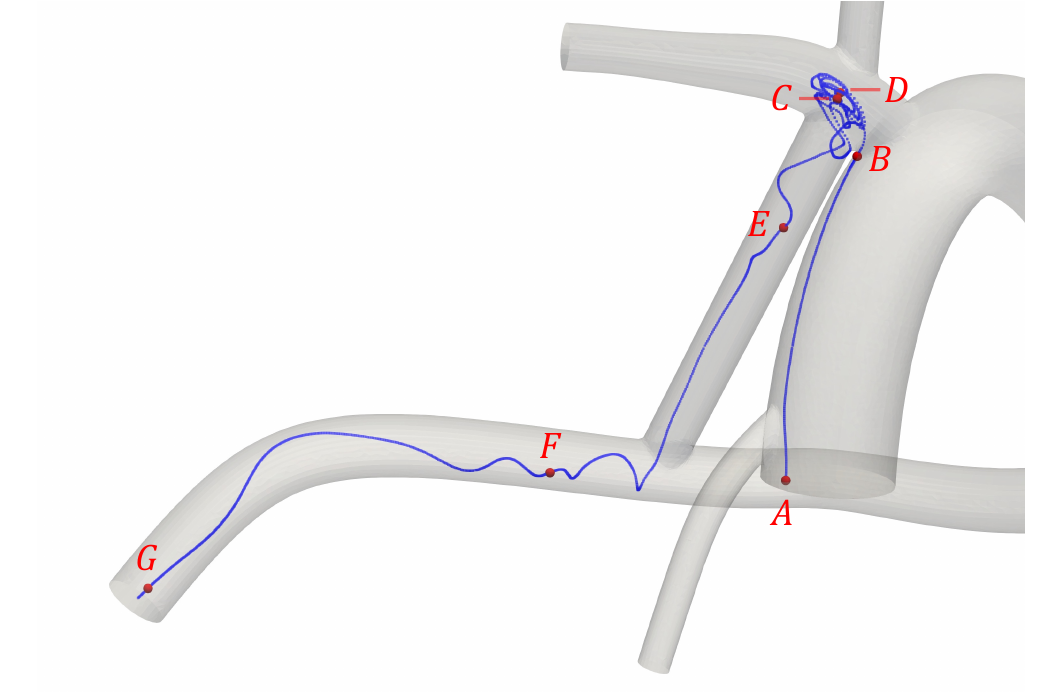}
  \vspace*{-2mm}
  \caption{Particle location}
  \label{fig_g:sub2}
\end{subfigure}

\begin{subfigure}{.48\textwidth}
  \centering
  \includegraphics[width=0.95\linewidth]{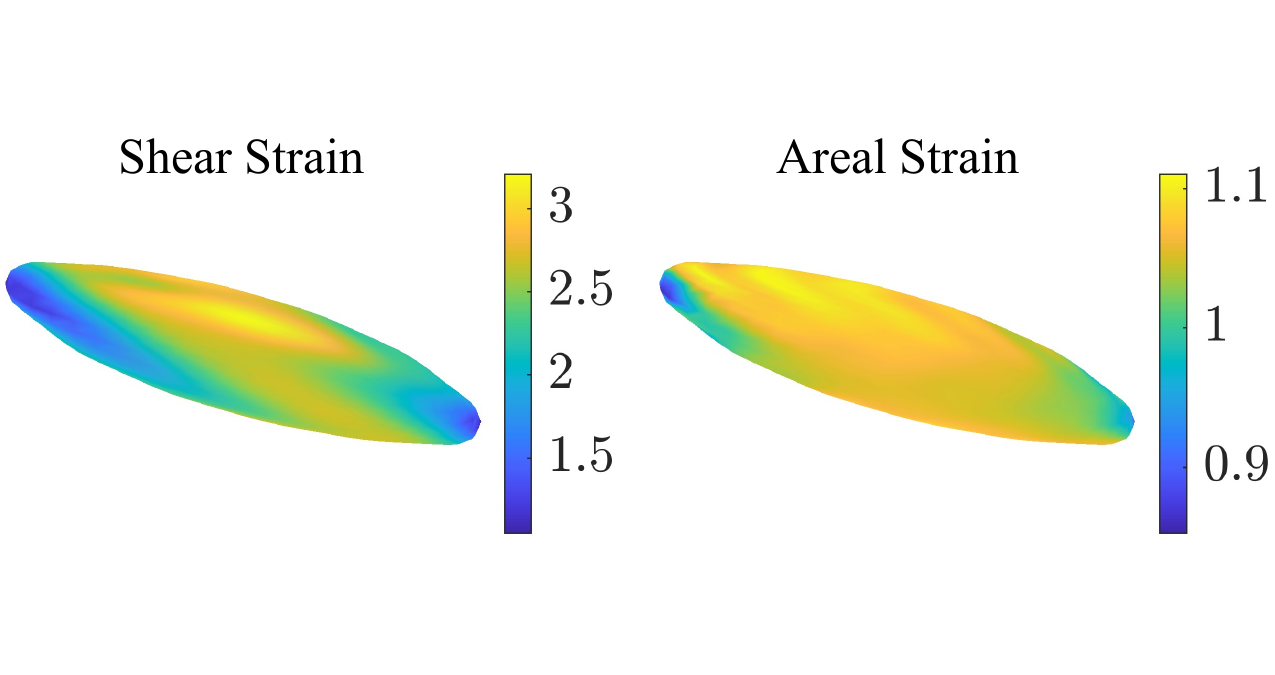}
  
  \caption{Distribution of shear and areal strains at instance B}
  \label{fig_g:sub4}
\end{subfigure}
\begin{subfigure}{.48\textwidth}
  \centering
  \includegraphics[width=1\linewidth]{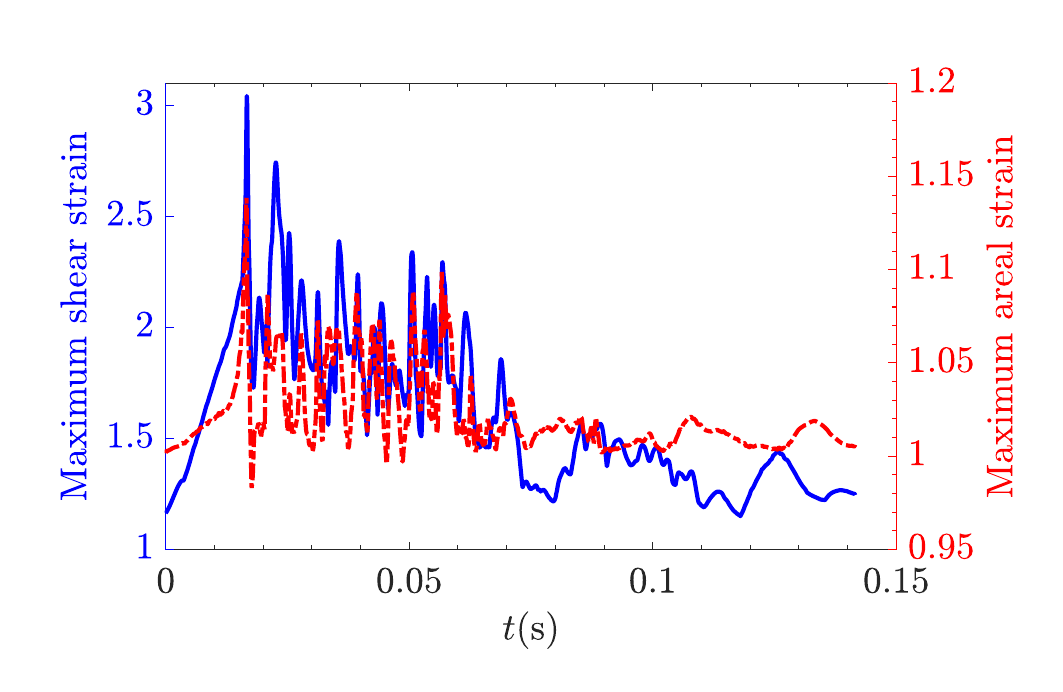}
  \vspace*{-8mm}
  \caption{Maximum shear and areal strain}
  \label{fig_g:sub5}
\end{subfigure}

\begin{subfigure}{.48\textwidth}
  \centering
  \includegraphics[width=1\linewidth]{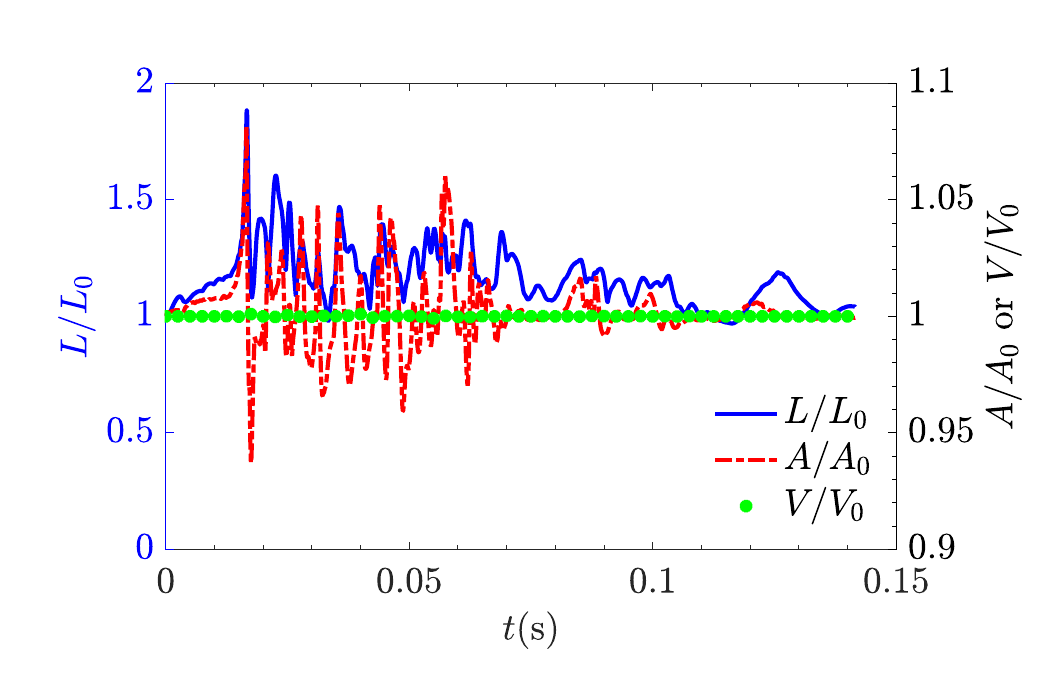}
  \vspace*{-8mm}
  \caption{Normalized length, area, and volume of RBC}
  \label{fig_g:sub3}
\end{subfigure}
\begin{subfigure}{.48\textwidth}
  \centering
  \includegraphics[width=0.95\linewidth]{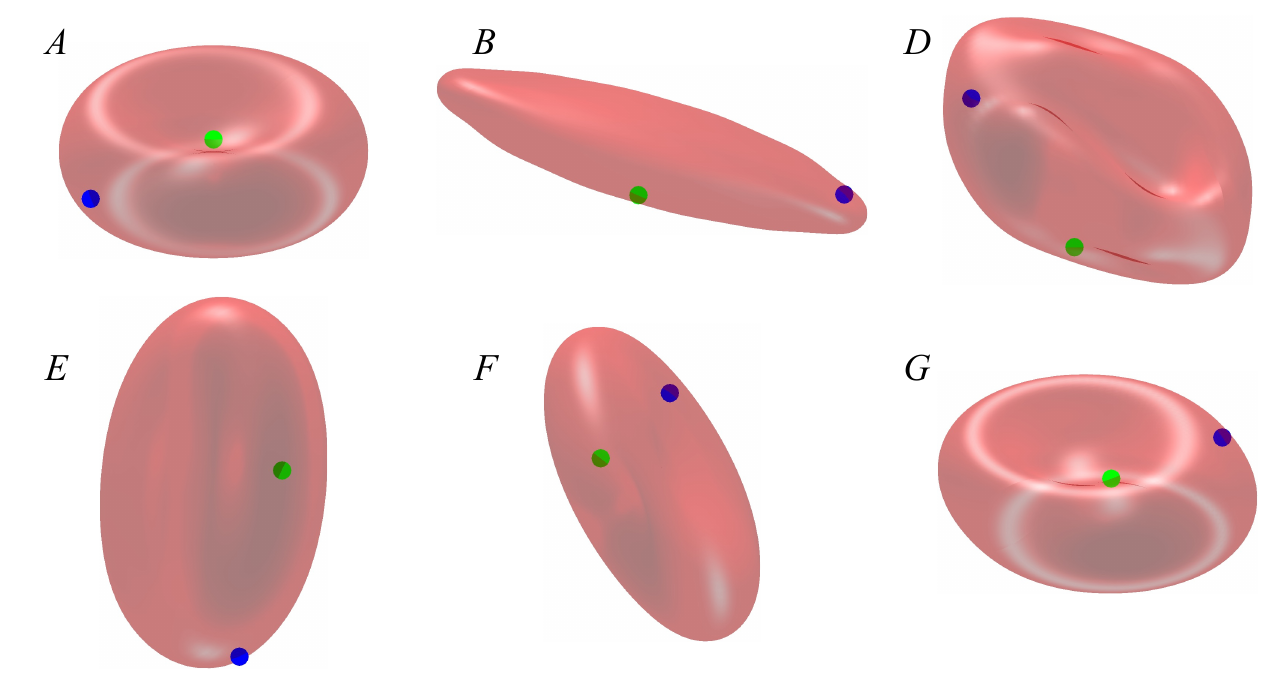}
  \vspace{2mm}
  \caption{Two material points at instances A,B,D,E,F,G}
  \label{fig_g:sub6}
\end{subfigure}

\caption{Behavior of a representative RBC as it traverses through 4.0mm BT shunt configuration}
\label{fig:PrtGamma}
\end{figure}

To provide a physical perspective of the mechanical behavior of RBCs as they traverse through the shunt, the detailed dynamics of a representative RBC is studied in this section. RBCs' ability to deform and adapt their shape in response to dynamically changing flow conditions not only will reduce their resistance to flow but also enable them to pass through small capillaries. 
For shear stress below the critical threshold of 10 Pa (i.e., shear rate of 2500 s$^{-1}$), the RBC undergoes a tumbling solid body rotation  \cite{Fischer2013}. At higher shear rates, the cell experiences a tank-treading motion. 
These behaviors are reproduced in our simulations, as illustrated in Fig. \ref{fig:PrtGamma}. 
In good agreement with previous studies \cite{Yazdani2011, Cordasco2014, Sinha2015, Tsubota2021}, we observe that the RBC shape strongly depends on the flow condition (Fig. \ref{fig:PrtGamma}(a)). The shear history, in turn, strongly depends on the trajectory of RBC, which is labeled with corresponding time points in Fig. \ref{fig:PrtGamma}(b).
At each time point, shear strain and areal strain distributions can be extracted on the RBC membrane (Fig. \ref{fig:PrtGamma}(c)).
The maximum of these distributions in space can be extracted at each time and plotted as a function of time (Fig. \ref{fig:PrtGamma}(d)). Each data point from this plot represents an ensemble, the statistics of which will be analyzed in the following section.
Note that the strains do not start from a value of 1 at $t = 0$ s. This is because the reference state of the RBC, which is an oblate spheroid close to a sphere, is utilized. The RBC's reference stress-free state differs from its experimentally observed biconcave shape \cite{Peng2014}. This results in values higher than 1 for both $\lambda_1 \lambda_2$ and $\lambda_1 /\lambda_2$ at the natural starting state for RBCs, viz. a biconcave rather than an oblate spheroid shape.

Elongation is another deformation metric commonly used to describe the extent of RBC deformation. For instance, Sutera and Mehrjardi \cite{Sutera1975}, reported a 100\% elongation for a shear stress of 250 Pa, while Sohrabi and Liu \cite{Sohrabi2017}, reported an elongation of a $\sim$100\% at a shear rate equal to 40,000 s$^{-1}$. The changes in the length of RBC on its stretch-axis, also called stretch ratio, as well as changes in the RBC surface area and volume, are displayed in Fig. \ref{fig:PrtGamma}(e). It is seen that the RBC's length increases by $\sim100\%$ at regions of high shear (around 40,000 s$^{-1}$) during the cardiac cycle, which is in good agreement with previous experimental \cite{Sutera1975} and computational results \cite{Sohrabi2017}.
Moreover, while the RBC's volume is constant owing to the conservation of cytoplasm mass, the membrane surface area will fluctuate as it deforms. Note, however, that the reported area change might be elevated from its physical baseline since the adopted area dilatation ratio was lower than its physical value.
Our observed value of 8\% area dilation around the shear rate of 40,000 s$^{-1}$ is well comparable to the previous computational studies reporting 6\% global areal strain under a simple shear flow with the same rate \cite{Vitale,Sohrabi2017}.
For the RBC analyzed in this section, the deformation index (DI) defined as $DI = (L+W)/(L-W)$ (with $L$ and $W$ denoting the length and width of the RBC) varies between 0 to 0.4 at its peak. Last but not least, the expected tank-treading motion of this RBC is visualized in time in Fig. \ref{fig:PrtGamma}(f) by highlighting two material points on its membrane.

\subsubsection{Statistical dynamics of red blood cells}
To understand the overall effect of flow on RBCs, we rely on statistics of RBCs' deformation. For each surgical configuration, more than 500 RBCs that pass through the shunt were randomly selected for cell-resolved simulations. We ensure the sampling is performed such that the probability distribution of $\dot{\gamma}$ is unchanged. While we analyze the maximum and average values of areal and shear strain, we consider the tails to be of higher significance as they signify the instances at which RBC damage is likely to occur.

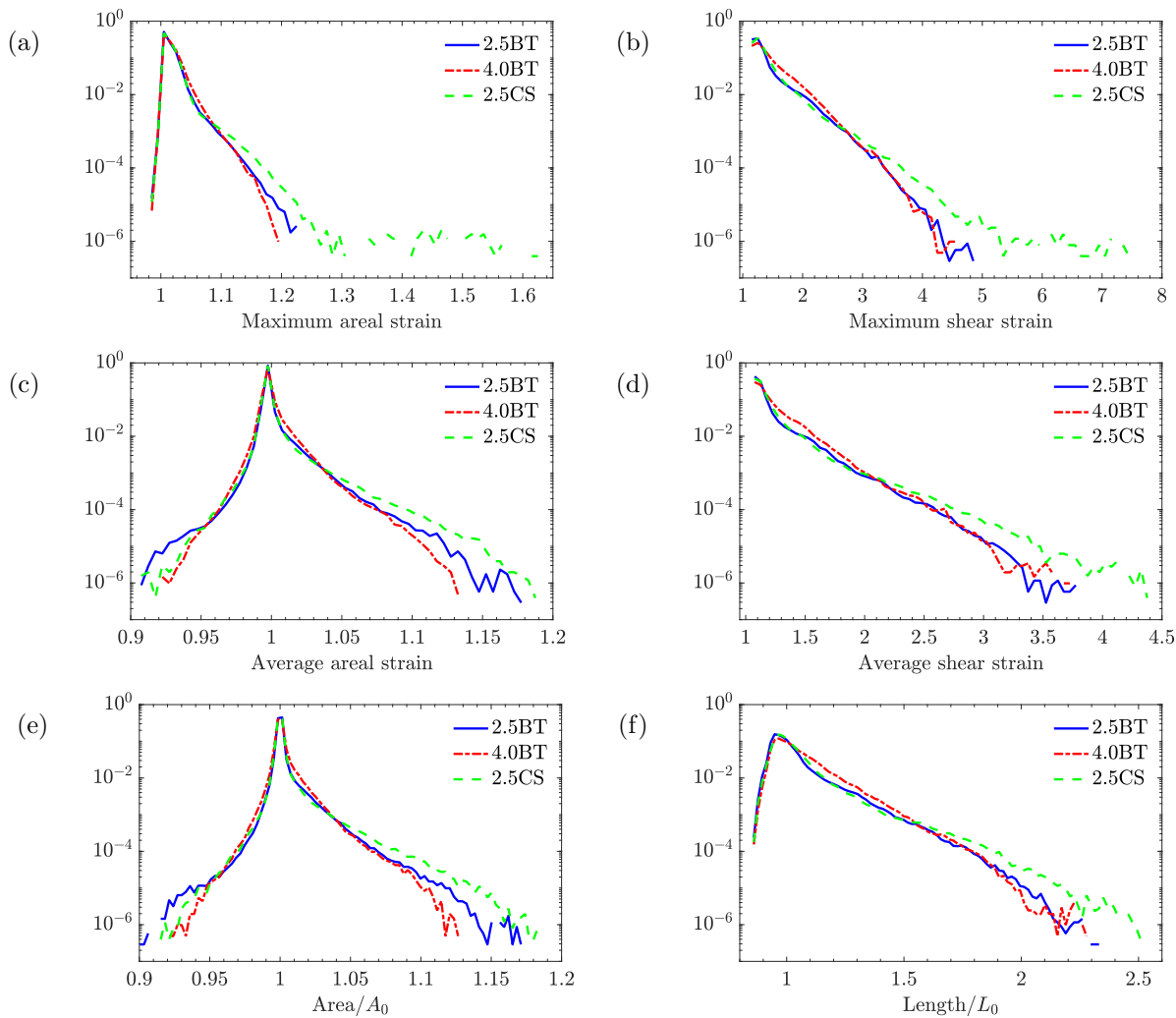
\begin{figure} [ht!]
\centering
\begin{subfigure}{.005\textwidth}
\begin{tikzpicture}
\node[anchor=south west] at (0,0) {};
\node[overlay] at (0pt,110pt) {(a)};
\end{tikzpicture}
\end{subfigure}
\begin{subfigure}{.48\textwidth}
  \centering
  \includegraphics[width=1\linewidth]{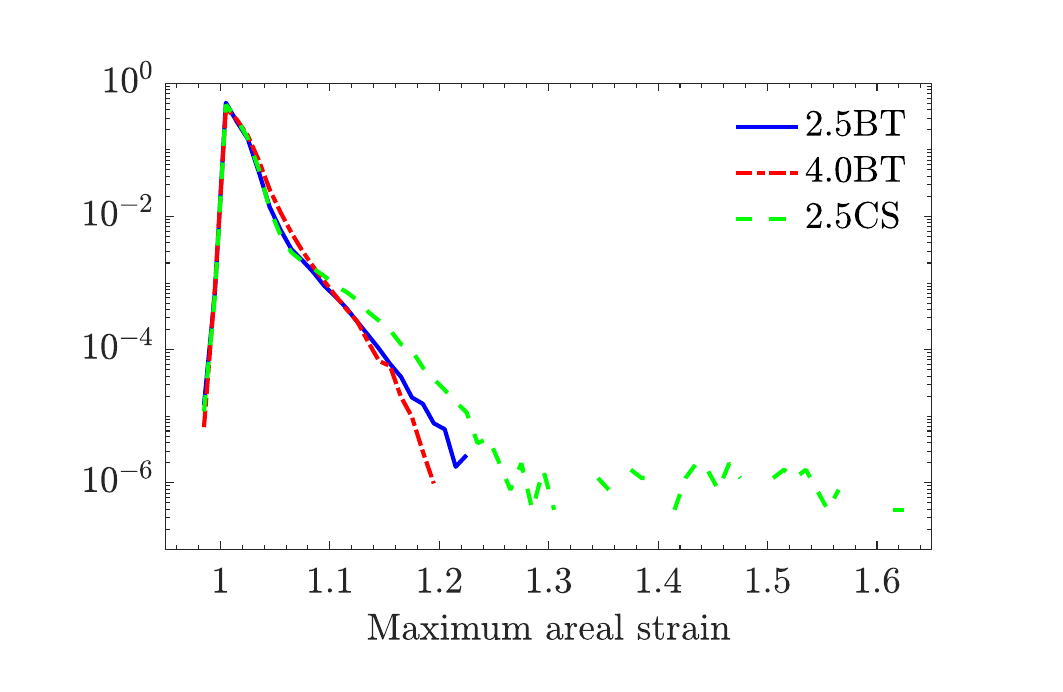}
  \vspace*{-8mm}
  \label{fig_pdf:sub1}
\end{subfigure}
\begin{subfigure}{.005\textwidth}
\begin{tikzpicture}
\node[anchor=south west] at (0,0) {};
\node[overlay] at (0pt,110pt) {(b)};
\end{tikzpicture}
\end{subfigure}
\begin{subfigure}{.48\textwidth}
  \centering
  \includegraphics[width=1\linewidth]{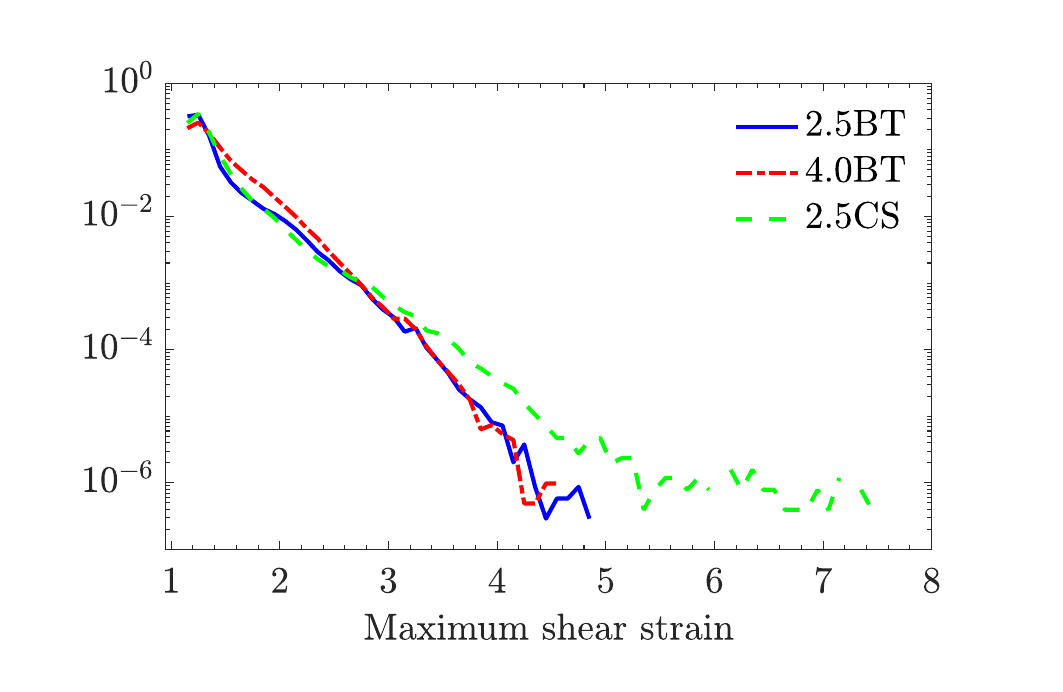}
  \vspace*{-8mm}
  \label{fig_pdf:sub2}
\end{subfigure}
\medskip
\vspace*{-4mm}
\begin{subfigure}{.005\textwidth}
\begin{tikzpicture}
\node[anchor=south west] at (0,0) {};
\node[overlay] at (0pt,110pt) {(c)};
\end{tikzpicture}
\end{subfigure}
\begin{subfigure}{.48\textwidth}
  \centering
  \includegraphics[width=1\linewidth]{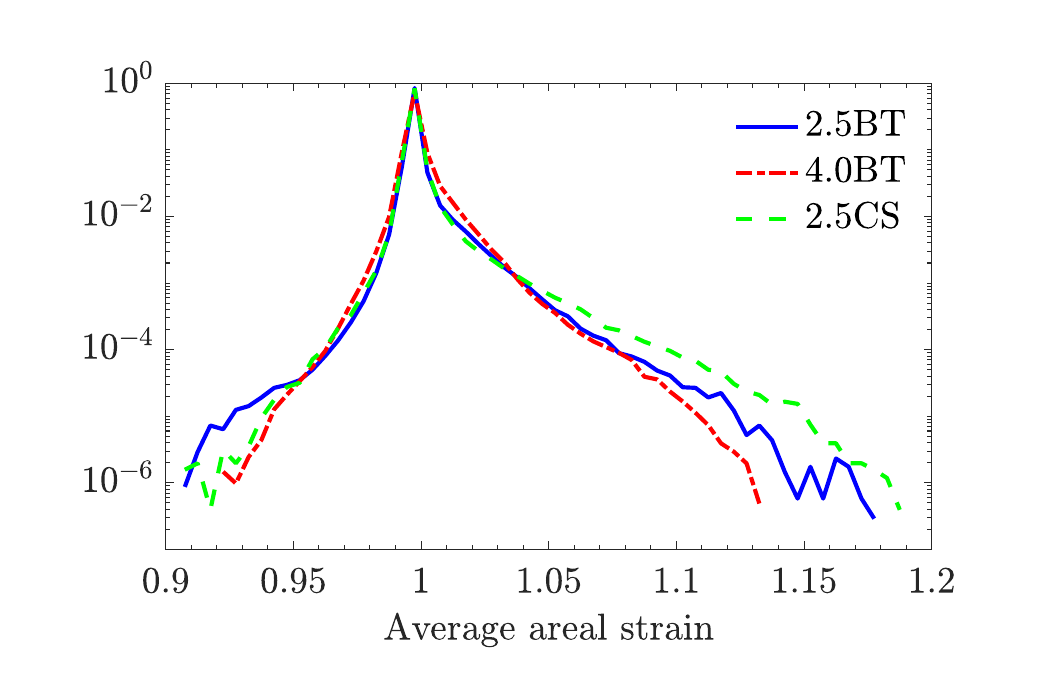}
  \vspace*{-8mm}
  \label{fig_pdf:sub3}
\end{subfigure}
\begin{subfigure}{.005\textwidth}
\begin{tikzpicture}
\node[anchor=south west] at (0,0) {};
\node[overlay] at (0pt,110pt) {(d)};
\end{tikzpicture}
\end{subfigure}
\begin{subfigure}{.48\textwidth}
  \centering
  \includegraphics[width=1\linewidth]{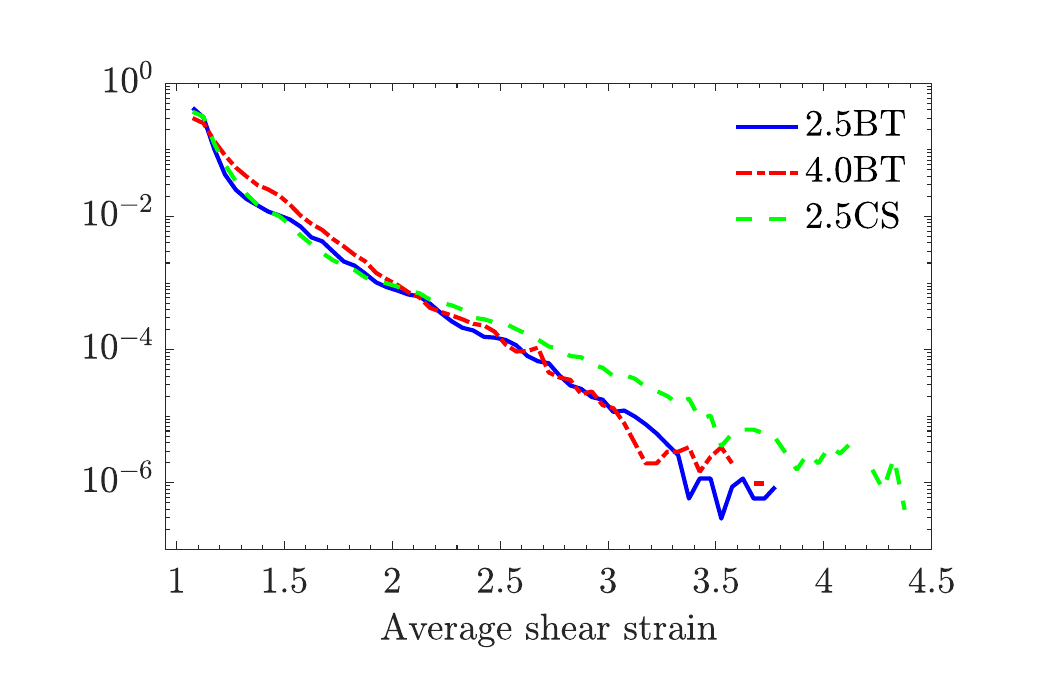}
  \vspace*{-8mm}
  \label{fig_pdf:sub4}
\end{subfigure}
\medskip
\vspace*{-4mm}
\begin{subfigure}{.005\textwidth}
\begin{tikzpicture}
\node[anchor=south west] at (0,0) {};
\node[overlay] at (0pt,110pt) {(e)};
\end{tikzpicture}
\end{subfigure}
\begin{subfigure}{.48\textwidth}
  \centering
  \includegraphics[width=1\linewidth]{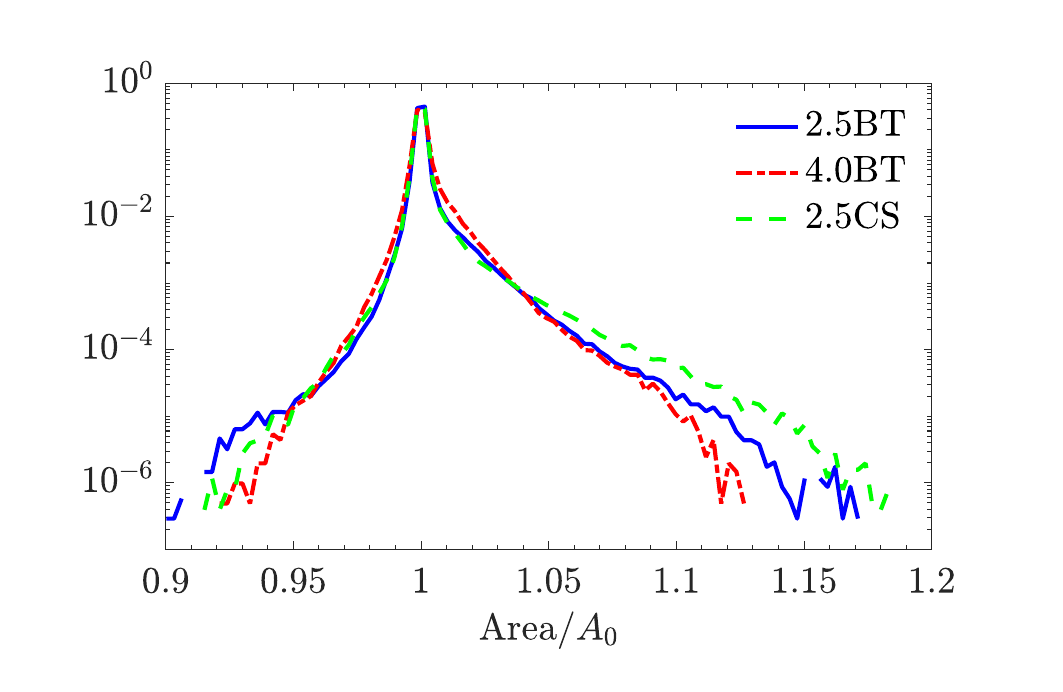}
  \label{fig_pdf:sub5}
  \vspace*{-8mm}
\end{subfigure}%
\begin{subfigure}{.005\textwidth}
\begin{tikzpicture}
\node[anchor=south west] at (0,0) {};
\node[overlay] at (0pt,110pt) {(f)};
\end{tikzpicture}
\end{subfigure}
\begin{subfigure}{.48\textwidth}
  \centering
  \includegraphics[width=1\linewidth]{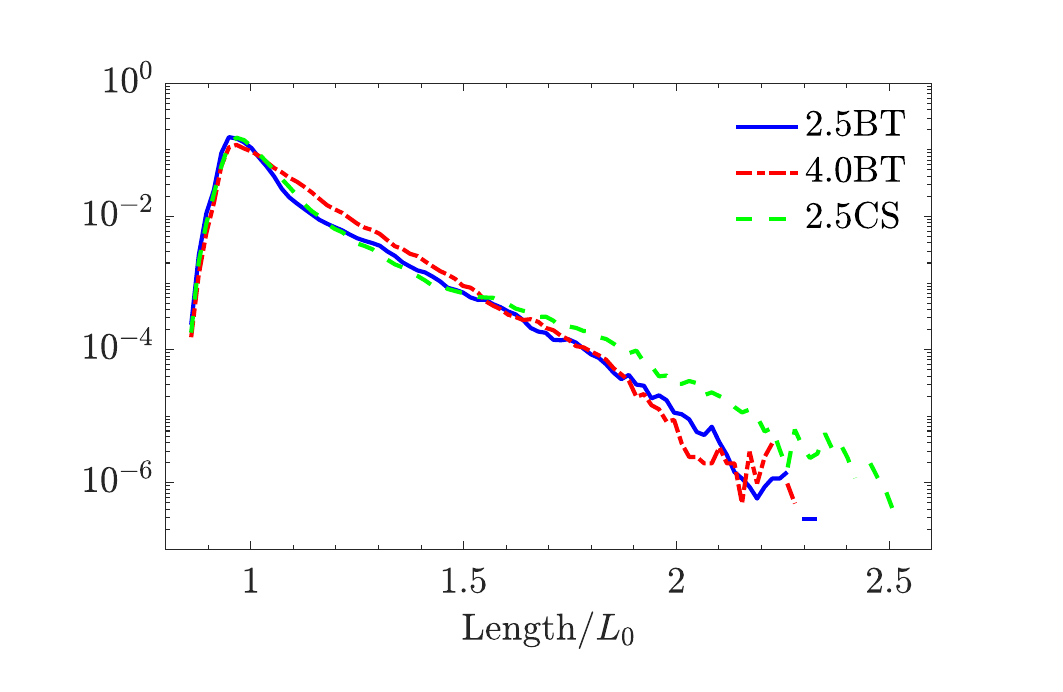}
  \label{fig_pdf:sub6}
  \vspace*{-8mm}
\end{subfigure}%
\caption{Probability density functions of different measures of RBC deformation: (a) and (b) maximum strains, (c) and (d) average strains, (e) and (f) overall RBC shape.}
\label{fig:PDF}
\end{figure}

 Figure \ref{fig:PDF} is one of the most important results of this study as it shows the PDFs of maximum (a,b) and average  (c,d) strains on the RBCs membrane as well as the cell's area, corresponding to global areal strain (e) and length (f). These PDFs are extracted for the three studied geometries considering all RBCs at all time steps. The cell's surface area and length are normalized by $A_0 = 16.8$ and $L_0 = 2.82$ in agreement with the literature \cite{Zhao2010}. The right tails of these PDFs are highly important as they mark the instances of highest deformation where hemolysis can occur.

The central shunt, having a shorter connection and thus a significant pressure drop per unit length compared to the mBTSs, creates instances of highest deformation. Furthermore, we anticipate that a smaller shunt causes higher RBC deformations compared to a larger shunt since a larger fraction of the cross-section is near the wall where the shear is the highest. Furthermore, the pressure drop across the shunt and hence the wall shear stress is larger for the smaller shunt. 
As observed in Fig. \ref{fig:PDF}(a), it is evident that the CS PDF tail reaches much higher values compared to the BT shunts. Differently put, the maximum areal strain that a RBC experiences as it goes through the CS can reach as high as 1.65, i.e., 65\% increase, whereas for the 2.5mm mBTS, this never exceeds 23\% (that upper bound figure is yet lower for 4.0mm mBTS at 20\%). Even limiting our attention to instances of 10-25\% area dilatation, the CS is more likely to subject RBCs to these less extreme events. Comparing the two modified BT shunt configurations, the smaller shunt causes more instances of high areal strain. The shear strain in Fig. \ref{fig:PDF}(b) tells a similar story with the CS producing the most extreme events followed by 2.5mm and 4.0mm mBTSs (7.9 relative to 4.8 and 4.5, respectively). 

A large difference is also observed between the right tail of the average areal strain, also referred to as $\langle{\lambda_1}{\lambda_2}\rangle$, PDFs (Fig. \ref{fig:PDF}(c)). Not only do RBCs in CS undergo the highest areal strain of up to 19\%, but also the probability of a RBC experiencing any area dilatation larger than 5\% is highest for CS, followed by 2.5mm mBTS and then 4.0mm mBTS. As for average shear strain, i.e., $\langle{\lambda_1}/{\lambda_2}\rangle$, (Fig. \ref{fig:PDF}(d)), the two mBTSs produce similar PDFs with the 4.0BT being slightly worse in terms of most extreme events. Again, the CS develops the highest chance of any average shear strain above 2. Note that, unlike PDFs of areal strain, shear strain probabilities do not have a left tail, since by definition it is always larger than one. 

Displayed in Fig. \ref{fig:PDF}(e) is the probability of the normalized area of RBCs, which is mainly similar to the average (areal) strain as expected. Thus, the same interpretation holds for this figure as well. The normalized length of RBCs along their elongation axis is shown in Fig. \ref{fig:PDF}(f). Like what was seen for other deformation measures, the central shunt relative to the two BT shunts generates the most incidents at which RBC is elongated by more than two-fold. However, in terms of that metric, the two studied diameters of mBTS do not differ significantly. 
Table \ref{table:PDFStat} in Appendix \ref{appendix:Extras} lists the mean and standard deviation (SD) of PDFs shown in Fig. \ref{fig:PDF}.

\subsubsection{Red blood cell deformation and mechanical damage}
Earlier experimental studies reported that extensive rupture of RBCs happens after a critical shear rate of around 42,000 s$^{-1}$ \cite{Leverett1972}. Furthermore, studies suggested critical area expansion for RBC damage ranges from 2\% to 10\% under steady shear flow conditions \cite{Mohandas1994,Razizadeh2020}. For instance, area dilatation of 6\% is numerically observed for RBCs subjected to shear flow as the shear rate approaches its critical limit in different studies \cite{Vitale,Sohrabi2017}. Razizadeh et al. \cite{Razizadeh2020} studied molecular dynamics of pore formation, growth, and recovery, for RBC membranes under large deformation. They showed that under large deformations pores will form on RBCs membrane and there exists a critical areal strain above which the pores will start forming. The onset of this mechanical pore formation is found to be 10.2\% and 12.5\% areal strains for two lipid bi-layers attached to cytoskeletons with junction-to-junction distances similar to that of the RBCs and twice denser cytoskeleton compared to that of the RBC at its normal state. Moreover, it has been shown that elongation of RBCs can cause structural damage \cite{Mohandas1994, Evans1975}, with critical shear strain $\lambda_1/\lambda_2$ being 2.4 to 3. The highest values reported in the literature for the ratio of elongated length to initial length ranges from 2 to 2.3 \cite{Sutera1975,Mohandas1994,Sohrabi2017}. Since higher elongations were not observed, this range is considered to be the critical elongation after which cells rupture.

Even though probability density functions of various deformation indices in Fig.  \ref{fig:PDF} provide a good statistical understanding of the overall mechanical behavior of RBCs in different surgical designs, it is still unclear how or to what extent such deformations cause blood damage. This is due to the fact that neither the number nor the location of RBCs experiencing high deformation can be read from the observed PDFs.
Based on this, the effect of critical global areal strain, local shear strain, and elongation for RBC damage is explored in Table \ref{table:DamagePer}, where a range of critical values are selected based on the discussed literature. The reason behind this is that for each determinant, the percentage of RBCs that pass the damage test is highly sensitive to the selected threshold. It is seen that using different critical elongation ratios, area expansion values, and shear strains, the CS always has a much higher percentage of mechanically damaged RBCs compared to the two mBTSs with the 2.5mm shunt being slightly more damaging than the 4.0mm one in most cases. For instance, for critical stretch ratio $\left({L}/{L_0} \right)_c = 2$, damaged RBCs for the central shunt are 2.8 times more than that of the 2.5BT and 6.3 times relative to 4.0BT. Based on $\langle{\lambda_1}/{\lambda_2}\rangle_c = 3$, the CS is 3 to 3.8 times more damaging when it is compared against 2.5BT and 4.0BT, respectively.
\vspace{1mm}

\begin{table}[ht!]
\renewcommand{\arraystretch}{1.0}
\centering
\caption{Damaged RBC percentage using different measures}
\begin{tabular}{ p{7.cm} p{1.9cm} p{2cm} p{2cm} p{1.9cm} }
\hline
\textbf{Criteria} & \textbf{Threshold} & \textbf{2.5BT (\%)} & \textbf{4.0BT (\%)} & \textbf{2.5CS (\%)} \\

\hline

 & 1.75 & 19.7 & 16.7 & 36.0 \\
 Critical elongation \cite{Sutera1975,Mohandas1994,Sohrabi2017}: $\left( {L}/{L_0} \right)_c$ & 2 & 3.1 & 1.4 & 8.8 \\
 & 2.25 & 0.4 & 0.2  & 2.1\\

 \hline 

 & 2.5 & 16.9 & 14.5 & 30.8 \\
 Critical mean shear strain \cite{Mohandas1994, Evans1975}: $\langle{\lambda_1}/{\lambda_2}\rangle_c$ & 3.0 & 3.1 & 2.4 & 9.2 \\
 & 3.5 & 0.4 & 0.6 & 2.1  \\

 \hline 

 & 10\% & 12.3 & 2.3 & 22.9 \\
 Critical area increment \cite{Razizadeh2020}: $\left( {A}/{A_0} \right)_c$ & 12\% & 5.7 & 0.6 & 11.1 \\
 & 14\% & 1.0 & 0.0  & 4.3\\

\hline
\end{tabular}
\label{table:DamagePer}
\end{table}

\begin{figure} [ht!]
\centering

\begin{subfigure}{.08\textwidth}
\centering
  \begin{tikzpicture}
\node[anchor=south west] at (0,0) {\includegraphics[width=0.8\linewidth]{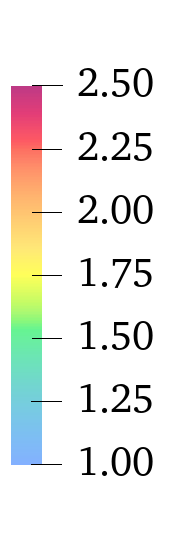}
};
\node[overlay] at (20pt,100pt) {$L/L_0$};
\end{tikzpicture}
\vspace{4mm}
\end{subfigure}
\begin{subfigure}{.28\textwidth}
  \centering
  \includegraphics[width=1\textwidth]{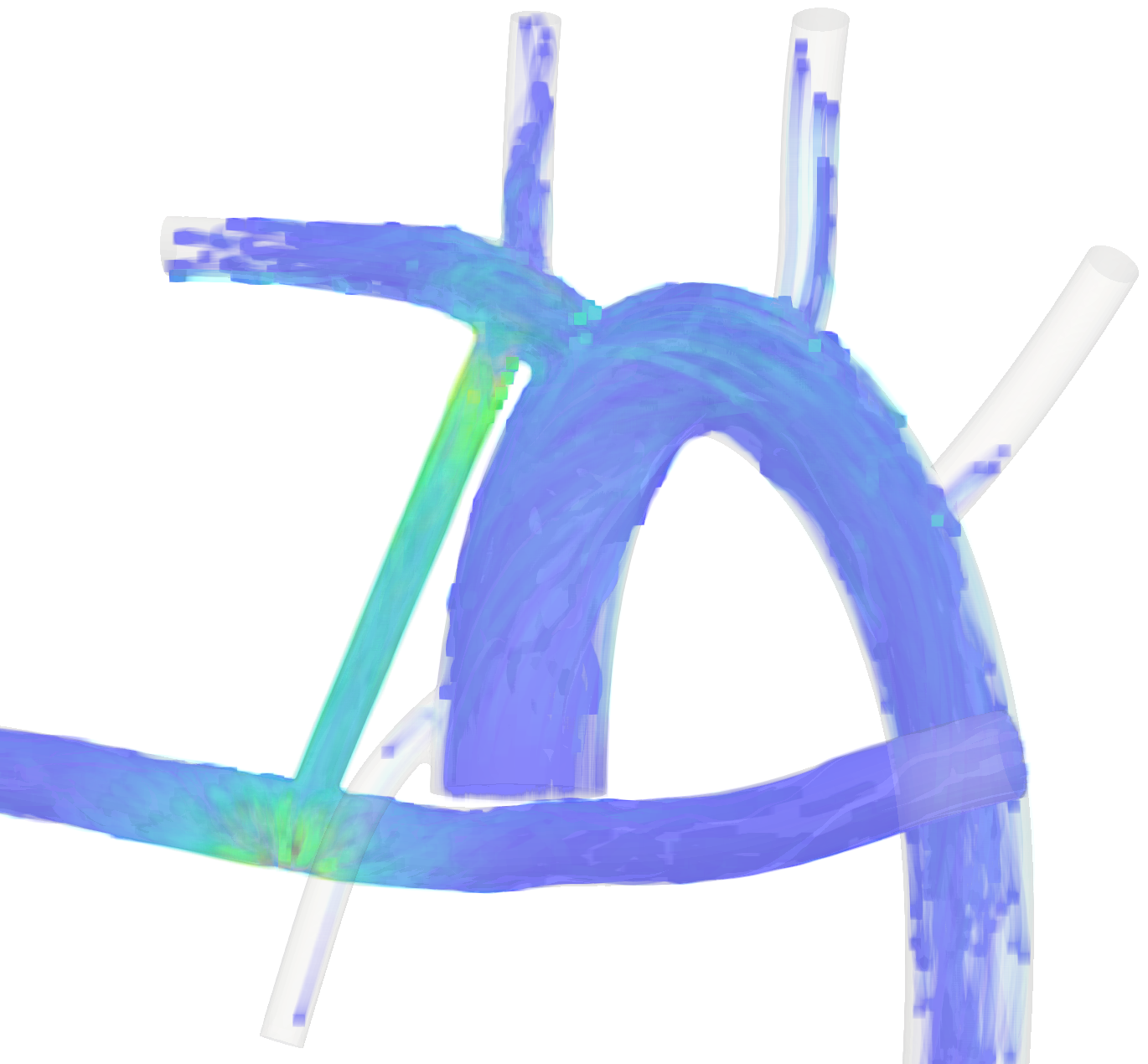}
  \label{fig_b:sub1}
\end{subfigure}
\begin{subfigure}{.28\textwidth}
  \centering
  \includegraphics[width=1\linewidth]{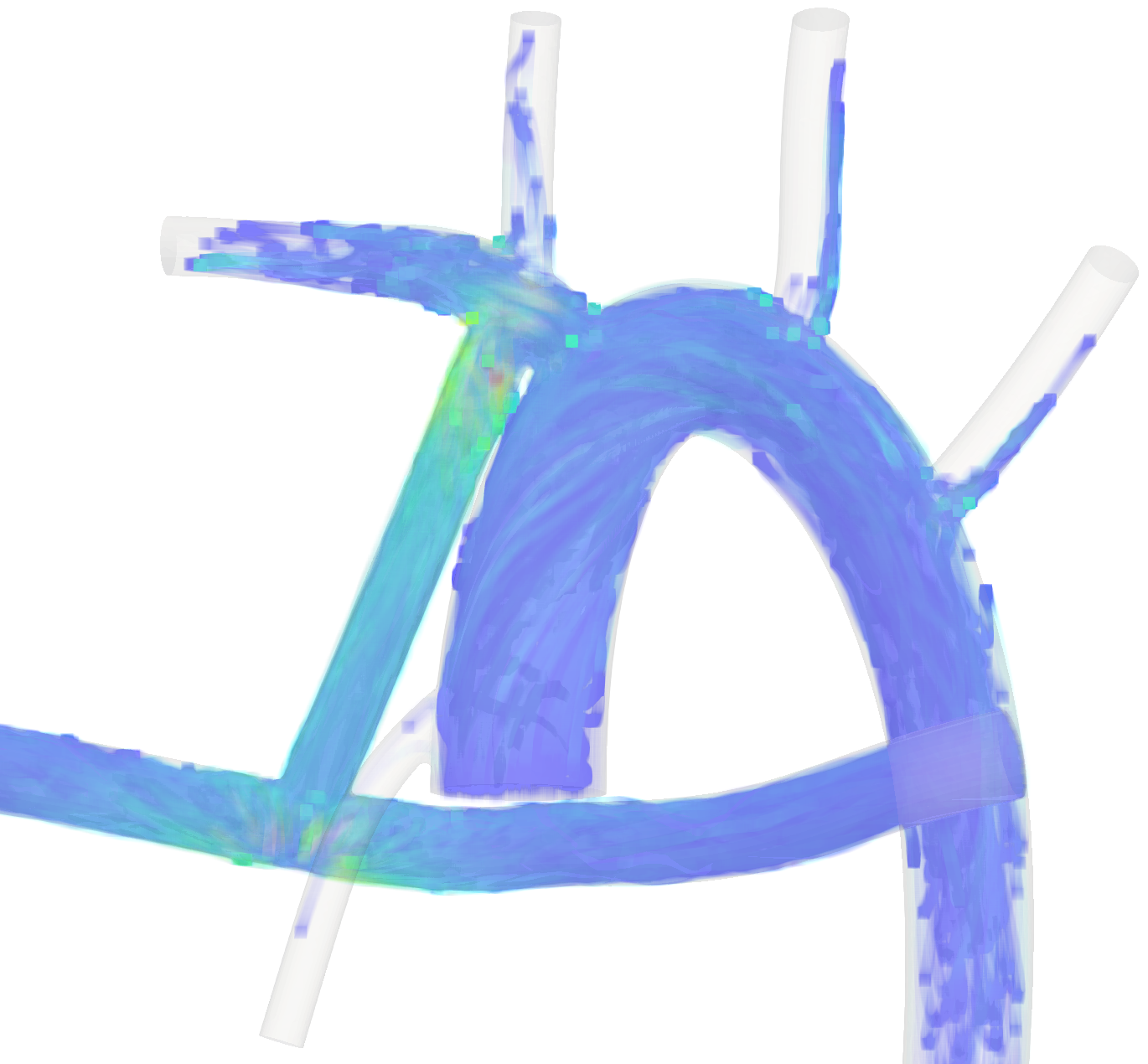}
  \label{fig_b:sub2}
\end{subfigure}
\begin{subfigure}{.28\textwidth}
  \centering
  \includegraphics[width=1\linewidth]{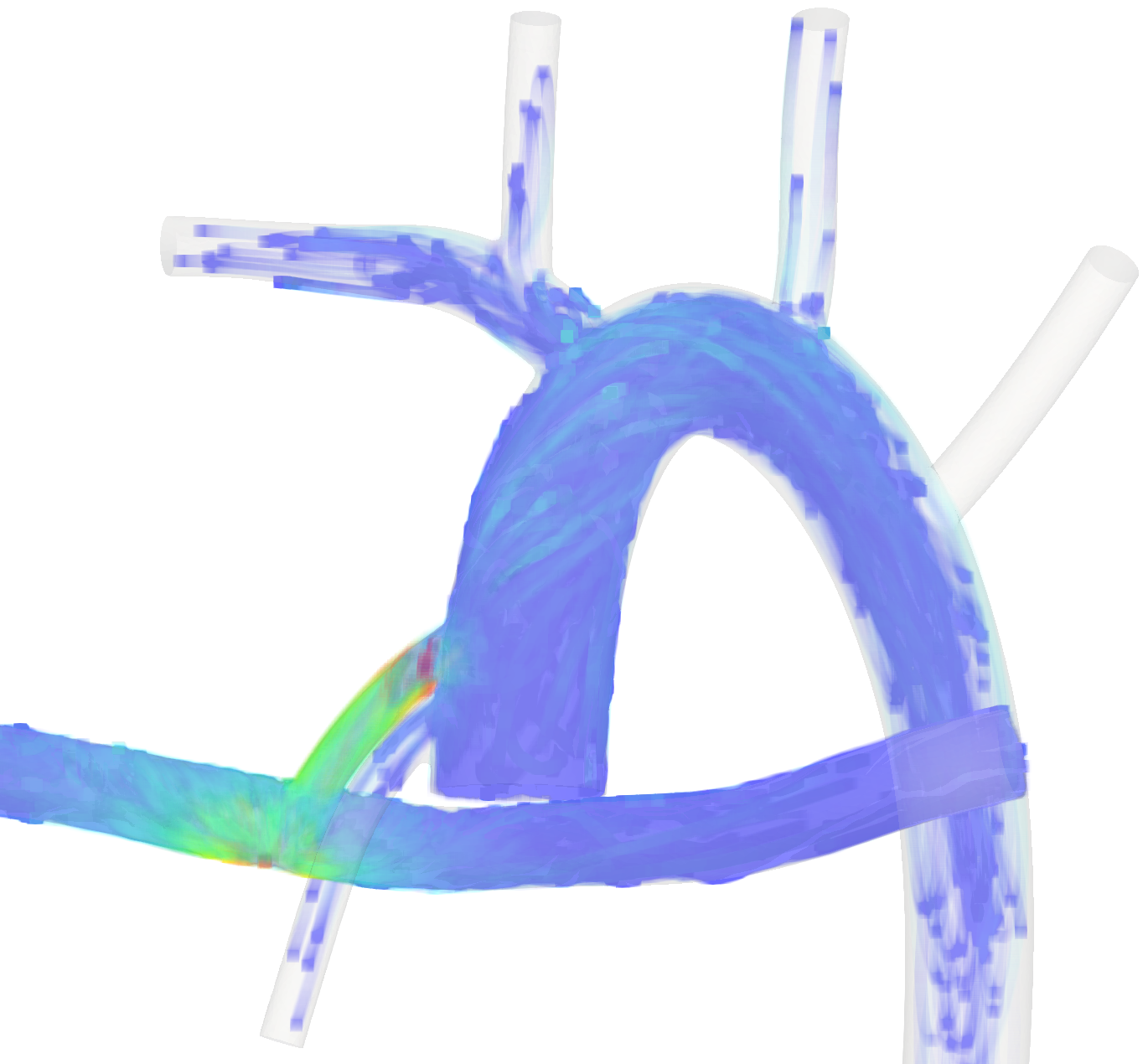}
  \label{fig_b:sub3}
\end{subfigure}%

\begin{subfigure}{.08\textwidth}
\centering
\begin{tikzpicture}
\node[anchor=south west] at (0,0) {\includegraphics[width=0.8\linewidth]{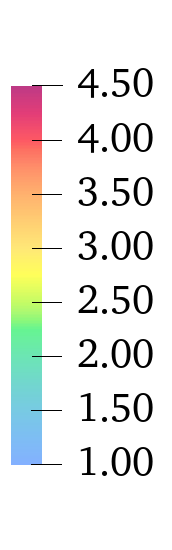}
};
\node[overlay] at (20pt,100pt) {$\lambda_1/\lambda_2$};
\end{tikzpicture}
\vspace{4mm}
\end{subfigure}
\begin{subfigure}{.28\textwidth}
  \centering
  \includegraphics[width=1\linewidth]{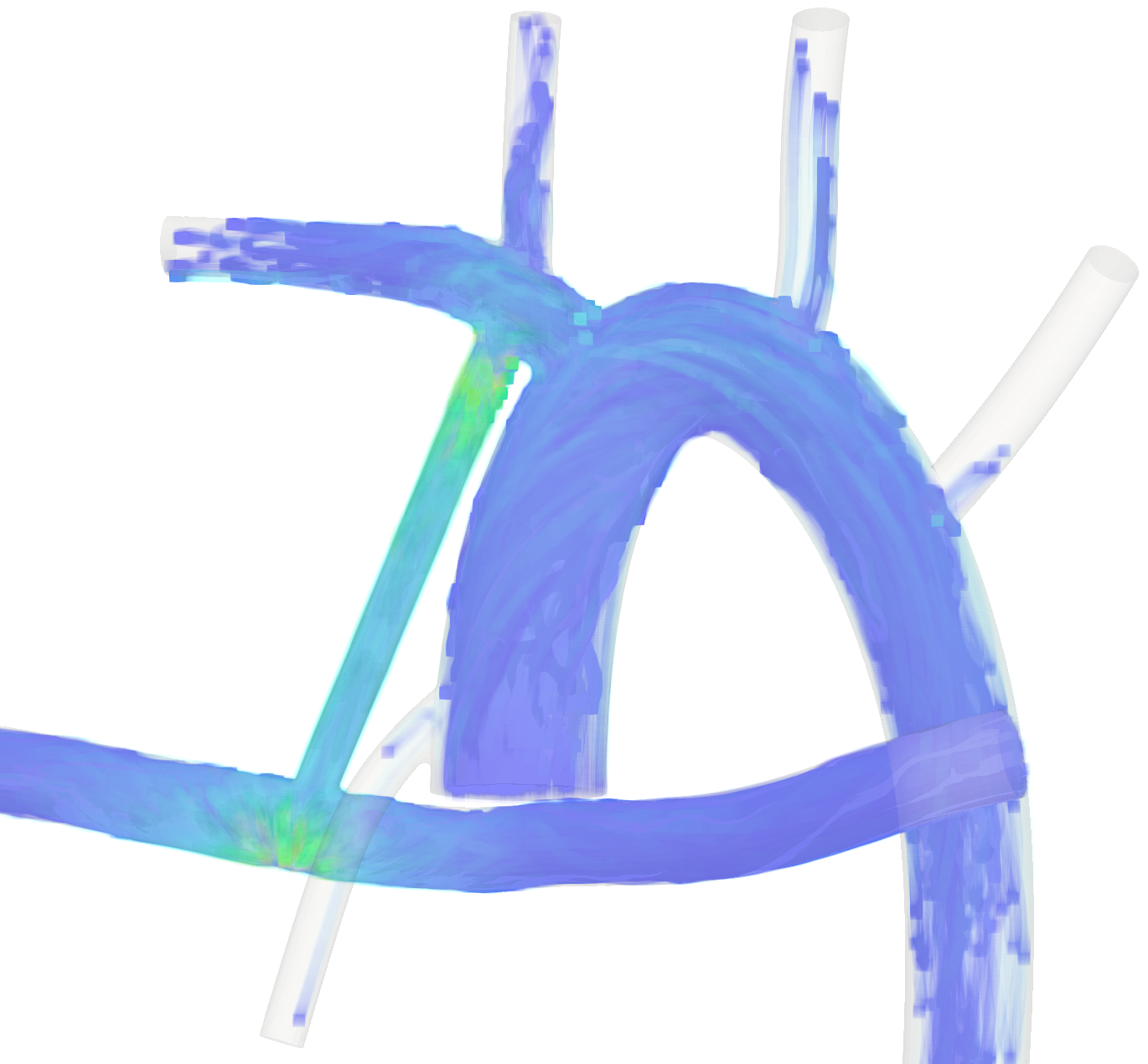}
  \label{fig_c:sub1}
\end{subfigure}
\begin{subfigure}{.28\textwidth}
  \centering
  \includegraphics[width=1\linewidth]{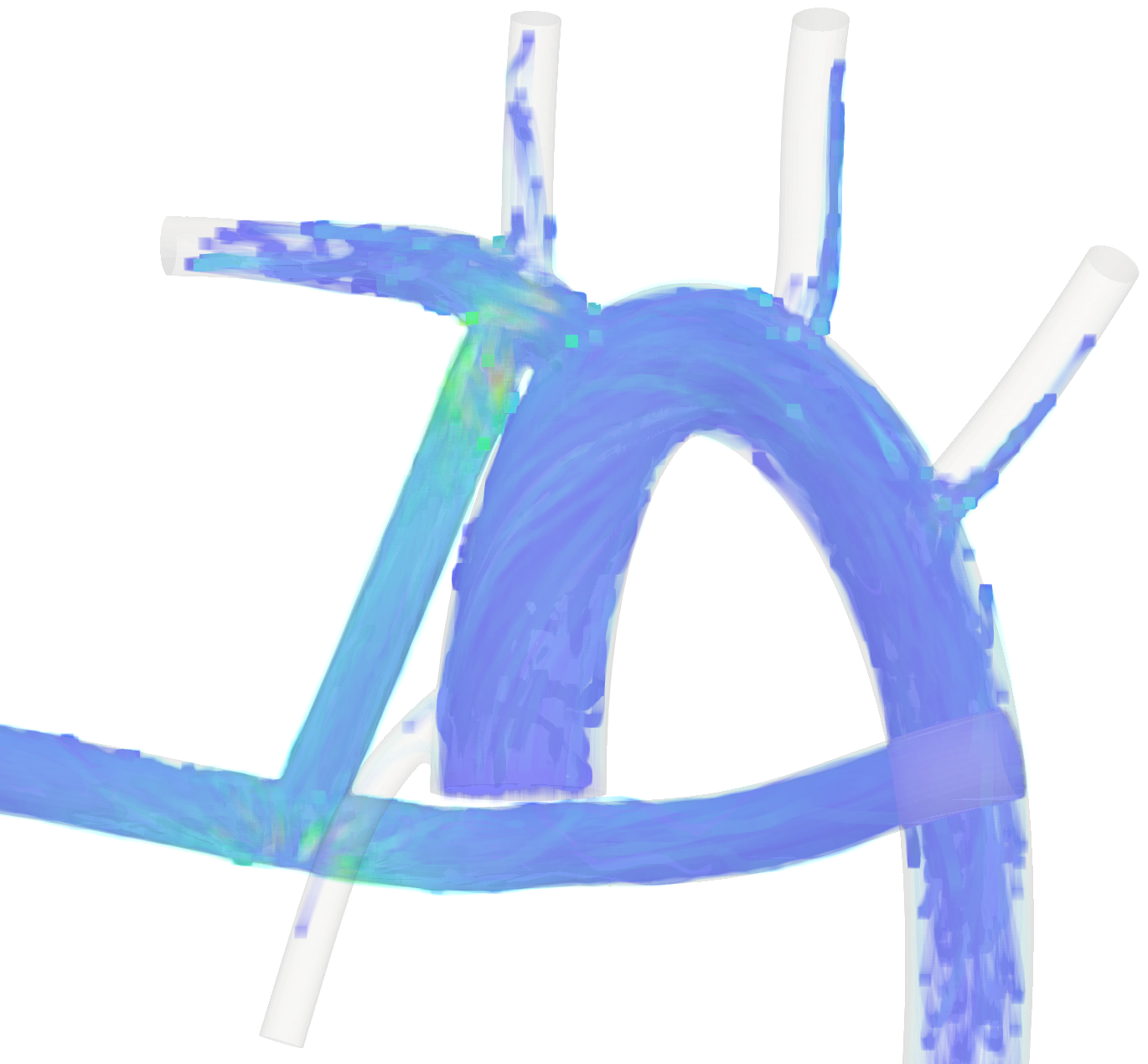}
  \label{fig_c:sub2}
\end{subfigure}
\begin{subfigure}{.28\textwidth}
  \centering
  \includegraphics[width=1\linewidth]{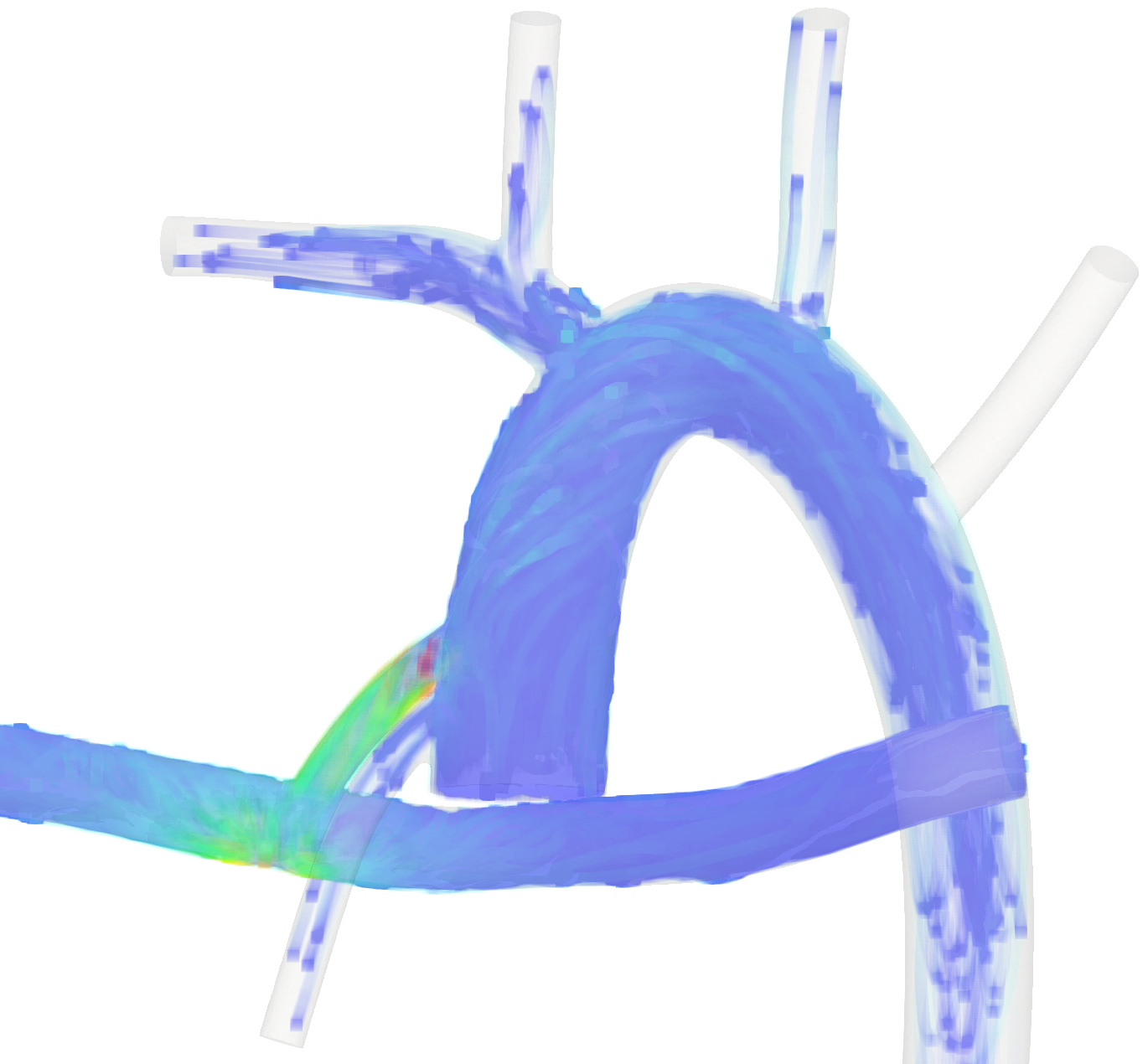}
  \label{fig_c:sub3}
\end{subfigure}%

\begin{subfigure}{.08\textwidth}
\centering
  \begin{tikzpicture}
\node[anchor=south west] at (0,0) {\includegraphics[width=0.8\linewidth]{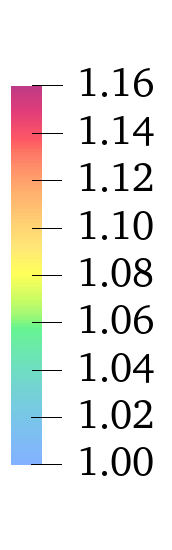}
};
\node[overlay] at (20pt,100pt) {$A/A_0$};
\end{tikzpicture}
\vspace{4mm}
\end{subfigure}
\begin{subfigure}{.28\textwidth}
  \centering
  \includegraphics[width=1\linewidth]{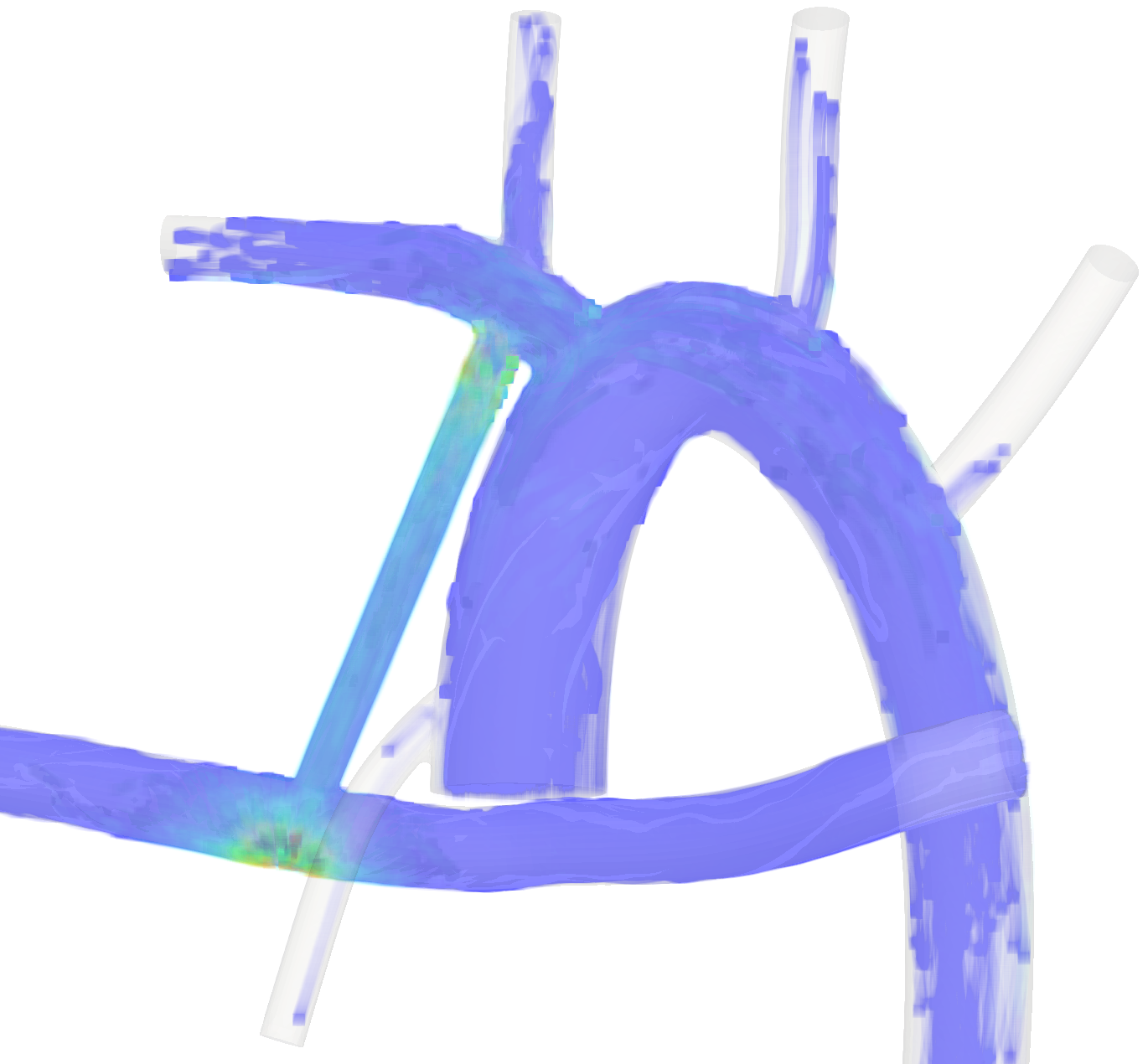}
  \caption*{2.5mm modified BT shunt}
  \label{fig_d:sub1}
\end{subfigure}
\begin{subfigure}{.28\textwidth}
  \centering
  \includegraphics[width=1\linewidth]{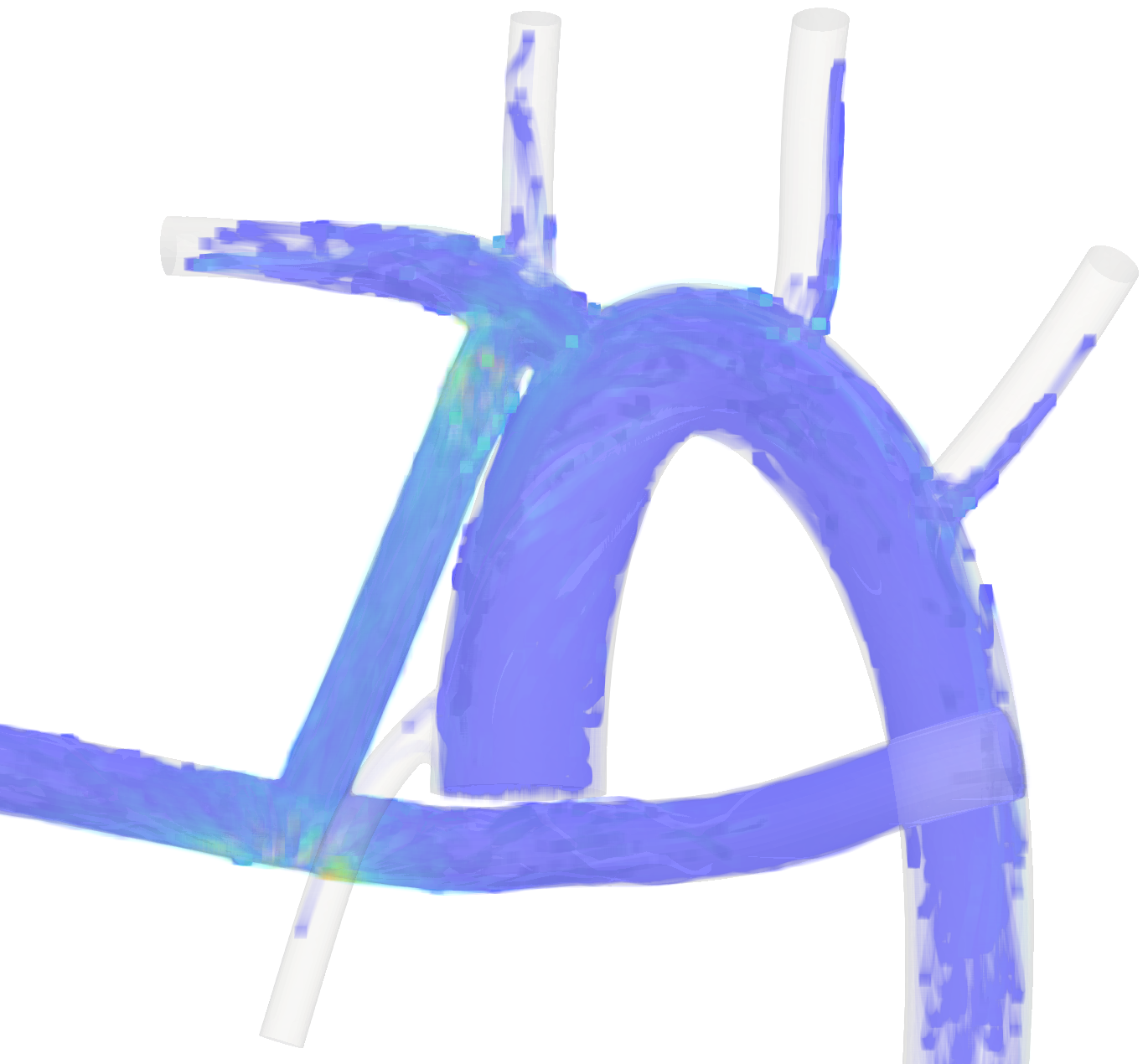}
  \caption*{4.0mm modified BT shunt}
  \label{fig_d:sub2}
\end{subfigure}
\begin{subfigure}{.28\textwidth}
  \centering
  \includegraphics[width=1\linewidth]{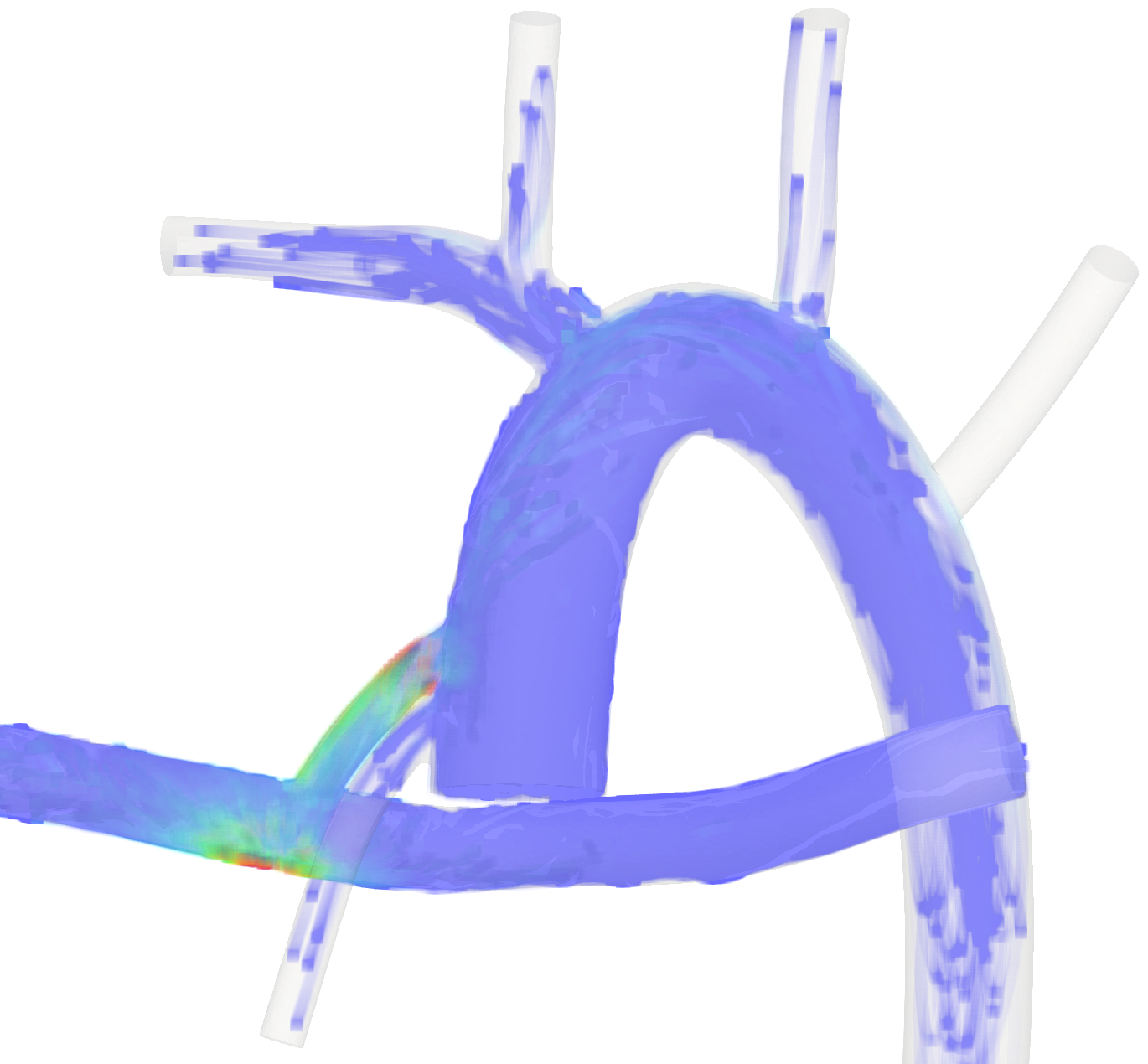}
  \caption*{2.5mm central shunt}
  \label{fig_d:sub3}
\end{subfigure}%

\caption{Damage maps}
\label{fig:damage}
\end{figure}
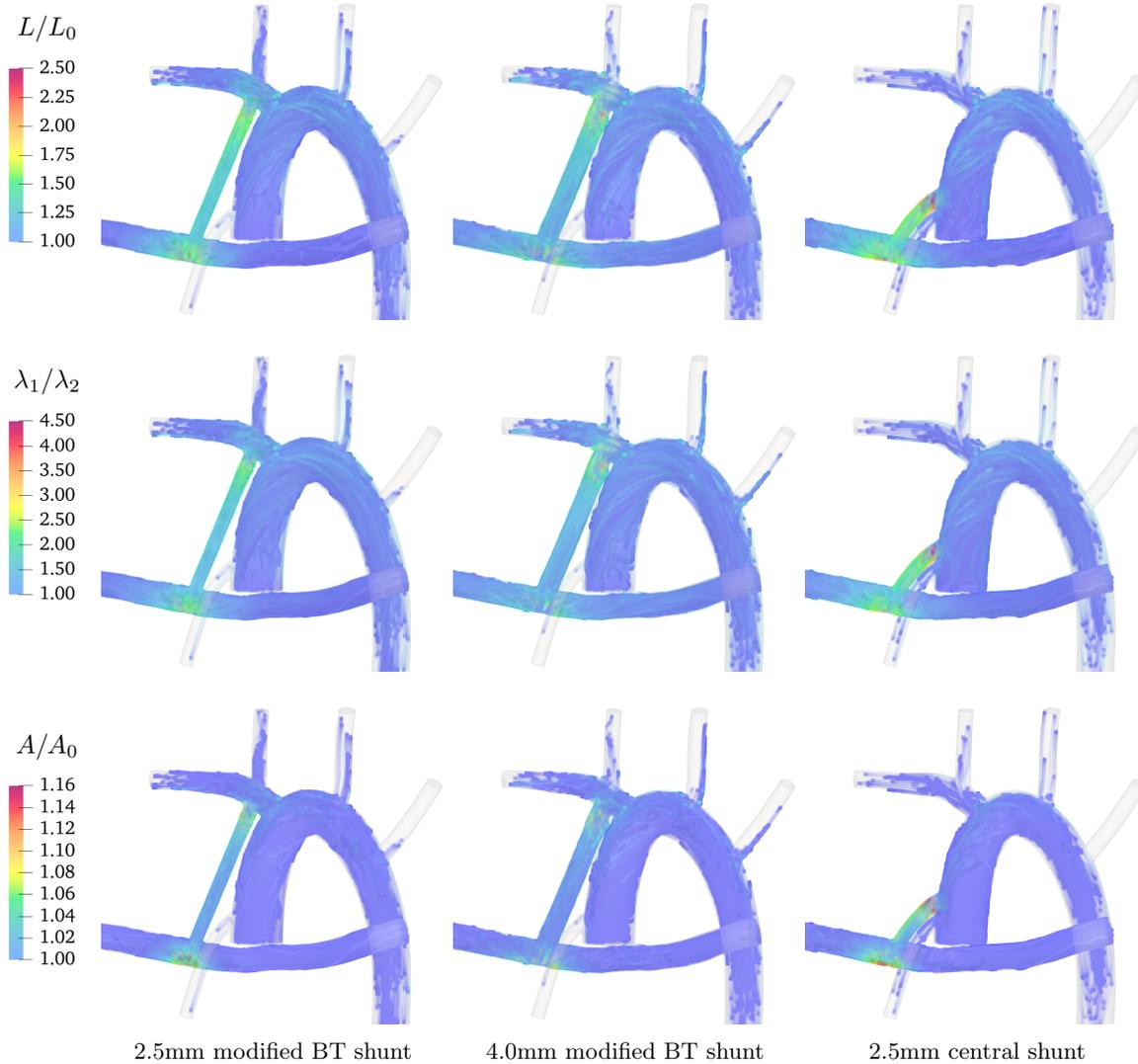

Beyond quantification of blood damage probability, having a spatial map of where RBC damage occurs can guide the design of surgeries more effectively by targeting the hot zones when anatomical modifications are made.
To this end, the distribution of RBC deformation is illustrated in Fig. \ref{fig:damage}, where RBCs' path lines are colored according to the extent of deformation. 
Here, rows correspond to the distribution of elongation, mean shear strain, and area expansion values from top to down. The extent to which RBCs are likely to get damaged through extreme deformation is captured by their color, with dark blue being the least deformed to red being the most deformed.
Columns of this figure from left to right represent 2.5mm mBTS, 4.0mm mBTS, and 2.5mm CS surgical configurations. Note that some parts of the right PA and AoD are trimmed in this figure to provide more space for important details.

It is evident from these damage maps that for the 2.5mm CS, the regions right after the shunt entrance (proximal) and the shunt exit (distal) are both high-risk. As for the 4.0mm mBTS, only after the shunt entrance do RBCs experience high deformation. It is interesting to see that that critical area shifts to the distal end of the shunt for 2.5mm mBTS in a trend that is similar to that of the WSS.
It is also seen that instances of large deformation are much more prominent in the case of CS compared to the two BT shunts. Additionally, the 2.5BT creates more instances of damaging large deformations as compared to the larger diameter of the same configuration. Comparing different rows of this figure shows even though the hot zones are slightly different from one damage criterion to the other, the conclusion would not differ in terms of the relative flow-induced damage associated with different configurations.

Few observations can be made by comparing these damage maps to Fig. \ref{fig:wss}. Firstly, in agreement with what the WSS maps suggest, RBCs get damaged mostly inside or in the vicinity of the shunt due to high velocity gradients that cause large deformations. Thus, locations with large WSS values are more susceptible to mechanical damage. Secondly, despite the similarities between these maps, the cell-resolved simulations remain an indispensable tool by providing quantitative measures of deformation-induced RBC damage. 

\begin{figure} [ht!]

\begin{subfigure}{.32\textwidth}
  \centering
  \includegraphics[width=0.95\linewidth]{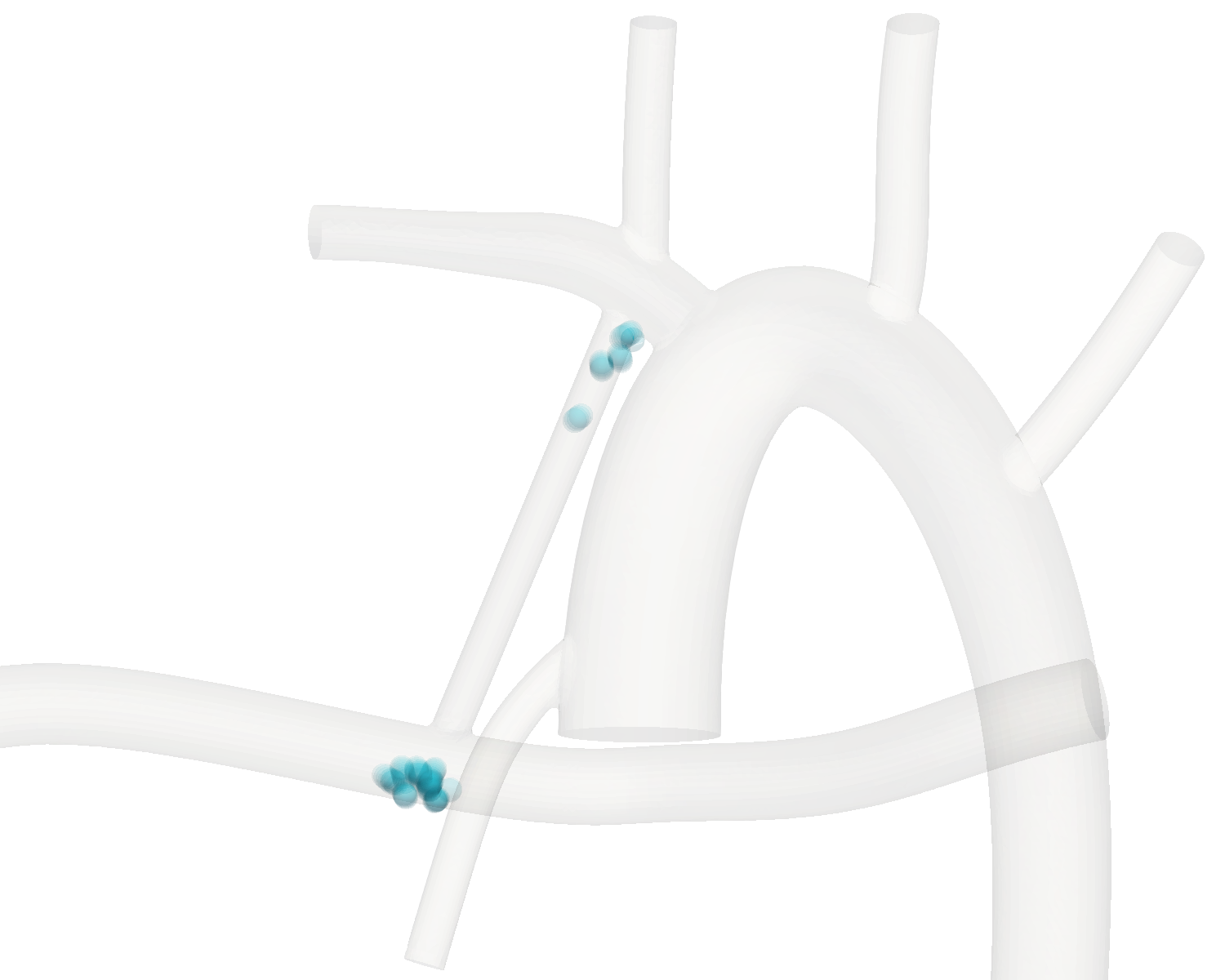}
  \caption{2.5mm modified BT shunt}
  \label{fig_a:sub1}
\end{subfigure}
\begin{subfigure}{.32\textwidth}
  \centering
  \includegraphics[width=0.95\linewidth]{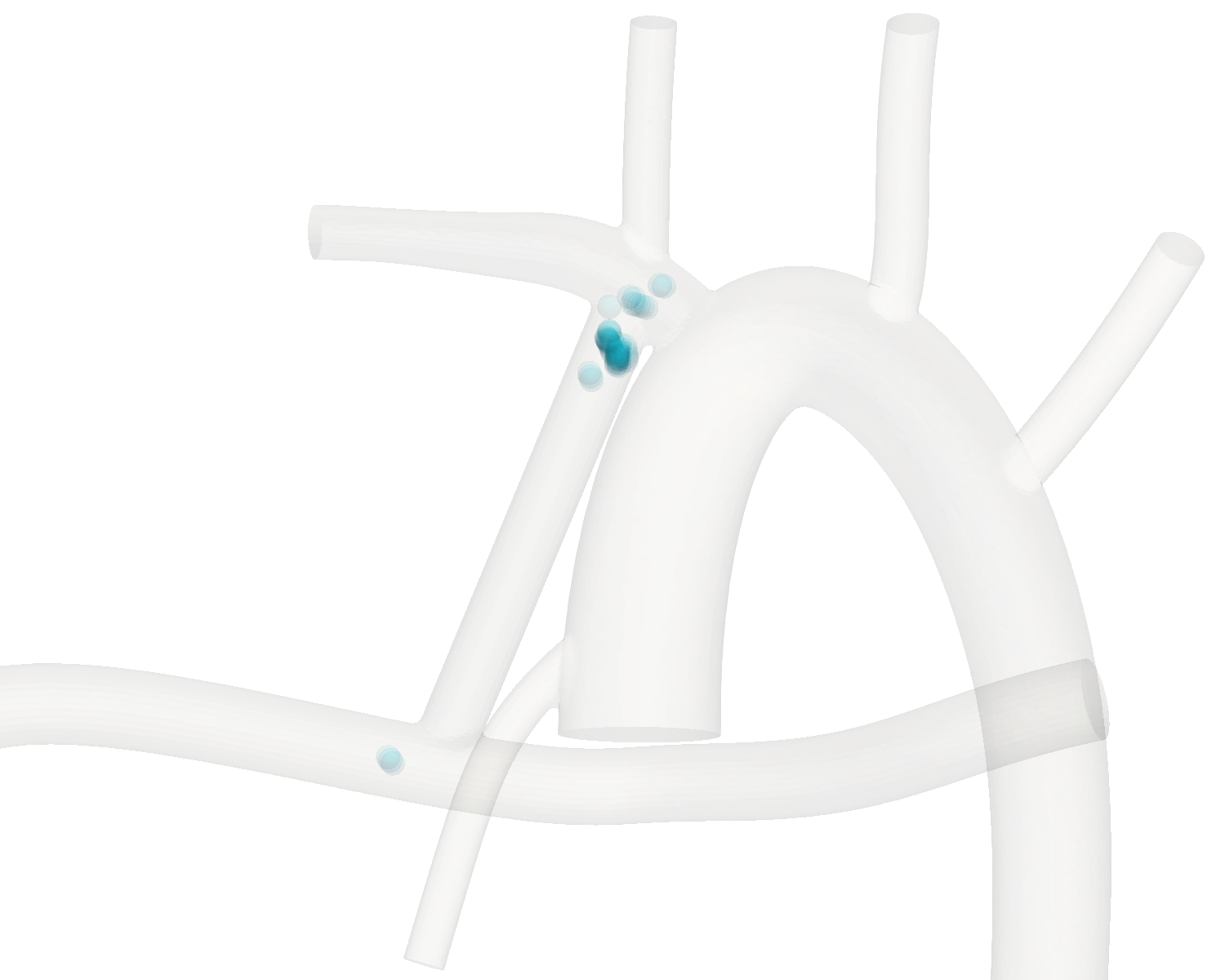}
  \caption{4.0mm modified BT shunt}
  \label{fig_a:sub2}
\end{subfigure}
\begin{subfigure}{.32\textwidth}
  \centering
  \includegraphics[width=0.9\linewidth]{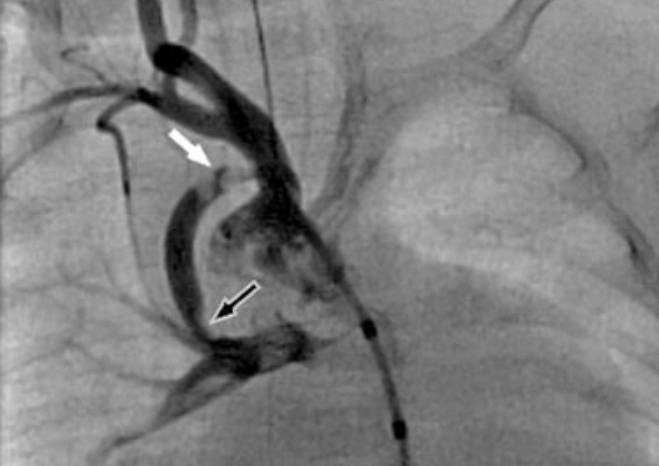}
  \caption{3.5mm modified BT shunt}
  \label{fig_a:sub3}
\end{subfigure}%

    \caption{Comparison of discrete damage maps obtained using $\lambda_1/\lambda_2 \geq 3$ for (a) 2.5mm mBTS, and (b) 4.0mm mBTS with (c) clinical image of a 3.5mm mBTS showing significant narrowing in the proximal part of the BT shunt with filling defect due to an elongated mass suggestive of thrombus formation (white arrow) as well as distal narrowing of the shunt (black arrow) reproduced from Ghawi et al. \cite{Ghawi2011} with permission from SNCSC.
}
\label{fig:DamageComp}
\end{figure}

Although clinical data on comparing CS and mBTS are scarce, there is some evidence that points to a lower level of complications in mBTS configuration. More specifically, clinical data suggest that the RV-PA Sano shunt, which one may argue produces flow conditions closer to that of the CS, poses more complications than mBTS that require interventions; dhat fact, higher short-term survival rate has been reported for patients receiving a Sano shunt than a mBTS \cite{Ohye2010,Ryerson2007,Arnaout2023}.  
Furthermore, the association of smaller shunts with an increased risk of failure is reported in clinical data \cite{Karpawich1985}.
The formation of thrombus has been clinically observed within the mBTS at the same locations identified by our in silico analysis. More specifically, Ghawi et al. \cite{Ghawi2011} observed significant narrowing of the proximal part of a 3.5mm modified BT shunt with a filling defect due to an elongated mass suggestive of thrombus formation. This is also the region where we have observed the cell elongation, shear strain, and area dilatation exceeds the damage threshold in our BT shunt simulations (see Fig. \ref{fig:DamageComp} for comparison of \textit{in silico} and \textit{in vivo} results).
Acute occlusion of the proximal part of the mBTS \cite{Moszura2010} and the proximal part of one pulmonary artery due to insertion of mBTS \cite{Pota2022} have also been observed in clinical studies.
Such similarity serves as further support for the suitability of the proposed approach for evaluating the risk of blood damage-related complications in various surgical options. 

Since the PDFs shown in Figure \ref{fig:PDF} were constructed by treating each time point as a separate ensemble, they do not directly capture the effect of cells' shear history on hemolysis. Accounting for this history is especially important as the accumulation of damage due to deformation affects hemolysis. 
To capture this effect, the number of instances of large deformation normalized by the number of damaged RBCs is reported in Table \ref{table:DamageIns}. This metric, which provides a measure of the duration by which RBCs are exposed to damaging conditions, leads us to the same conclusion as before. Namely, the central shunt is a riskier option compared to BT shunts, followed by the smaller mBTS.

\begin{table}[ht!]
\renewcommand{\arraystretch}{1.2}
\centering
\caption{Number of instances per damaged RBC}
\begin{tabular}{ p{2.4cm} p{1.4cm} p{1.4cm} p{1.1cm} }
\hline
\textbf{Criteria} & \textbf{2.5BT} & \textbf{4.0BT} & \textbf{2.5CS} \\

\hline

$\left( {L}/{L_0} \right)_c$ $\geq$ 2 & 11.9 & 9.7 & 15.3 \\
$\langle{\lambda_1}/{\lambda_2}\rangle_c$ $\geq$ 3.0 & 13.0 & 8.4 & 15.5 \\
$\left( {A}/{A_0} \right)_c$ $\geq$ 12\% & 4.1 & 2.0 & 6.3 \\

\hline
\end{tabular}
\label{table:DamageIns}
\end{table}

An alternative way to account for the history of cell deformation is to rely on a cell damage index (CDI), which is defined based on the cumulative deviation of RBC area \cite{Gusenbauer2018} as
\begin{equation}
   CDI = \frac{1}{n_{RBC} }\sum_{id=1}^{n_{RBC}} \sum_i \left( \frac{ |A_{id}(t_i) - A_0|}{A_0} \right)^\alpha \Delta t_i. 
\label{eqn:cdi}
\end{equation}
$\alpha$ in this relation controls the importance of extreme events such that $\alpha > 1$ assigns more weight to instances of larger deformation in determining the damage. This is in opposition to $\alpha=1$ \cite{Gusenbauer2018} that provides equal weight to all instances by simply averaging over all of them. Note that while Eq. \eqref{eqn:cdi} is defined for the area, similar metrics can be defined for the elongation ratio as well as shear strain. The result of these calculations is reported in Table \ref{table:CDI}.

Using $\alpha = 1$ in Eq. \ref{eqn:cdi}, results in CDIs of 1.28 are 1.08 for 2.5mm mBTS and 4.0mm mBTS when normalized with 2.5mm CS case as the reference, respectively. This is in disagreement with previous observations. The reason behind this is that the summation of very small deformations can contribute to CDI as much as a large deformation can.
In fact, the average residence time of all RBCs can be calculated and is 0.204, 0.330, and 0.239 s for 4.0BT, 2.5BT, and 2.5CS, respectively, explaining why the cell damage index is elevated for the 2.5mm mBTS.
Since very small deviations of RBC shape from its resting condition are rather elastic and do not contribute to permanent damage of RBCs as much as more extreme events, $\alpha > 1$ would provide a more realistic picture of damage.  This non-linear increase in the rate of damage accumulation with the stress level is also well supported and shown by experimental data \cite{Yeleswarapu1995}.
In fact, the corresponding empirical coefficient of $\alpha$ in the power law model ($HI = c \tau^\alpha t^\beta$; where $HI$ and $\tau$ are hemolysis index and shear stress) that best capture experimental results ranges from 1.99 to 2.42 \cite{Giersiepen1990,Song2003a}. Thus, relative cell damage indices using different values of this exponent are reported in Table \ref{table:CDI} for damage criteria that are defined based on elongation, shear strain, and area expansion. 

\begin{table}[ht!]
\renewcommand{\arraystretch}{1.0}
\centering
\caption{CDI of different configurations normalized with that of 2.5CS }
\begin{tabular}{p{5.4cm} p{1cm} p{1.4cm} p{1.4cm} p{1.2cm} }
\hline
\textbf{Criteria} & \textbf{$\bm{\alpha}$} & \textbf{2.5BT} & \textbf{4.0BT} & \textbf{2.5CS} \\
 \hline 
& 1.0 & 1.41 & 1.07 & 1.00 \\
Elongation: $\left( {(L - L_0)}/{L_0} \right)^\alpha$ 
& 2.5 & 1.18 & 1.00 & 1.00  \\
& 4.0 & 0.86 & 0.64 & 1.00  \\
\hline

& 1.0 & 1.32 & 1.04 & 1.00 \\
Shear strain: $\left( \lambda_1/\lambda_2 - 1 \right)^\alpha$ 
& 2.5 & 1.17 & 1.07 & 1.00  \\
& 4.0 & 0.83 & 0.66 & 1.00  \\
\hline

& 1.0 & 1.28 & 1.08 & 1.00 \\
Area expansion: $\left( {(A - A_0)}/{A_0} \right)^\alpha$ 
& 2.5 & 0.92 & 0.56 & 1.00  \\
& 4.0 & 0.64 & 0.26 & 1.00 \\
\hline

\end{tabular}
\label{table:CDI}
\end{table}

The first observation based on these results is that regardless of the utilized exponent and criterion, 4.0mm mBTS is always a safer choice compared to 2.5mm mBTS.
Secondly, while all CDIs suggest that the 2.5mBTS is more damaging compared to the others at $\alpha=1$, increasing the exponent results in 2.5CS having the highest CDIs followed by the 2.5BT, and 4.0BT, respectively. Lastly, the $\alpha$ at which a given level of damage is predicted varies between different criteria. 
For instance for $\alpha=2.5$ and area expansion, $CDI_{\rm 2.5CS}$ is 9\% larger than that of 2.5BT, and $CDI_{\rm 4.0TS}$ is 39\% smaller compared to 2.5BT. These figures are 20\% and 21\% for instance when considering $\alpha= 4.0$ and shear strain instead. Thus, the relative destructiveness of instances of larger deformation compared to lower ones for each criterion (the $\alpha$ exponent), should be studied and measured experimentally before applying CDI to obtain a quantitative measure of damage.

It should be noted that the studied anatomies are only compared based on the relative red cell damage in this study. Other factors in the choice of a design such as the heart load, oxygen delivery, thrombus formation, etc., must be considered before a choice is made. Furthermore, a more direct link must be established between the mechanical stresses and strains applied to RBCs and the sublethal RBC damage and hemolysis. Therefore, even though the conclusion of this study may remain valid for comparison between various reconstructive surgeries or blood-wetted devices on a relative basis, the results cannot be used as a quantitative basis for evaluating the absolute risk of an operation for direct comparison against experimental hemolysis index yet.
Moreover, idealized geometries are utilized in this study for a general comparison of different surgical configurations. Patient-specific anatomical models will be required to produce more accurate predictions and tailored surgical planning for individual patients.

\section{Conclusions}
In line with the need for better surgical options for pediatric patients with severe heart defects like HLHS, this work conducts a multi-scale cell-resolved in silico study of flow-induced red blood cell deformation as a consequence of altered organ-scale blood flow paths in Norwood or stage I procedure. RBC damage is selected due to its connection with post-operative complications observed in these patients. Two different shunt configurations, i.e., Blalock-Taussig shunt and central shunt, were simulated and compared. Moreover, the effect of shunt size was studied by considering 2.5mm and 4.0mm modified BT shunts. 
These surgical options were compared in terms of various deformation statistics as well as spatial damage maps, which allowed for the identification of hot zones in each configuration. 

The central shunt configuration produces more incidents in which RBCs undergo extreme deformation. Between the two mBTS sizes, it was the smaller diameter that produced those extreme events more frequently, but only sparingly. We hypothesized that those incidents, which can damage RBCs, are caused by the steep pressure gradient within the central shunt that in turn exposes RBCs to a high level of shear. We also used various metrics to relate those deformation statistics to a measure of RBC damage. The metrics that did not directly account for the history of deformation produced similar predictions as those obtained from tails of distributions. Incorporating the history in such metrics, on the other hand, may or may not change those conclusions depending on the weight given to the extreme events relative to benign conditions observed along the RBC trajectory. For a reasonable weighting, however, the earlier conclusion of the central shunt being most damaging followed by the smaller and larger mBTSs is confirmed. The damage metrics could be translated to spatial maps, thus providing a powerful tool for guiding surgical design. The maps produced through this process not only are consistent with the wall shear stress maps, but also provide a more refined quantitative measure of damage. In fact, they highlight hot zones in the shunt that coincide with the reported clinical images of shunt thrombosis on an anecdotal basis. Further refinement of the damage model and its coupling with the coagulation cascade can make the proposed multi-scale framework an indispensable tool for managing risk in the design of surgeries or medical devices.

\section*{Acknowledgement}
This work was supported by the National Heart, Lung, and Blood Institute of the National Institutes of Health under award number R01HL089456.
The authors wish to thank Grant Rydquist for sharing his expertise on the cell-resolved modeling of RBCs.

\section*{CRediT author contributions statement}
\textbf{Saba Mansour}: Conceptualization, Data curation, Formal analysis, Investigation, Methodology, Software, Validation, Visualization, Writing-original draft. \textbf{Emily Logan}: Investigation, Software. \textbf{James F. Antaki}: Funding acquisition, Writing- review \& editing.
\textbf{Mahdi Esmaily}: Conceptualization, Funding acquisition, Supervision, Resources, Writing-review \& editing.



\begin{appendices}
\counterwithin{figure}{section}
\counterwithin{equation}{section}
\section{Lumped parameter network: heart model and parameters}\label{appendix:LPN}

\begin{figure} [ht!]
\centering
    \includegraphics[width=0.6\linewidth]{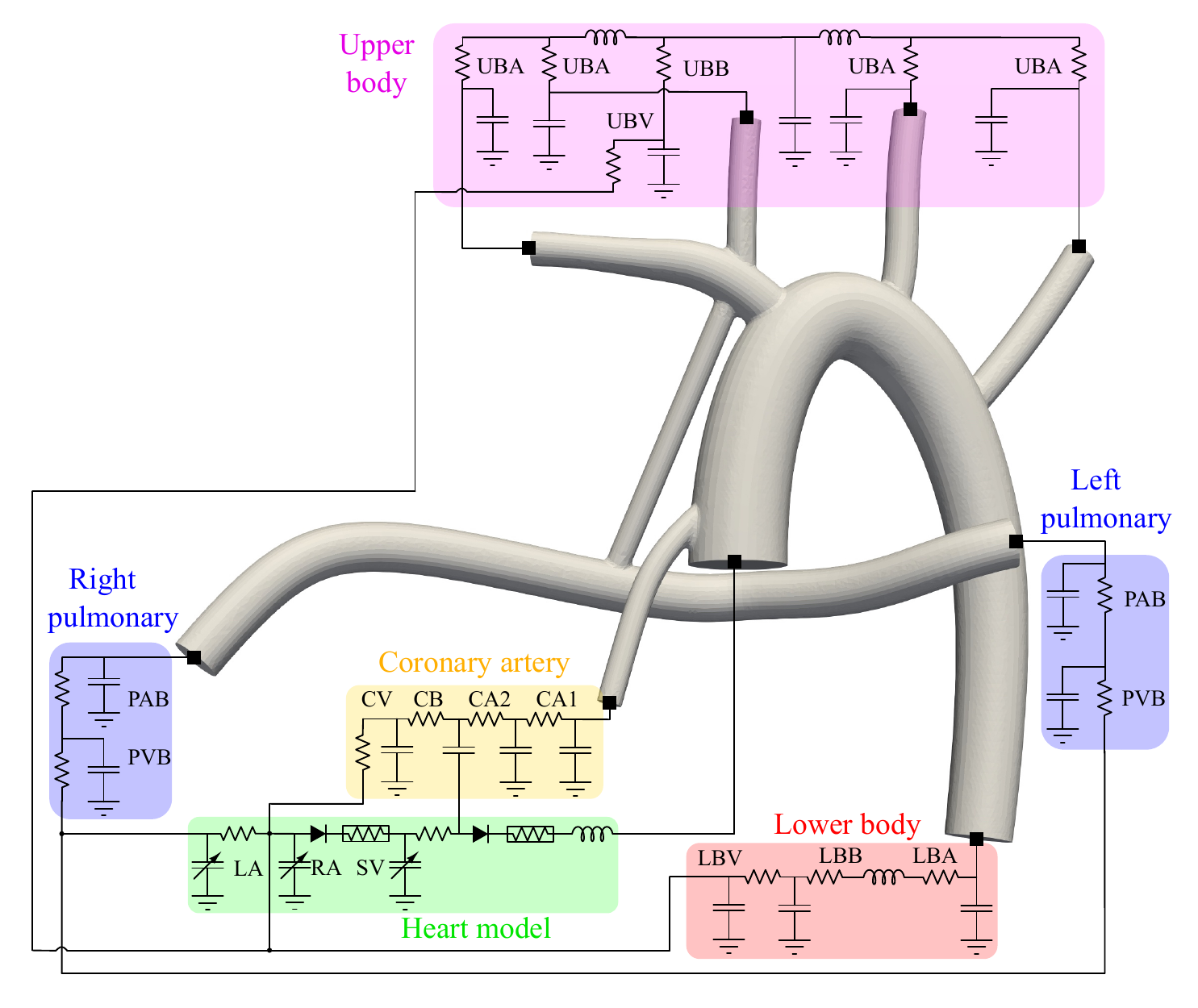}
    \caption{Lumped parameter network for Norwood operation. 
}
\label{fig:LPNBT}
\end{figure}

Figure \ref{fig:LPNBT} shows the LPN structure, composed of six main blocks, that is used in the current study (UBA: upper body artery, UBB: upper body bed, UBV: upper body vein, 
LBA: lower body artery, LBB: lower body bed, LBV: lower body vein, 
PAB: pulmonary artery bed, PAV: pulmonary artery vein, 
CA: coronary artery, CB: coronary bed, CV: coronary vein, 
LA: left atrium, RA: right atrium, and SV: single ventricle).

As observed in this figure, outside of the heart, the viscous dissipation, vessel wall distensibility, and flow inertia are modeled via linear resistors, capacitors, and inductors. Table \ref{table:LPNparam} lists all the parameters of the LPN that appear in Fig. \ref{fig:LPNBT} as well as their values and units.

Due to the turbulence effects, the pressure drop across the heart valves is modeled by two non-linear resistors. The pressure of heart chambers is modeled as the sum of an active and a passive components, which for the ventricles takes the form
\begin{equation}
    P_v = A_v \left[ E_{v_1} (V_v -V_{v_u}) + E_{v_2} (V_v -V_{v_u})^2 \right] + P_{v_0}\left(e^{K_v(V_v-V_{v_u})}-1\right),
    \label{eqn:P_vent}
\end{equation}
and for the atrium takes the form
\begin{equation}
    P_a = A_a E_{a} (V_a -V_{a_u}) + P_{a_0}\left(e^{K_a(V_a-V_{a_u})}-1\right).
    \label{eqn:P_atri}
\end{equation}
In the above equations, $A_v$ and $A_a$ are sinusoidal functions with non-zero values during ventricle and atrium contraction, respectively. The constant value of remaining parameters, e.g., $E_{v_1}, V_v, P_{a_0},$ and $K_a$, is reported in Table \ref{table:LPNparam}.


\begin{center}
\renewcommand*{\arraystretch}{1}
\begin{longtable}{ p{2cm} p{2cm} p{2cm} p{2.2cm} } 
\label{table:LPNparam} \\
\caption{Values of LPN parameters } \label{tab:long} \\

\hline \multicolumn{1}{l}{\textbf{Block}} & \multicolumn{1}{l}{\textbf{Parameters}} & \multicolumn{1}{l}{\textbf{Values}} & \multicolumn{1}{l}{\textbf{Unit}} \\ \hline 
\endfirsthead

\multicolumn{4}{c}%
{{\bfseries \tablename\ \thetable{} -- continued from previous page}} \\
\multicolumn{1}{l}{\textbf{Block}} & \multicolumn{1}{l}{\textbf{Parameters}} & \multicolumn{1}{l}{\textbf{Values}} & \multicolumn{1}{l}{\textbf{Unit}}\\ \hline 
\endhead

\hline \multicolumn{4}{r}{{Continued on next page}} \\ \hline
\endfoot

\hline \hline
\endlastfoot

Upper body & $R_{\text{UBA}}$ & $28.0899$ & mmHg.s/ml \\
    & $C_{\text{UBA}}$ & $0.04430$ & ml/mmHg \\
    & $L_{\text{UBA}}$ & $0.02138$ & mmHg.s\textsuperscript{2}/ml \\
    & $R_{\text{UBB}}$ & $0.64510$ & mmHg.s/ml \\
    & $C_{\text{UBB}}$ & $0.15515$ & ml/mmHg \\
    & $R_{\text{UBV}}$ & $0.16529$ & mmHg.s/ml \\
    & $C_{\text{UBV}}$ & $2.03945$ & ml/mmHg \\
Lower body & $R_{\text{LBA}}$ & $7.02239$ & mmHg.s/ml \\
    & $C_{\text{LBA}}$ & $0.07758$ & ml/mmHg \\
    & $L_{\text{LBA}}$ & $0.01069$ & mmHg.s\textsuperscript{2}/ml \\
    & $R_{\text{LBB}}$ & $0.64510$ & mmHg.s/ml \\
    & $C_{\text{LBB}}$ & $0.07758$ & ml/mmHg \\
    & $R_{\text{LBV}}$ & $0.16529$ & mmHg.s/ml \\
    & $C_{\text{LBV}}$ & $2.03945$ & ml/mmHg \\
Pulmonary & $R_{\text{PAB}}$ & $0.83376$ & mmHg.s/ml \\
    & $C_{\text{PAB}}$ & $0.02039$ & ml/mmHg \\
    & $R_{\text{PVB}}$ & $0.02194$ & mmHg.s/ml \\
    & $C_{\text{PVB}}$ & $0.44375$ & ml/mmHg \\
Coronary & $R_{\text{CA1}}$ & $10.6739$ & mmHg.s/ml \\
    & $C_{\text{CA1}}$ & $1.9435\times10^{-3}$ & ml/mmHg \\
    & $R_{\text{CA2}}$ & $10.6739$ & mmHg.s/ml \\
    & $C_{\text{CA2}}$ & $5.1827\times10^{-3}$ & ml/mmHg \\
    & $R_{\text{CB}}$ & $21.3477$ & mmHg.s/ml \\
    & $C_{\text{CB}}$ & $7.7741\times10^{-3}$ & ml/mmHg \\
    & $R_{\text{CV}}$ & $10.6739$ & mmHg.s/ml \\
    & $C_{\text{CV}}$ & $0.05\times10^{-3}$ & ml/mmHg \\
Heart & $E_{v_1}$ & $18.5$ & mmHg/ml \\
    & $E_{v_2}$ & $-0.042$ & mmHg/ml$^2$ \\
    & $V_{v_u}$ & $4.0$ & ml \\
    & $P_{v_0}$ & $0.9$ & mmHg \\
    & $K_{v}$ & $0.062$ & 1/ml \\
    & $E_{a}$ & $7.35$ & mmHg/ml \\
    & $V_{a_u}$ & $1$ & ml \\
    & $P_{a_0}$ & $0.17$ & mmHg \\
    & $K_{a}$ & $0.484$ & 1/ml \\
    & $\hat{R}_{tric}$ & $4\times10^{-5}$ & mmHg.s$^2$/ml$^2$ \\
    & $\hat{R}_{ao}$ & $4\times10^{-4}$ & mmHg.s$^2$/ml$^2$ \\
    & $R_{v}$ & $0.09$ & mmHg.s/ml \\
    & $R_{asd}$ & $0.001$ & mmHg.s/ml \\
    & $C_{ao}$ & $0.041555$ & ml/mmHg \\

\end{longtable}
\end{center}

\section{Additional material}\label{appendix:Extras}
\begin{table}[ht!]
\renewcommand{\arraystretch}{1.0}
\centering
\caption{Mean $\pm$ standard deviation of RBC deformation statistics}
\begin{tabular}{ p{3.4cm} p{3cm} p{3cm} p{3cm} }
\hline
\textbf{Parameter} & \textbf{2.5BT} & \textbf{4.0BT} & \textbf{2.5CS} \\
\hline
 Ave. areal strain & 0.9981 $\pm$ 0.0056 & 0.9984 $\pm$ 0.0060 & 0.9981 $\pm$ 0.0063 \\
 Ave. shear strain & 1.1557 $\pm$ 0.1464 & 1.1983 $\pm$ 0.1675 & 1.1630 $\pm$ 0.1546 \\
 Max. areal strain & 1.0136 $\pm$ 0.0123 & 1.0158 $\pm$ 0.0134 & 1.0143 $\pm$ 0.0134 \\
 Max. shear strain & 1.3200 $\pm$ 0.2402 & 1.4018 $\pm$ 0.2752 & 1.3357 $\pm$ 0.2414 \\
 Area$/A_0$ & 1.0007 $\pm$ 0.0055 & 1.0010 $\pm$ 0.0059 & 1.0007 $\pm$ 0.0063 \\
 Length$/L_0$ & 1.0109 $\pm$ 0.0987 & 1.0403 $\pm$ 0.1097 & 1.0185 $\pm$ 0.1013 \\
\hline
\end{tabular}
\label{table:PDFStat}
\end{table}

\end{appendices}

\printbibliography

\end{document}